\documentclass[12pt,preprint]{emulateapj} 
  
\usepackage{hyperref}

\shortauthors{Song et al.}

\newcommand{\sol}{$_{\odot}$}
\newcommand{\OIII}{[O\,{\sc iii}]}
\newcommand{\OII}{[O\,{\sc ii}]}
\newcommand{\NII}{[N\,{\sc ii}]}

\newcommand{\HII}{H\,{\sc ii}}
\newcommand{\HI}{H\,{\sc i}}

\begin{document}
\slugcomment{Accepted for publication in ApJ, June 12, 2014}

\title{The HETDEX Pilot Survey V: The Physical Origin of Lyman-alpha Emitters Probed by Near-infrared Spectroscopy}
\author{Mimi Song\altaffilmark{1},
	Steven L. Finkelstein\altaffilmark{1},
	Karl Gebhardt\altaffilmark{1},
	Gary J. Hill\altaffilmark{1},
	Niv Drory\altaffilmark{1},	
	Matthew L. N. Ashby\altaffilmark{2},
	Guillermo A. Blanc\altaffilmark{3},
	Joanna Bridge\altaffilmark{4,5},
	Taylor Chonis\altaffilmark{1},
	Robin Ciardullo\altaffilmark{4,5},
	Maximilian Fabricius\altaffilmark{6},
	Giovanni G. Fazio\altaffilmark{2},
	Eric Gawiser\altaffilmark{6},
	Caryl Gronwall\altaffilmark{4,5},
	Alex Hagen\altaffilmark{4,5},
	Jia-Sheng Huang\altaffilmark{2},
	Shardha Jogee\altaffilmark{1},
	Rachael Livermore\altaffilmark{1},
	Brett Salmon\altaffilmark{7},
	Donald P. Schneider\altaffilmark{4,5},
	S. P. Willner\altaffilmark{2},
	Gregory R. Zeimann\altaffilmark{4,5}}
\email{mmsong@astro.as.utexas.edu}

\altaffiltext{1}{Department of Astronomy, The University of Texas at Austin, 2515 Speedway, Stop C1400, Austin, TX 78712, USA}
\altaffiltext{2}{Harvard-Smithsonian Center for Astrophysics, 60 Garden St., Cambridge, MA 02138, USA}
\altaffiltext{3}{Observatories of the Carnegie Institution of Washington, 813 Santa Barbara Street, Pasadena, CA 91101, USA}
\altaffiltext{4}{Department of Astronomy \& Astrophysics, The Pennsylvania State University, University Park, PA 16802, USA}
\altaffiltext{5}{Institute for Gravitation and the Cosmos, The Pennsylvania State University, University Park, PA 16802, USA}
\altaffiltext{6}{Department of Physics and Astronomy, Rutgers University, Piscataway, NJ 08854, USA}
\altaffiltext{7}{Department of Physics and Astronomy, Texas A\&M University, College Station, TX 77843, USA}

\begin{abstract}

We present the results from a VLT/SINFONI and Keck/NIRSPEC near-infrared spectroscopic survey of 16 Lyman-alpha emitters (LAEs) at $z$ = 2.1 -- 2.5 in the COSMOS and GOODS-N fields discovered from the HETDEX Pilot Survey. We detect rest-frame optical nebular lines (H$\alpha$ and/or \OIII$\lambda$5007) for 10 of the LAEs and measure physical properties, including the star formation rate (SFR), gas-phase metallicity, gas-mass fraction, and Ly$\alpha$ velocity offset. We find that LAEs may lie below the mass-metallicity relation for continuum-selected star-forming galaxies at the same redshift. 
The LAEs all show velocity shifts of Ly$\alpha$ relative to the systemic redshift ranging between +85 and +296 km s$^{-1}$ with a mean of +180 km s$^{-1}$. This value is smaller than measured for continuum-selected star-forming galaxies at similar redshifts. The Ly$\alpha$ velocity offsets show a moderate correlation with the measured star formation rate (2.5$\sigma$), but no significant correlations are seen with the SFR surface density, specific SFR, stellar mass, or dynamical mass ($\lesssim$ 1.5$\sigma$). Exploring the role of dust, kinematics of the interstellar medium (ISM), and geometry on the escape of Ly$\alpha$ photons, we find no signature of selective quenching of resonantly scattered Ly$\alpha$ photons. However, we also find no evidence that a clumpy ISM is enhancing the Ly$\alpha$ equivalent width. Our results suggest that the low metallicity in LAEs may be responsible for yielding an environment with a low neutral hydrogen column density as well as less dust, easing the escape of Ly$\alpha$ photons over that in continuum-selected star-forming galaxies.

\end{abstract}

\section{Introduction}

Empirical relations among fundamental galaxy parameters provide stringent constraints on the physical processes driving galaxy evolution.  One such well-established scaling relation seen in nearby galaxies is the ``mass-metallicity relation'' (MZR).  Using $\sim$ 53,000 galaxies from the Sloan Digital Sky Survey (SDSS; \citealt{york00}), \citet{tremonti04} showed that there exists a tight correlation between stellar mass and gas-phase metallicity among local ($z \sim$ 0.1) star-forming galaxies (MZR), with a scatter of only about 0.1 dex. Subsequent studies using continuum-selected star-forming galaxies (e.g., Lyman-break galaxies) found that this MZR exists at redshifts up to $\sim$ 3.5 \citep{erb06a,maiolino08} 
and evolves smoothly with redshift, in that galaxies at higher redshift are more metal-poor than those at lower redshift at a given stellar mass. This is a record of the chemical enrichment history of galaxies, which is in principle governed by star formation and modulated by inflows of pristine gas and metal ejection by outflows (e.g., \citealt{dave11}).

In contrast to the typical star-forming galaxies selected by their ultraviolet (UV) continuum light (i.e., the ``dropout technique''; \citealt{steidel93}) which have been utilized for probing the MZR, another method commonly used to select high-redshift galaxies is via their strong Ly$\alpha$ emission lines.  Early studies of these Lyman-alpha emitters (LAEs) using broad-band spectral energy distribution modeling reported that LAEs appeared to be predominantly young, low-mass, and low in dust extinction (e.g., \citealt{gawiser06,gawiser07,fin07,nilsson07,gronwall07,ouchi08}), although more recent works have reported that the LAE population does contain a subset of more evolved systems containing a moderate amount of dust (e.g., \citealt{fin09a,nilsson09,pentericci09, guaita11}). It is interesting that we are \textit{observing} strong Ly$\alpha$ emission from dusty systems despite the fact that in a static homogeneous medium the resonant nature of Ly$\alpha$ should make its escape practically impossible if even a small amount of dust exists (e.g., \citealt{charlot93}).

One way to enable the escape of Ly$\alpha$ even with the presence of dust is the existence of outflows. Galactic-scale starburst-driven winds have been observed to be ubiquitous in both local starbursts and high redshift star-forming galaxies (e.g., \citealt{heckman90,shapley03,martin05}).  This bulk motion of neutral gas can in principle help the escape of Ly$\alpha$ photons by shifting the Ly$\alpha$ photons out of resonance and reducing the number of resonant scatterings that they undergo before escape.  For example, numerical modeling of Ly$\alpha$ radiative transfer in a simplified expanding shell scenario predicts that Ly$\alpha$ will preferentially escape redshifted with respect to the systemic redshift (which can be measured from nebular lines such as H$\alpha$ or \OIII\  originating from the \HII\ regions), as the red wing of the Ly$\alpha$ line can escape by backscattering off the receding side of the expanding shell \citep{verhamme06, verhamme08,schaerer11}.  This prediction can explain (although not exclusively) what is found by observational studies using continuum-selected star-forming galaxies with Ly$\alpha$ emission (e.g., \citealt{shapley03,steidel10}), which find that Ly$\alpha$ is commonly redshifted  by $\sim$ 450 km s$^{-1}$ on average, and up to $\sim$ 800 km s$^{-1}$.

Another scenario proposed in addition to kinematics to enhance the chance of the escape of Ly$\alpha$ photons is a multi-phase ISM with an inhomogeneous dust distribution \citep{neufeld91,hansen06}.  In this scenario the dust is confined in clumps of neutral gas, and Ly$\alpha$ photons suffer little dust attenuation compared to the continuum photons by resonantly scattering off of the clump surfaces, and thus have a higher probability of escape.  This was first observationally studied by \citet{fin08,fin09a}, and these studies along with subsequent investigations (e.g., \citealt{blanc11}) support a ``quasi-clumpy'' ISM, where dust does not preferentially attenuate Ly$\alpha$ more than the UV continuum.

Another observable that appears to be an important factor in governing the presence of Ly$\alpha$ is the metallicity.  This property has been, however, relatively poorly understood because the metallicity inferred from broad-band imaging data is highly uncertain, and near-infrared (near-IR) spectroscopy is required to directly measure the metallicity using rest-frame optical nebular lines for galaxies at significant redshift.  In this context, it is interesting that there have been recent reports that LAEs at low redshift ($z \sim$ 0.3) may lie below the empirical relation between stellar mass and metallicity that holds for typical star-forming galaxies at the same epoch \citep{cowie10,fin11a}. At higher redshift, \citet{fin11b} found a massive ($M_{*}\sim10^{10}$ M\sol) but significantly more metal-poor LAE at $z \sim$ 2.3 than continuum-selected star-forming galaxies with the same stellar mass.  \citet{nakajima13} also reported two LAEs at similar redshift that are offset towards lower metallicity in the stellar mass -- gas-phase metallicity plane.

Obtaining a better understanding of how Ly$\alpha$ emission escapes, and how LAEs are different from continuum-selected star-forming galaxies with little or no Ly$\alpha$ emission thus requires a large suite of datasets, including multi-wavelength imaging (to derive stellar mass and dust attenuation), optical spectroscopy (to measure Ly$\alpha$), and near-infrared spectroscopy (to measure the metallicity and systemic redshift).
However, only a few LAEs at $z \gtrsim$ 2 have measured metallicities \citep{fin11b,nakajima13,guaita13}.  This is primarily due to observational difficulties: the bright night sky still poses difficulties for near-infrared spectroscopy, although new instruments are rapidly becoming more sensitive.  Additionally, most known LAEs are discovered via the narrowband imaging technique (e.g., \citealt{cowie98,rhoads00}), which probes a narrow redshift range, and thus a relatively small volume, to deep line flux limits. These studies discovered numerous faint LAEs but not many bright ones suitable for detailed spectroscopic observation.  An integral field unit survey can probe a large volume and is thus able to provide a bright LAE sample for near-IR spectroscopic observations.  In this study, we utilize LAEs discovered from a blind integral-field unit survey, the Hobby Eberly Telescope Dark Energy Experiment (HETDEX) Pilot Survey (HPS), which discovered 104 LAEs at 1.9 $< z <$ 3.8 from a comoving volume of $\sim$ 10$^6$ Mpc$^3$ over a 169 arcmin$^2$ area \citep{adams11}; several times larger than a typical narrowband LAE survey (e.g., $\sim$ 1$\times$10$^5$ Mpc$^3$ by \citealt{gronwall07} and \citealt{guaita10}, $\sim$ 3$\times$10$^5$ Mpc$^3$ by \citealt{nilsson09}).

Here we present a near-IR spectroscopic study of LAEs at $z =$ 2.1 -- 2.5 discovered from the HETDEX Pilot Survey to directly measure their metallicities using rest-frame optical emission lines.  We will use previous Ly$\alpha$ spectroscopy and multiwavelength imaging data to also investigate ISM kinematics by comparing the redshift inferred from Ly$\alpha$ to that from H$\alpha$ and/or \OIII$\lambda$5007 and explore correlations between the velocity offset of the Ly$\alpha$ line with other physical properties. Also, the flux ratio of Ly$\alpha$ to optically thin nebular lines (e.g., H$\alpha$) and the derived dust extinction will provide insights into their ISM geometry.
These data will allow an unprecedented exploration into the physical properties of LAEs, which has previously been probed mainly through spectral energy distribution fitting techniques, 
as well as the nature of LAEs by enabling the comparison with continuum-selected star-forming galaxies with comparable physical properties (e.g., stellar mass) at the same redshift.

In Section \ref{sec:data}, we describe our near-IR spectroscopic observations and data reduction for our sample of LAEs at $z =$ 2.1 -- 2.5.  Combining these data with ancillary datasets, we present our measurement of physical properties in Section \ref{sec:prop}.  In Section \ref{sec:discussion}, we explore the mass--metallicity relation, study the role of dust, ISM kinematics, and geometry on the escape of Ly$\alpha$ photons, 
and discuss the nature of LAEs. Lastly, we summarize our results in Section \ref{sec:conclusion}.  Throughout the paper, we assume a standard $\Lambda$CDM cosmology with $H_0$ = 70 km s$^{-1}$ Mpc$^{-1}$, $\Omega_M$ = 0.3, and $\Omega_{\Lambda}$ = 0.7.  Magnitudes are in AB magnitude system \citep{oke83}, and a Salpeter initial mass function \citep{salpeter55} is assumed thoughout the paper unless otherwise specified.


\section{Data}\label{sec:data}

\subsection{Sample Selection}\label{sec:sample}

HETDEX \citep{hill08a} is a blind integral-field spectrograph (IFS) survey, which, starting in 2014, will probe dark energy using baryonic acoustic oscillations traced by LAEs at $z = 1.9-3.8$. Our sample is selected from the 104 $1.9 < z < 3.8$ LAEs spectroscopically-discovered from the HETDEX pilot survey \citep{adams11}, which utilized a prototype IFS (the Mitchell spectrograph; formerly called VIRUS-P; \citealt{hill08b}) mounted on the 2.7-m Harlan J. Smith Telescope at the McDonald Observatory. 
The data used to select the LAE sample have a resolution FWHM of $\sim$ 5\AA, 
corresponding to FWHM of $\sim$ 400 km s$^{-1}$ for Ly$\alpha$ at the median redshift $z$ = 2.3 of our sample in this study.

For our follow-up observations, we selected LAEs from the HETDEX Pilot Survey sample that have bright Ly$\alpha$ emission ($f_{\rm Ly\alpha} >$ 10$^{-16}$ erg cm$^{-2}$ s$^{-1}$) and a redshift such that the H$\alpha$ line falls in the $K$-band (2 $\lesssim z \lesssim$ 2.6).  Using these criteria, we selected 16 LAEs (15 in the COSMOS and 1 in the GOODS-N field, including HPS\,194 and HPS\,256, which were originally published in \citet{fin11b} but re-analyzed and included in this study) at $z = 2.1-2.5$ ($\langle z \rangle = 2.33$) as suitable for our study.  The range of $r^+_{\rm AB}$ magnitude of LAEs in our sample is 22.9 -- 25.4, bright enough to satisfy the selection criteria for $z \sim$ 2 BX galaxies ($z \sim 2$ counterpart of Lyman-break galaxies at $z \sim 3$) in the apparent magnitude cut (\textit{R} $<$ 25.5; \citealt{adelberger04}) and rest-frame UV color probed by $(\textit{g}^{+}-\textit{r}^{+})$. About half of our sample, however, have bluer $(\textit{u}^{+}-\textit{g}^{+})$ colors than the BX criteria. At the known redshifts, Ly$\alpha$ emission would contribute flux to the \textit{u}$^{+}$-band.
Ly$\alpha$ luminosities of our LAEs range from log(L$_{\rm Ly\alpha}$/erg s$^{-1}$)= 42.8 -- 43.4, about ten times brighter than the median Ly$\alpha$ luminosity of log(L$_{\rm Ly\alpha}$/erg s$^{-1}$)= 42.1, or a few times than the characteristic luminosity of log(L$^*$/erg s$^{-1}$)= 42.3 of the Ly$\alpha$ luminosity function \citep{ciardullo12}, at $z \sim$ 2.1 of narrowband survey by \citet{guaita10}.

\subsection{Observations} \label{sec:observation}

\subsubsection{SINFONI} 
 
Observations for 10 of the LAEs in our sample were performed with the Spectrograph for Integral Field Observations in the Near Infrared (SINFONI; \citealt{eisenhauer03}) mounted on the Very Large Telescope (VLT) UT4 between 2010 December and 2012 February. Each object in the sample was observed with both the \textit{H}- ($\lambda$ = 1.45--1.85 $\mu$m) and \textit{K}-band ($\lambda$ = 1.95--2.45 $\mu$m) gratings, with spectral resolutions of $R \sim$ 3,000 and 4,000, respectively. The observations were conducted in seeing-limited mode, with a point-spread function (PSF) full-width at half-maximum (FWHM) range of 0\farcs4--2\farcs1 in the NIR (mean $=$ 1\farcs0).  This corresponds to a physical of 3.3--17.5 kpc (mean $=$ 8.3 kpc) at $z \sim 2.3$, which is much larger than the typical size of $\lesssim$ 2 kpc in rest-frame UV for LAEs at similar redshifts (e.g., \citealt{bond12}), thus none of our LAEs are spatially resolved. Considering the large uncertainties in the positions of our sample 
inherited from the large fiber diameter of the Mitchell Spectrograph (4\farcs1) used for Ly$\alpha$ detection, we used the 250 mas pixel$^{-1}$ scale to produce a field of view (FOV) of 8\arcsec $\times$ 8\arcsec.

Each observing block (OB) typically consisted of 16 $\times$ 150 s individual exposures, with 3\arcsec\  on-source dithering of an ABAB pattern (i.e., our targets were always kept in the FOV). Depending mainly on the expected H$\alpha$ or \OIII\ flux from the observed Ly$\alpha$ flux and broad-band fluxes, 1 -- 3 OBs were obtained for each object.  The mean integration time was $\sim$ 70 minutes for the \textit{H}-band and $\sim$ 90 minutes for the \textit{K}-band.
For telluric absorption correction, as well as flux calibration, we observed one to six telluric standard stars with spectral types of B2V -- B8V in each filter per night with an object -- sky -- object -- sky pattern.  

\subsubsection{NIRSPEC}

We observed with the Near Infrared Spectrograph (NIRSPEC; \citealt{mclean98}) on the Keck II 10-m telescope on 15 and 17 of April 2011 (UT).  Two LAEs from the HETDEX Pilot Survey (HPS\,251 and HPS\,306) were observed on the first night, while on the second night we observed three LAEs (HPS\,269, HPS\,286, and HPS\,419).  Each LAE was observed in the \textit{K}-band with a spectral resolution of $R \sim$ 1,500, using six $\times$ 15 minute exposures, dithering with an ABBAAB pattern for the removal of night-sky lines.  We obtained flat field and arc lamp calibrations in the afternoon, and we observed 1--2 telluric standards each night.

Table \ref{tab:target} lists the total on-source integration time in each bandpass obtained for our LAEs, as well as their celestial coordinates, Ly$\alpha$ flux, Ly$\alpha$ equivalent width (EW), Ly$\alpha$ redshift.


\subsection{Data Reduction} \label{sec:data_reduction}

\begin{deluxetable*}{cccllccc}
\tabletypesize{\scriptsize}
\tablecaption{\label{tab:target} Summary of target LAEs}
\tablewidth{0pt}
\tablehead{
\colhead{Object} & \colhead{R.A.\tablenotemark{a}} & \colhead{Dec.\tablenotemark{a}} &  \colhead{F$_{\rm Ly\alpha}$} & 
\colhead{EW$_{\rm Ly\alpha}$\tablenotemark{b}}
& \colhead{$z_{\rm Ly\alpha}$\tablenotemark{c}} & \colhead{EXPTIME ($H$)\tablenotemark{d}} & \colhead{EXPTIME ($K$)\tablenotemark{d}}\\
\colhead{$ $} & \colhead{(J2000)} & \colhead{(J2000)} & \colhead{(10$^{-17}$ erg s$^{-1}$ cm$^{-2}$)}& \colhead{(\AA)} & \colhead{$ $} & \colhead{(min)} & \colhead{(min)}\\
}
\startdata
\smallskip
VLT/SINFONI &&&&&& \\
HPS\,145& 10:00:06.26 & 02:13:10.9 &  ~~~~~ 84.0 $^{+8.1}_{-14.8}$ & 155 $_{-22}^{+35}$ & 2.1765 $\pm$ 0.0004 & \phantom{1}40 & \phantom{1}40 \\
HPS\,160& 10:00:08.61 & 02:17:38.6 &  ~~~~~ 17.1 $^{+6.4}_{-10.5}$ & 698 $_{-334}^{+1000}$ & 2.4361 $\pm$ 0.0004 & \phantom{1}40 & 160 \\
HPS\,182&  10:00:12.33 & 02:14:16.0 & ~~~~~  25.6 $^{+5.2}_{- 5.8}$ & 240 $_{-50}^{+56}$ & 2.4352 $\pm$ 0.0004 & 120 & \phantom{1}80 \\ 
HPS\,183& 10:00:12.44  & 02:17:53.0  &  ~~~~~ 27.8 $^{+11.3}_{-23.1}$  & 206 $_{-85}^{+173}$ & 2.1638 $\pm$ 0.0004 & \phantom{1}70 & \phantom{1}65 \\
HPS\,189& 10:00:13.11 & 02:18:56.3  &  ~~~~~ 12.9 $^{+6.7}_{- 8.7}$ &  \phantom{1}90 $_{-47}^{+61}$ & 2.4531 $\pm$ 0.0004 & \phantom{1}80 & \phantom{1}80 \\
HPS\,194& 10:00:14.16 & 02:14:28.3  &  ~~~~~ 61.0 $^{+ 4.3 }_{-4.9}$  & 175 $_{-18}^{+18}$ & 2.2897 $\pm$ 0.0004 & \phantom{1}80 & 120 \\
HPS\,223& 10:00:18.56 & 02:14:59.8  &  ~~~~~ 39.0 $^{+ 9.4}_{-11.5}$ & 268 $_{-87}^{+157}$ & 2.3071 $\pm$ 0.0004 & \phantom{1}80 & \phantom{1}80 \\
HPS\,263&  10:00:29.06 & 02:14:09.2  &  ~~~~~ 24.1 $^{+ 7.7}_{- 8.0}$  &  \phantom{1}52 $_{-17}^{+19}$ & 2.4338 $\pm$ 0.0004 & \phantom{1}60 & 100 \\
HPS\,313& 10:00:40.78 & 02:18:23.6  & ~~~~~  25.1 $^{+10.1}_{-12.4 }$ &  \phantom{1}23 $_{-9}^{+11}$ & 2.0989 $\pm$ 0.0004 & \phantom{1}65 & \phantom{1}80 \\
HPS\,318&  10:00:44.13 & 02:15:58.9 &  ~~~~~ 30.3 $^{+11.1}_{- 8.9}$  &  \phantom{1}70 $_{-26}^{+21}$ & 2.4574 $\pm$ 0.0004 & \phantom{1}80 & \phantom{1}95 \\
 &&&&&& \\
\smallskip
Keck/NIRSPEC &&&&&& \\
\phantom{1}HPS\,194\tablenotemark{e}& 10:00:14.16 & 02:14:28.3  &  ~~~~~ 61.0 $^{+ 4.3 }_{-4.9}$  & 175 $_{-18}^{+18}$ & 2.2897 $\pm$ 0.0004 & \phantom{1}90 & \phantom{1}60  \\
HPS\,251& 10:00:27.28 & 02:17:31.3 &  ~~~~~ 45.0 $^{+ 11.6}_{-13.7}$ & 208 $_{-54}^{+64}$ & 2.2866 $\pm$ 0.0004 & \phantom{1}--- & \phantom{1}90 \\
\phantom{1}HPS\,256\tablenotemark{e}& 10:00:28.25 & 02:17:58.4 &  ~~~~~ 31.4 $^{+ 6.5 }_{ -9.3}$ & 185 $_{-38}^{+56}$ &  2.4922 $\pm$ 0.0004 & \phantom{1}20 & \phantom{1}60 \\   
HPS\,269& 10:00:30.60 & 02:17:43.9 & ~~~~~ 13.9 $^{+ 4.4}_{-2.9}$ &  \phantom{1}95 $^{+39}_{-24}$ & 2.5672 $\pm$ 0.0004 & \phantom{1}--- & \phantom{1}90 \\  
HPS\,286& 10:00:33.91 & 02:13:17.9 &  ~~~~~ 28.4 $^{+ 4.3}_{-8.1}$ &  \phantom{1}79 $_{-12}^{+23}$ & 2.2307 $\pm$ 0.0004 & \phantom{1}--- & \phantom{1}90 \\
HPS\,306& 10:00:39.61 & 02:13:38.6 &  ~~~~~ 38.3 $^{+ 9.2}_{-5.8}$ &  \phantom{1}85 $_{-21}^{+13}$ & 2.4405 $\pm$ 0.0004 & \phantom{1}--- & \phantom{1}90 \\
HPS\,419& 12:36:50.04 & 62:14:00.7 & ~~~~~ 24.4 $^{+ 3.3}_{-5.1}$ & \phantom{1}72 $^{+19}_{-18}$ & 2.2363 $\pm$ 0.0004 & \phantom{1}--- & \phantom{1}90 \\ 
\enddata
\tablenotetext{a}{R.A. \& Dec. of Ly$\alpha$ emission. Taken from \citet{adams11}, along with column 4.}
\tablenotetext{b}{Rest-frame Ly$\alpha$ EWs calculated from the observed Ly$\alpha$ flux and the mean best-fit model continuum from SED fitting in a $\Delta \lambda_{\rm rest}$= 100 \AA\ region at wavelengths redward of the Ly$\alpha$ line (see Section \ref{sec:ew}). For LAEs not detected in our NIR observation  (i.e., HPS\,145, HPS\,160, HPS\,223, HPS\,263, HPS\,269, HPS\,419), we present values from \citet{adams11}.}
\tablenotetext{c}{Ly$\alpha$ redshift in the heliocentric frame. Corrected for n$_{\rm atm}$ from the values in \citet{adams11}. See \S 3.1 in \citet{chonis13} for details.}
\tablenotetext{d}{Total on-source integration time.}
\tablenotetext{e}{Originally published in \citet{fin11b}, but re-analyzed and included in our analysis. }
\end{deluxetable*}

\subsubsection{SINFONI}\label{sec:data_reduction_sinfo}

Basic data reduction was performed using the SINFONI pipeline and \textit{Gasgano} application package.\footnote{\url{www.eso.org/sci/software/gasgano/}}\ This data processing includes dark subtraction, flat fielding, distortion correction, cosmic ray rejection, sky subtraction, wavelength calibration, and cube reconstruction. Sky background was subtracted using two consecutive science frames (for our LAE samples) or sky frames (for telluric standards).  Residual sky lines were removed by modeling a scaling function at each wavelength that was used to generate a modified sky cube as described in \citet{davies07}.  Cosmic rays were eliminated by first removing pixels with the highest 5\% and the lowest 5\% values and then applying 10 iterative 2$\sigma$ clippings (rejecting the highest $\sim$ 2.5\% and the lowest $\sim$ 2.5\%).  Data cubes were reconstructed for each OB, and then, additionally, for those where we could identify H$\alpha$ or \OIII\ emission, every possible combination of co-added data cubes were constructed using the spatial shifts determined from central positions in the smoothed H$\alpha$ or \OIII\ images.
These individual and co-added data cubes for each filter and object were examined as described in the next section, to maximize the signal-to-noise ratio (SNR) for further analysis.

Subsequent data reduction was conducted with in-house custom IDL codes. To extract the 1D spectrum from each cube, we utilized an aperture box of approximately 1.5$\times$ the seeing (PSF FWHM) on a side, which corresponds to a typical box size of 1\farcs4 $\times$ 1\farcs4.  For our LAEs, the center of the extraction box was determined by finding the peak position in a 3-pixel boxcar smoothed H$\alpha$ or \OIII\ image, starting at an initial estimate determined from visual inspection.  For our telluric standards, we determined the position of the extraction box by performing 2D Gaussian surface fitting.  Two boxes with the same size as the extraction box, located two times the extraction box size apart from the source in a direction perpendicular to the dithering, were used for additional residual sky subtraction.

Flux calibration and correction for telluric absorption were performed using standard star spectra as described in \citet{fin11b}.  Briefly, we first found a model stellar spectrum from the Kurucz library \citep{kurucz93} which has the same spectral type as the observed standard star.  Absorption features common in the model and standard spectra were removed by linear interpolation of the adjacent continuum.  We scaled the model to the flux-calibrated Two Micron All Sky Survey (2MASS; \citealt{skrutskie06}) \textit{H}- or \textit{K$_s$}-band flux, and the ratio of the 1D observed standard spectrum to the scaled model spectrum gives the calibration array.

In the case where there was more than one telluric standard observed in each night and filter, we first rejected outliers in the calibration arrays 
and utilized only those which were taken under similar seeing conditions with our sample. This ensures a more accurate aperture correction by adopting the same extraction box size for standards as that for our sample.

Errors for the final spectrum consist of a combination of photometric errors and the systematic error from the flux calibration.  First, we estimated photometric errors for each OB as follows: since the background sky is the dominant source of error, we started by extracting multiple independent (non-overlapping) sky regions selected within the FOV of the cube excluding the region where the object was extracted.  In the object OBs, individual frames with 3\arcsec\ ABAB dithering were stacked, thus the overlapping central ($\sim$ 8\arcsec $\times$ 5\arcsec) regions, where the source spectrum is also extracted, are less noisy. When at least 10 extraction boxes are possible, we limited the noise estimation to only these central regions, but we utilized the entire image otherwise.
Using these extracted off-source spectra, we calculated flux uncertainty at each wavelength as the standard deviation of pixel counts to create the resultant error spectrum.  Error spectra for standard stars were measured in the same manner but using the dedicated sky frames. Systematic errors at each wavelength due to the flux calibration were estimated as the standard deviation of the calibration arrays used for flux calibration. The resulting final error spectrum is the quadrature sum of photometric errors of object and systematic errors of flux calibration. The uncertainties in the final error spectrum, however, are found to be dominated by the photometric errors of the object ($>$ 99\%). 

\begin{figure*}
  \epsscale{0.38}
  \plotone{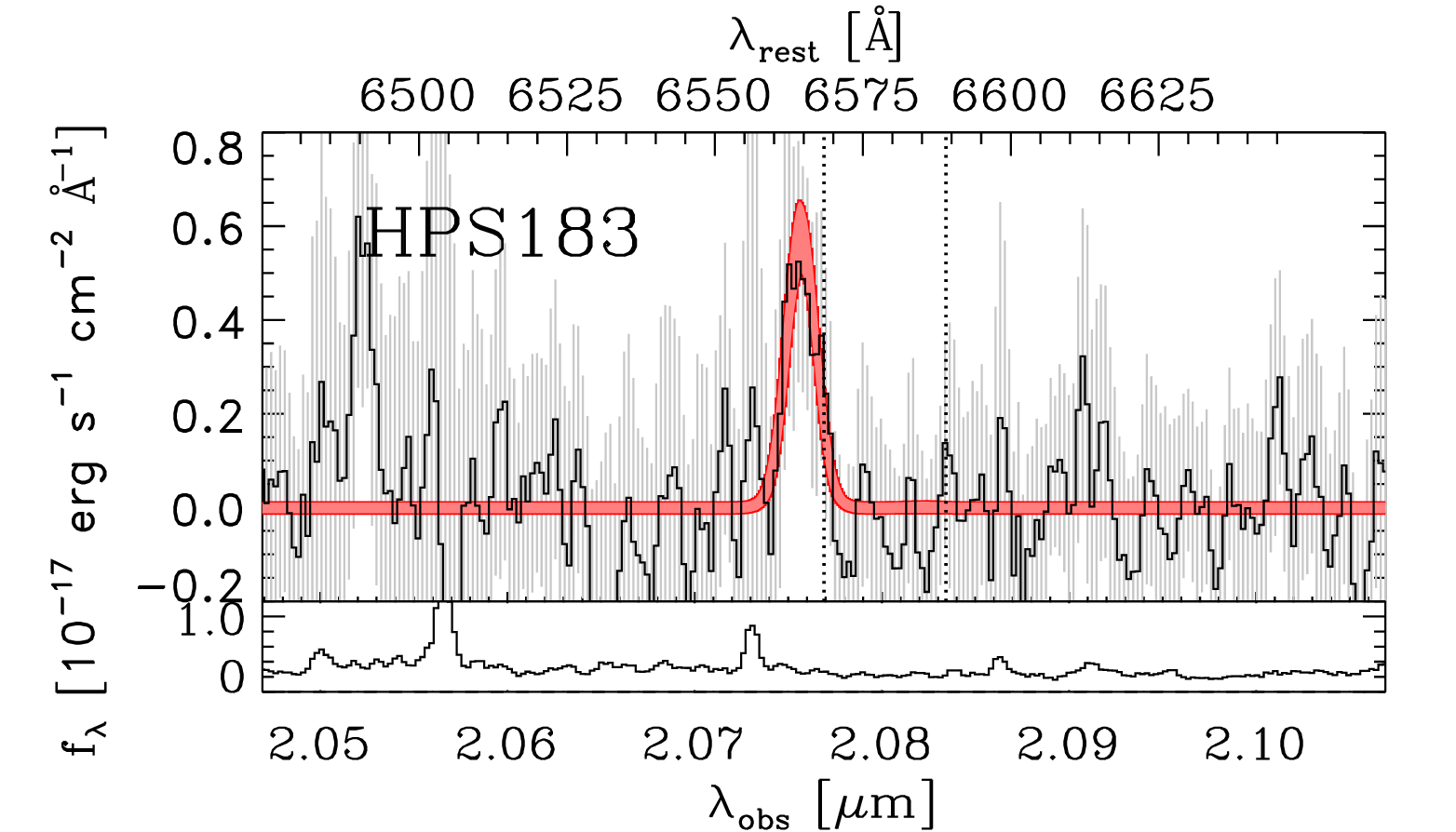}\plotone{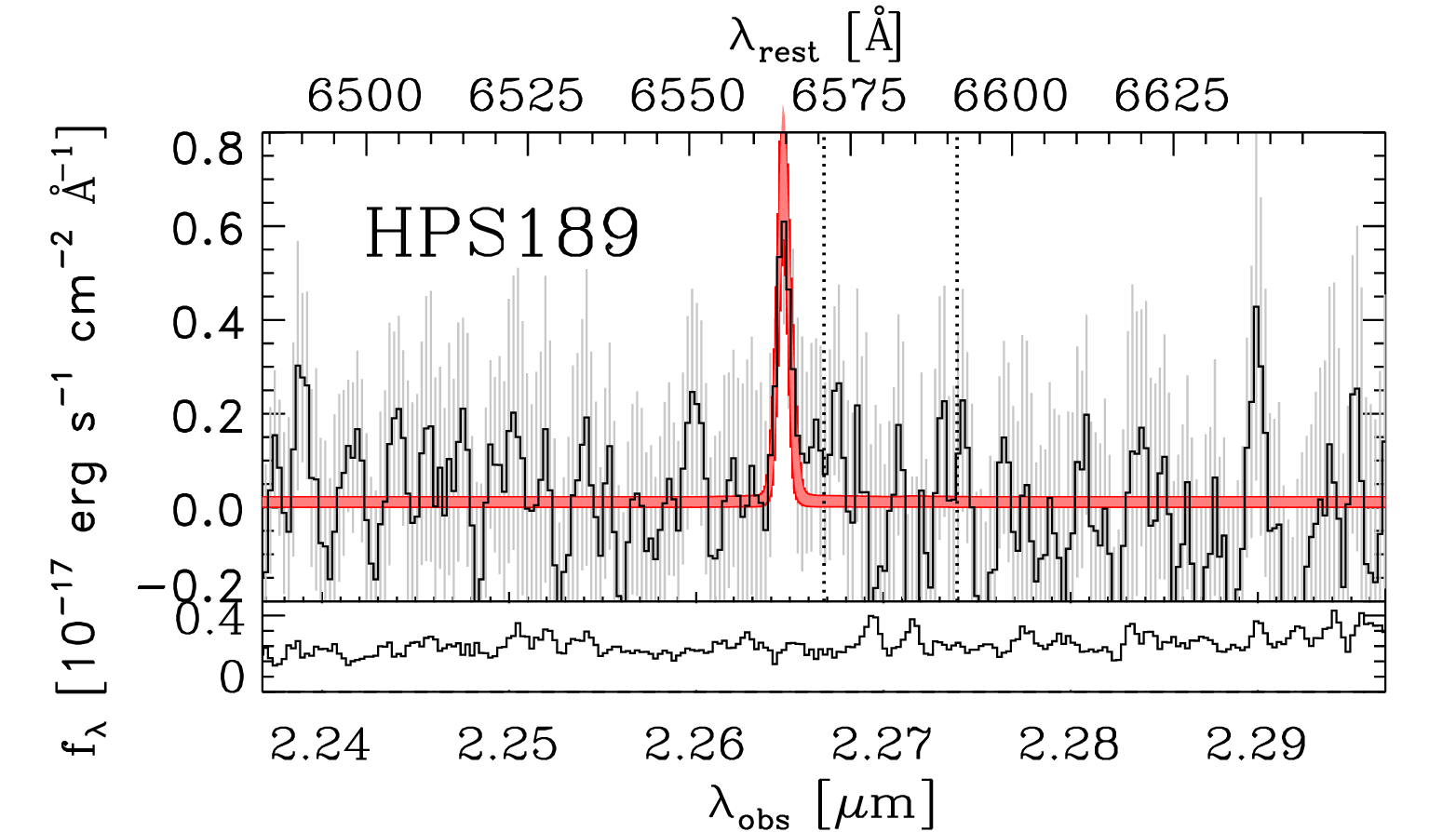}\plotone{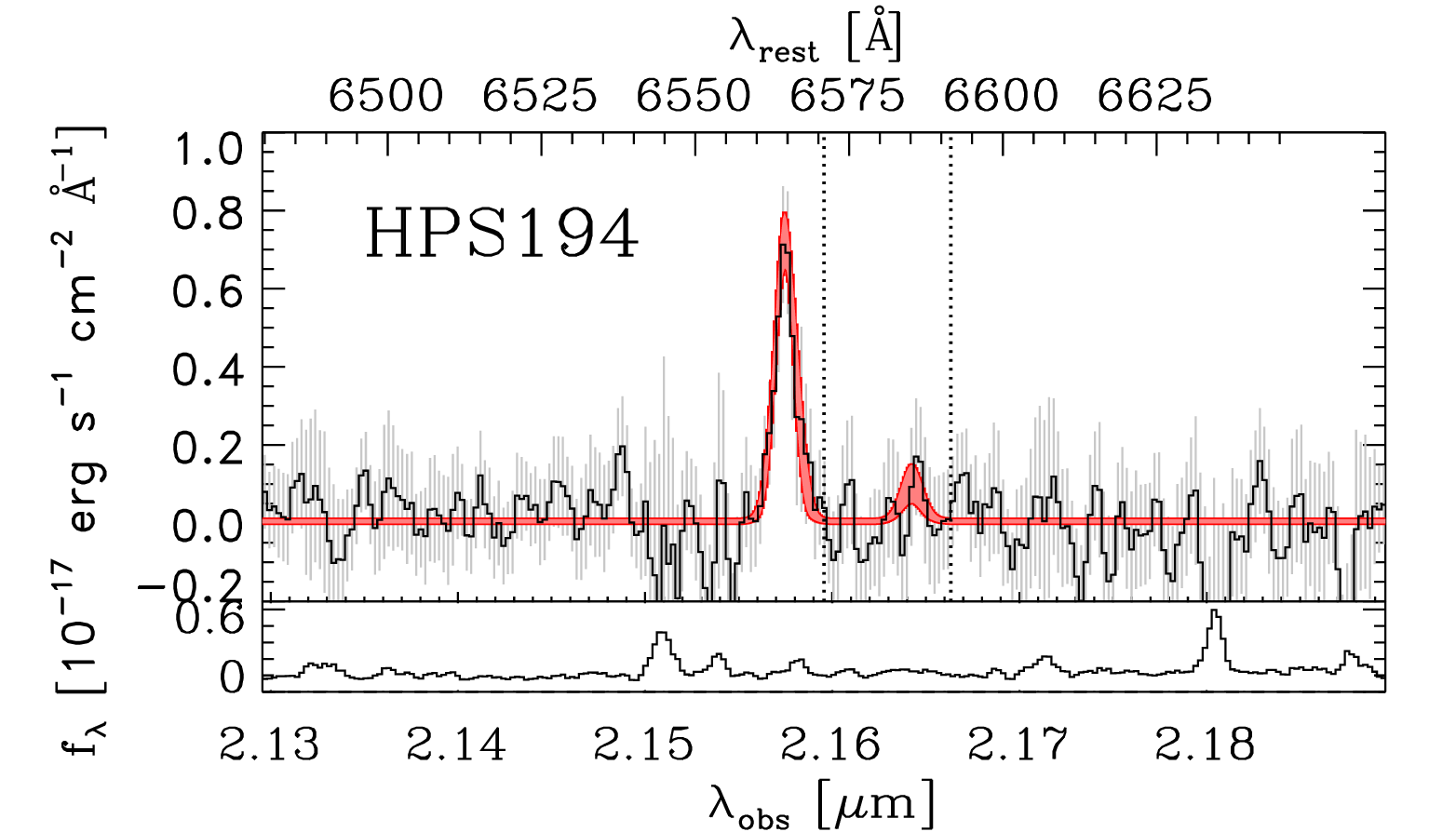}
  \plotone{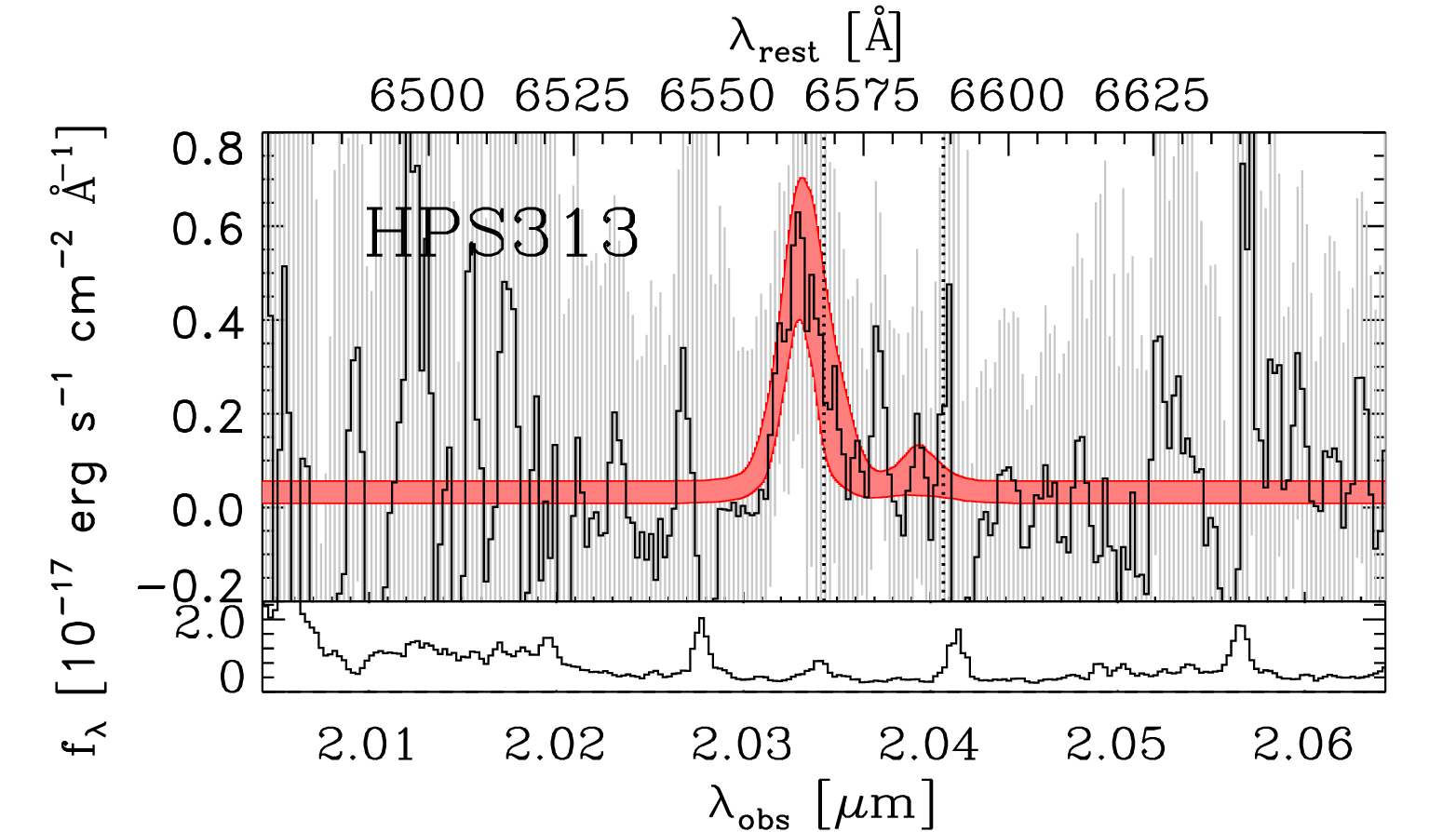}\plotone{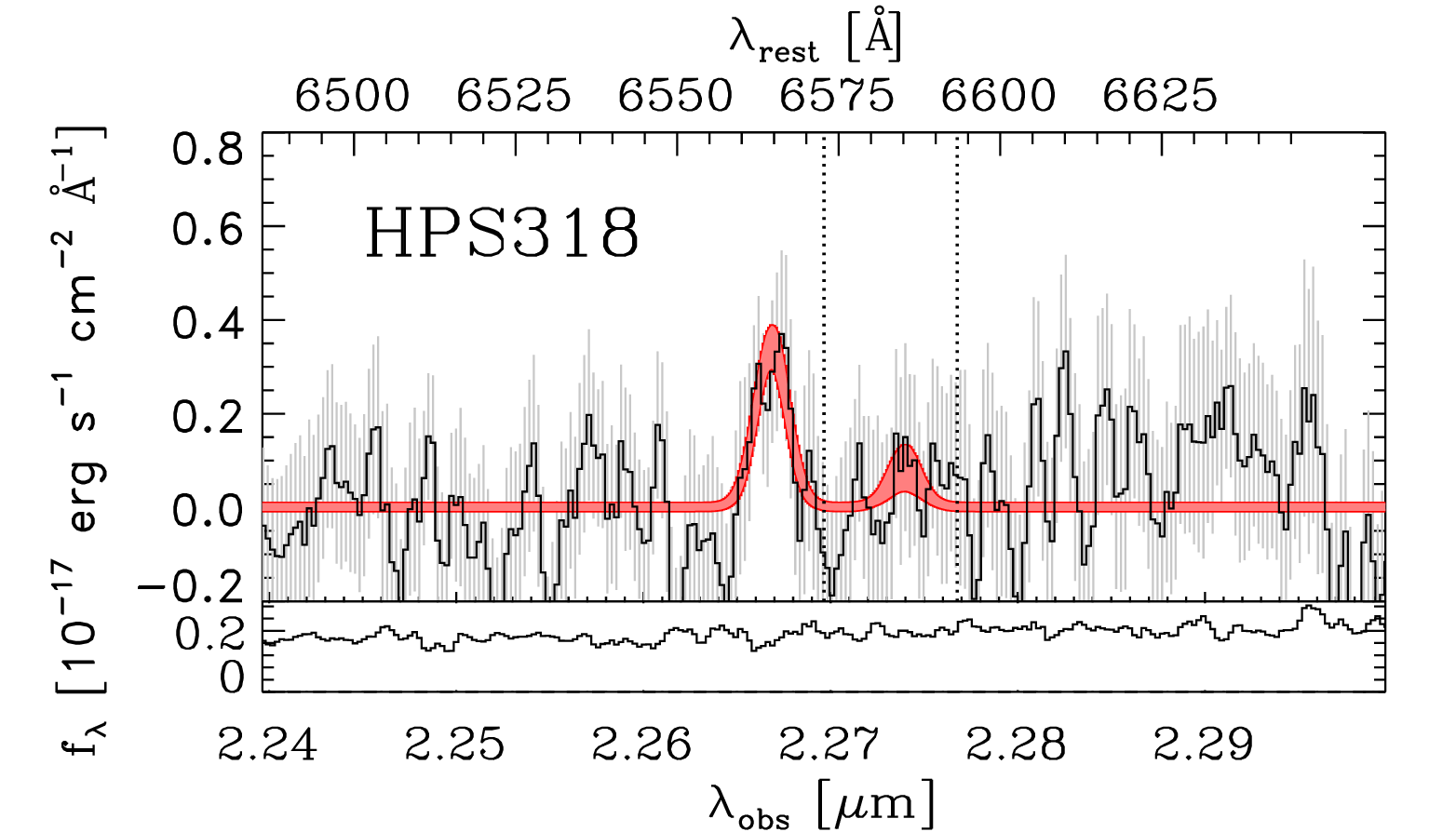}\\
  \plotone{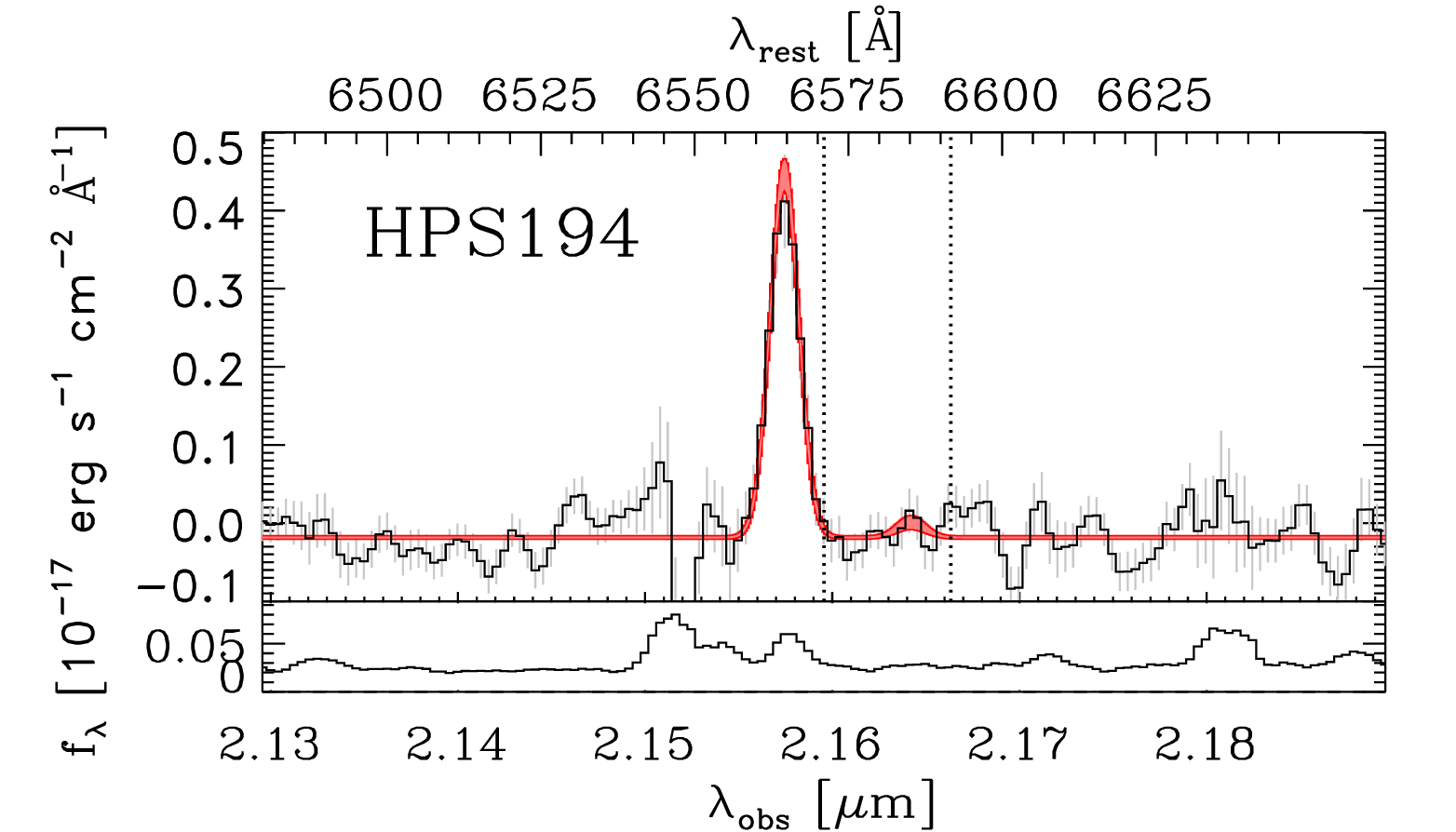}\plotone{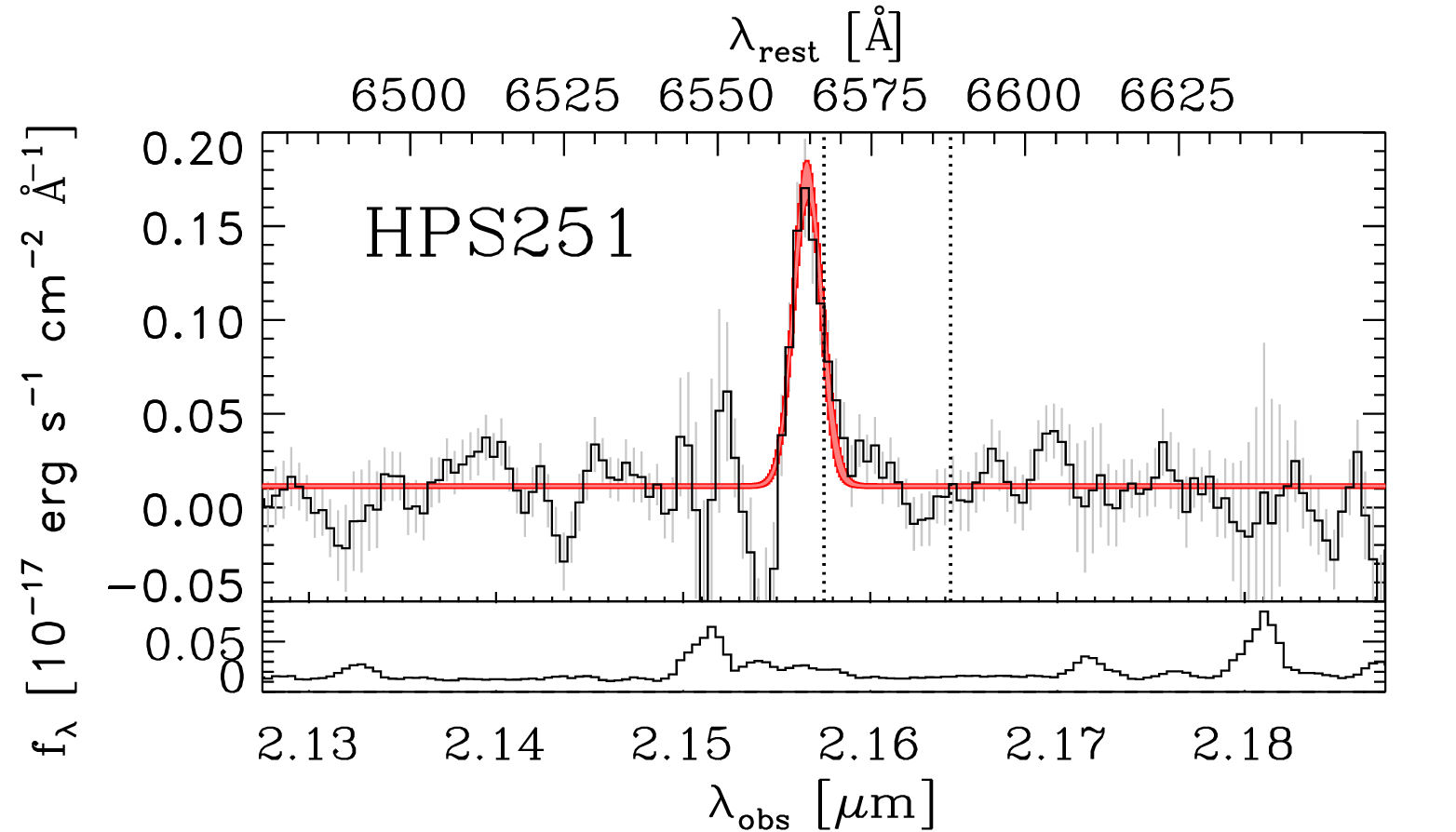}\\
  \plotone{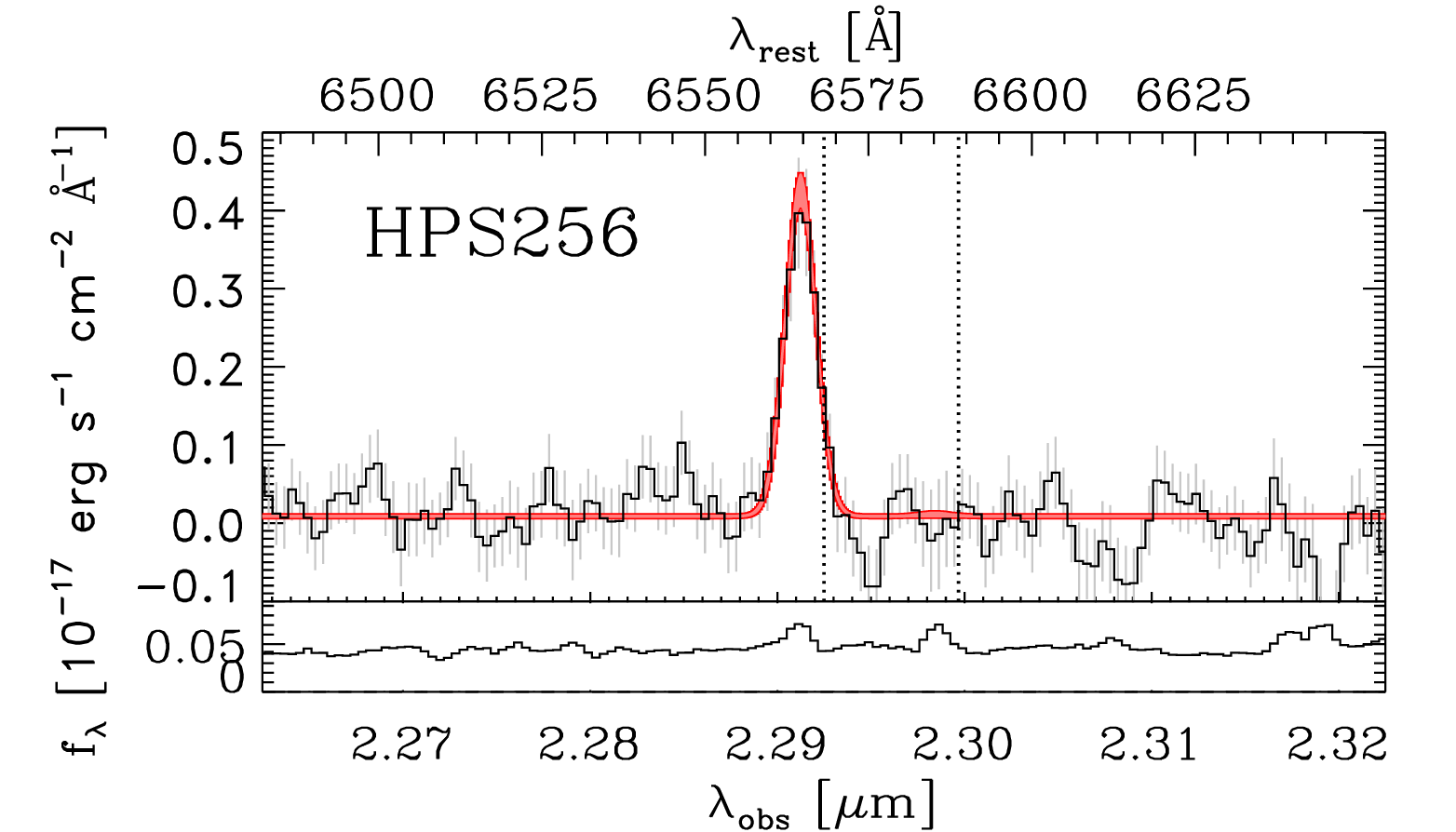}\plotone{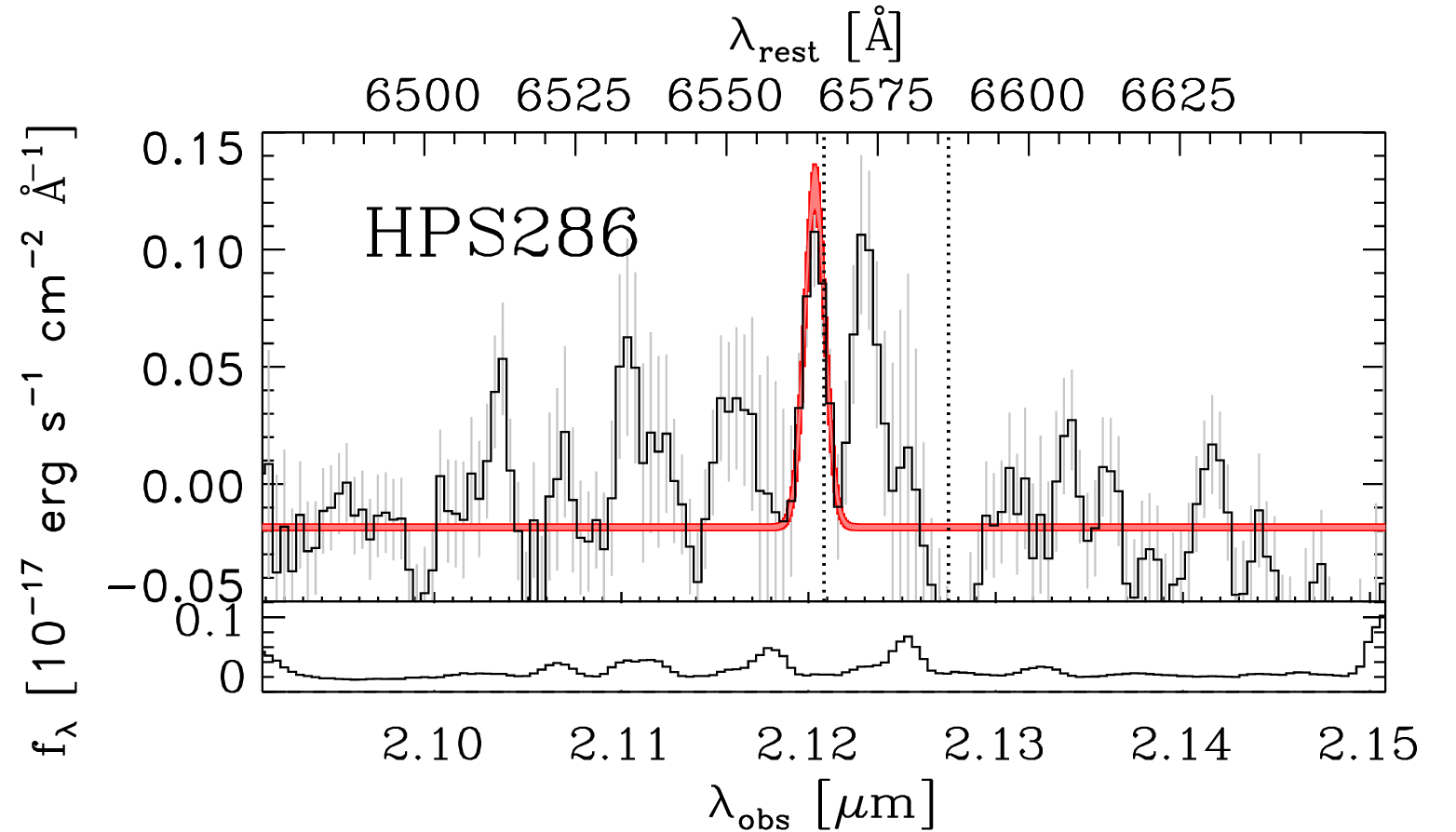}\plotone{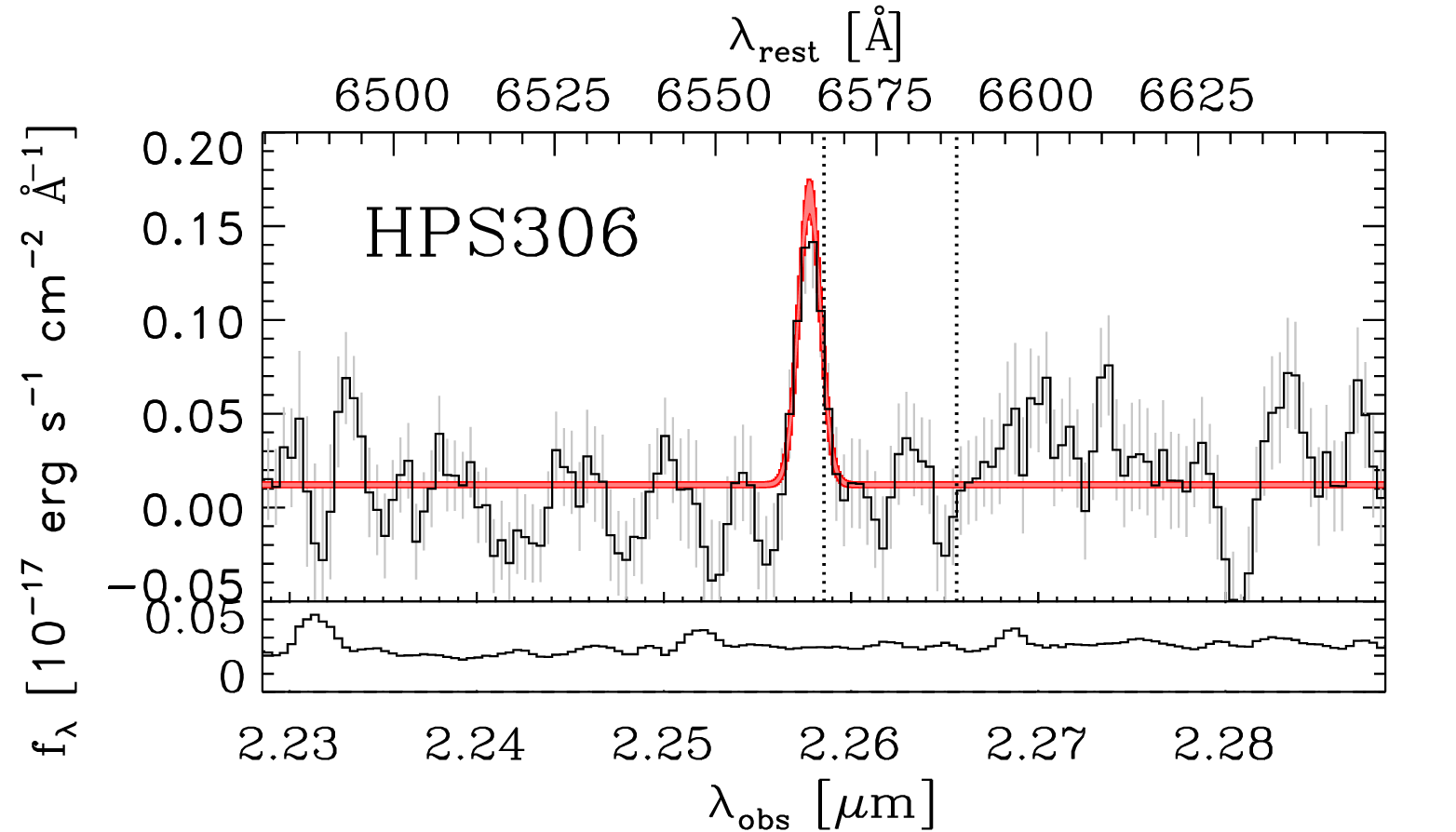}
  \caption{\label{fig:spec_K}
  \textit{K}-band spectra around the H$\alpha$ line for the 9 H$\alpha$-detected LAEs. The top two rows show our VLT/SINFONI observations, while the last two rows show our Keck/NIRSPEC observations.  HPS\,194 was observed (and detected) with both instruments.  The best-fit double Gaussian is overplotted in red, and the error spectrum is shown in the bottom panel.  Vertical dotted lines are the expected wavelengths of H$\alpha$ and \NII$\lambda$6583 from the observed Ly$\alpha$ line. 
 }
 \end{figure*}

 \begin{figure*}
   \epsscale{0.38}
   \plotone{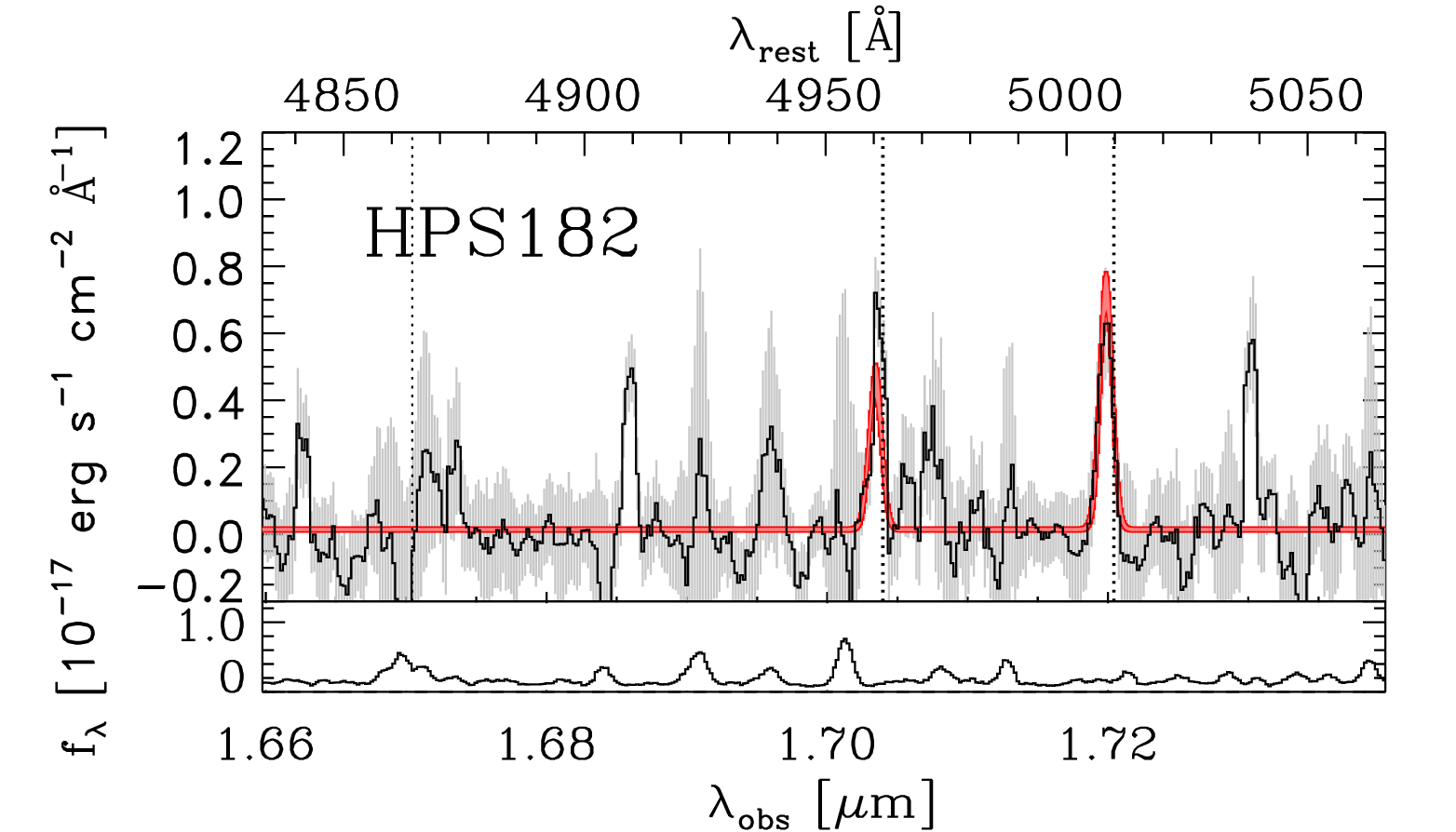}\plotone{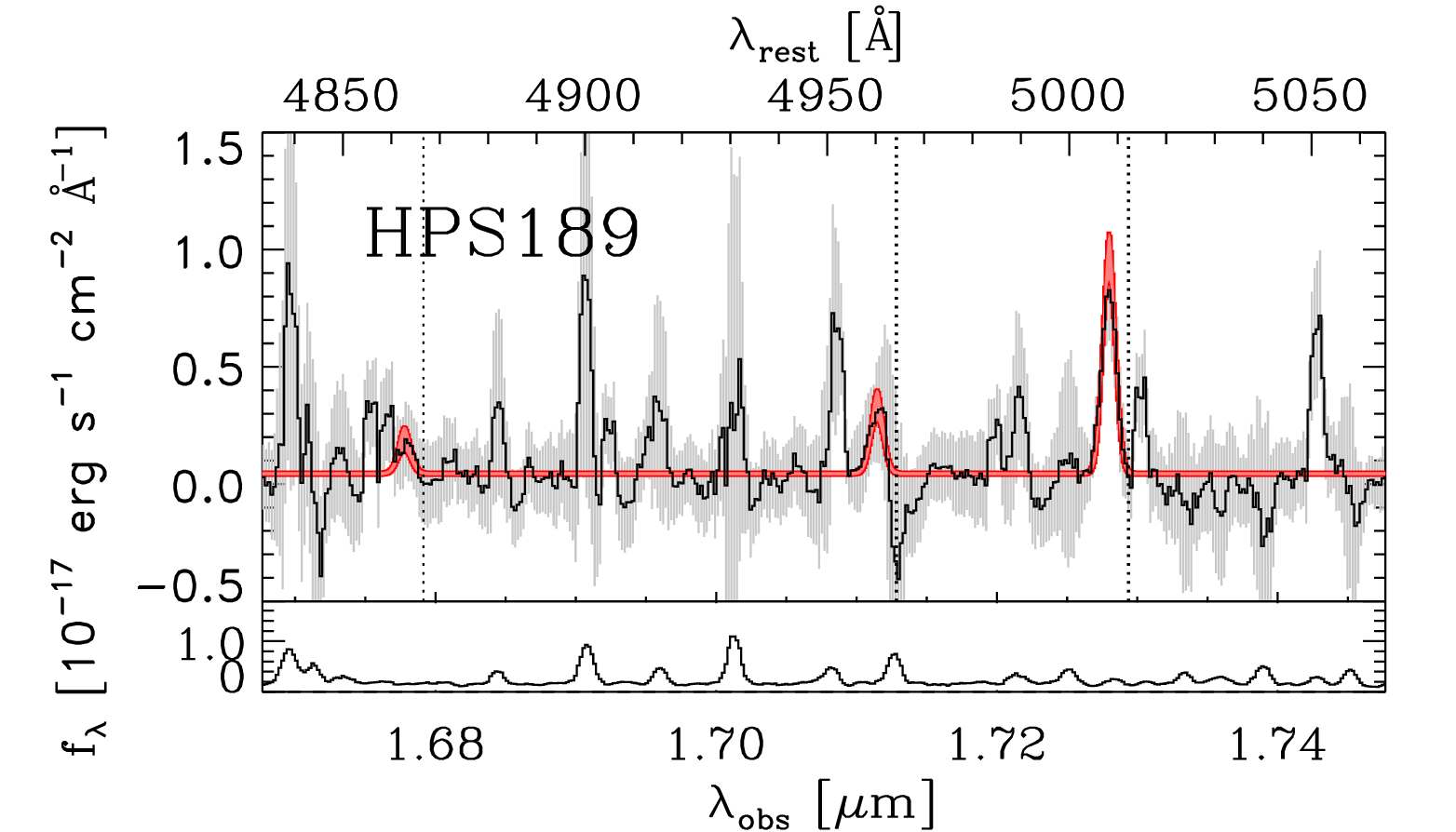}\plotone{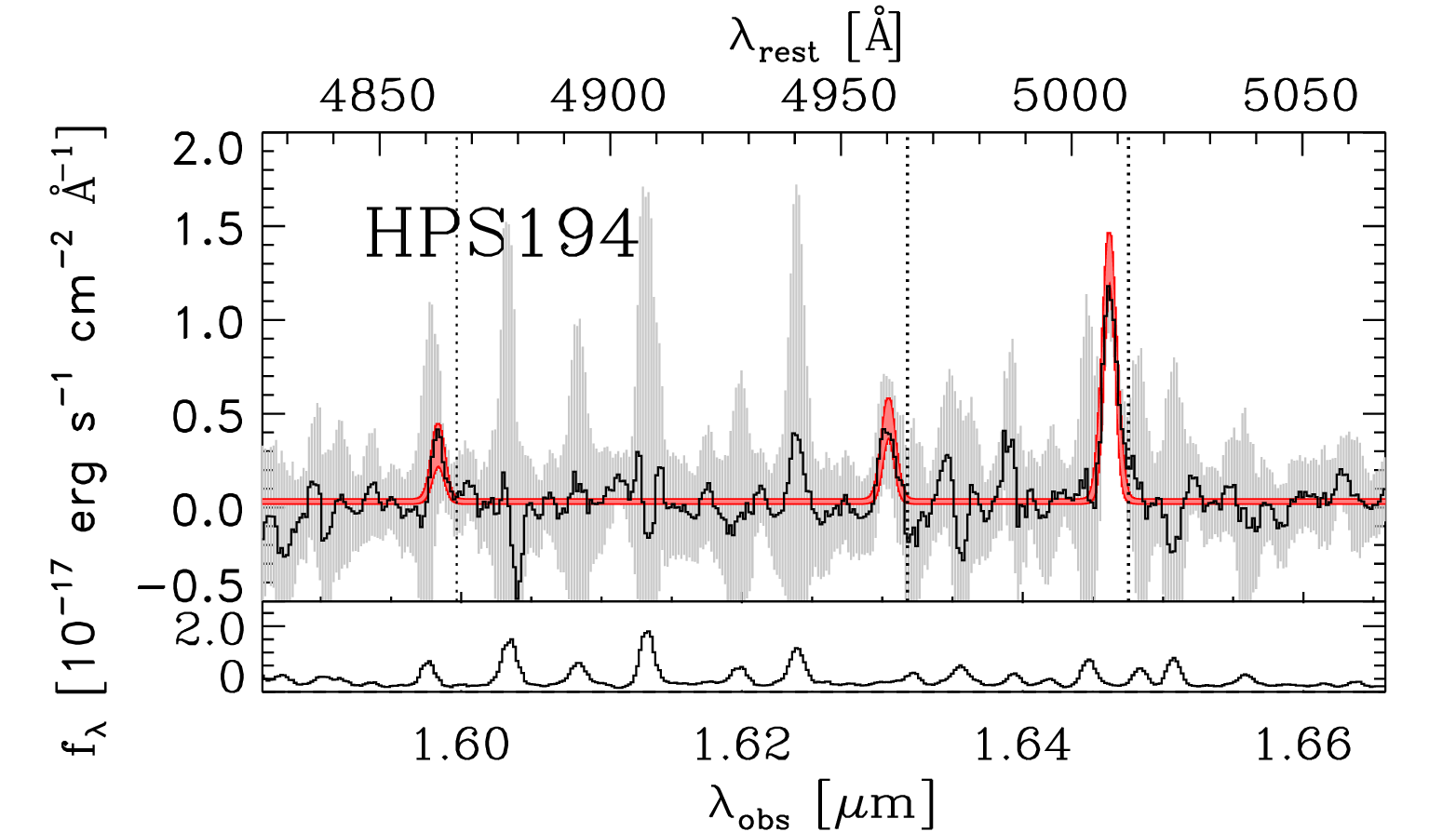}
   \plotone{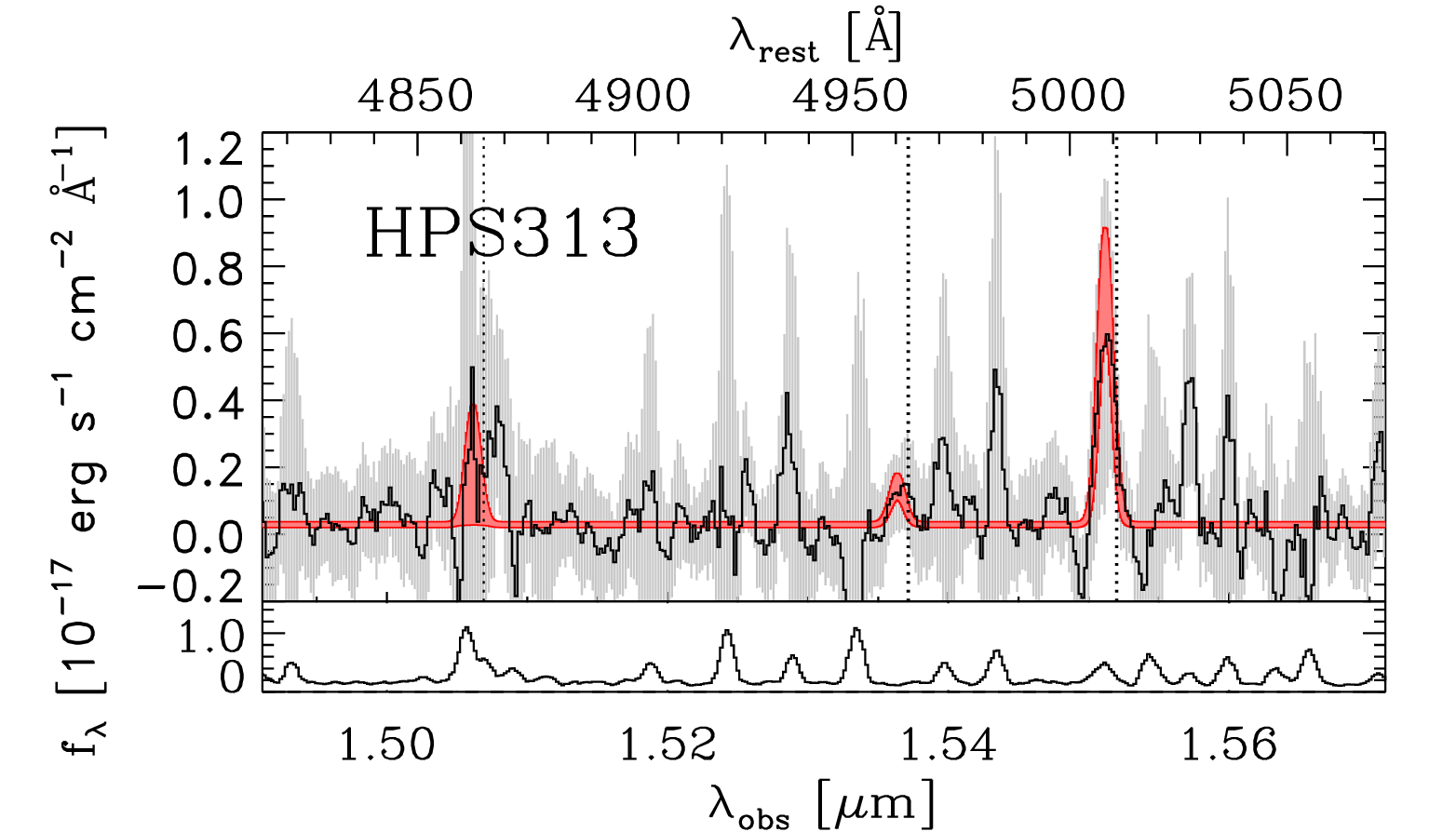}\plotone{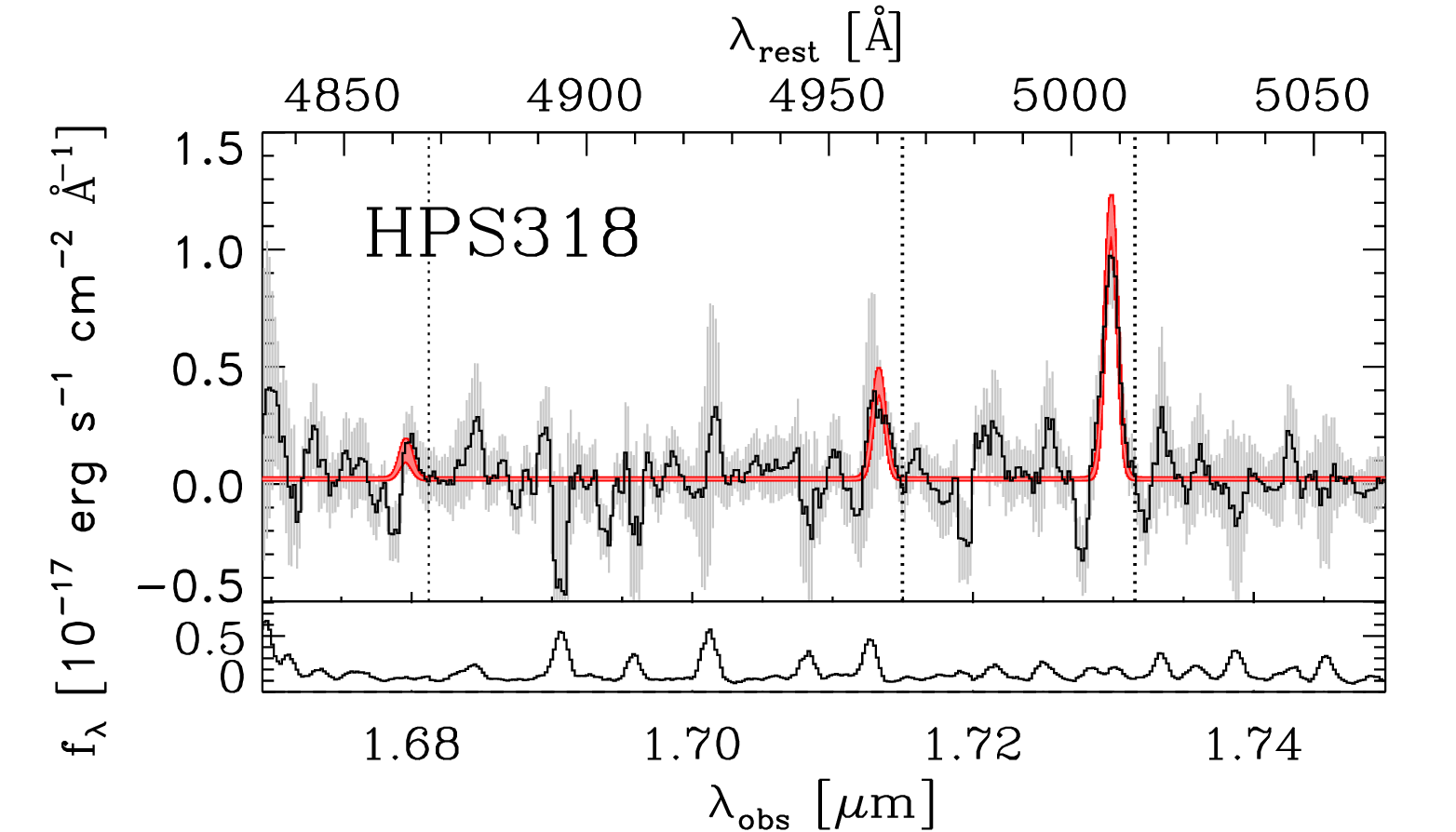}\\
   \plotone{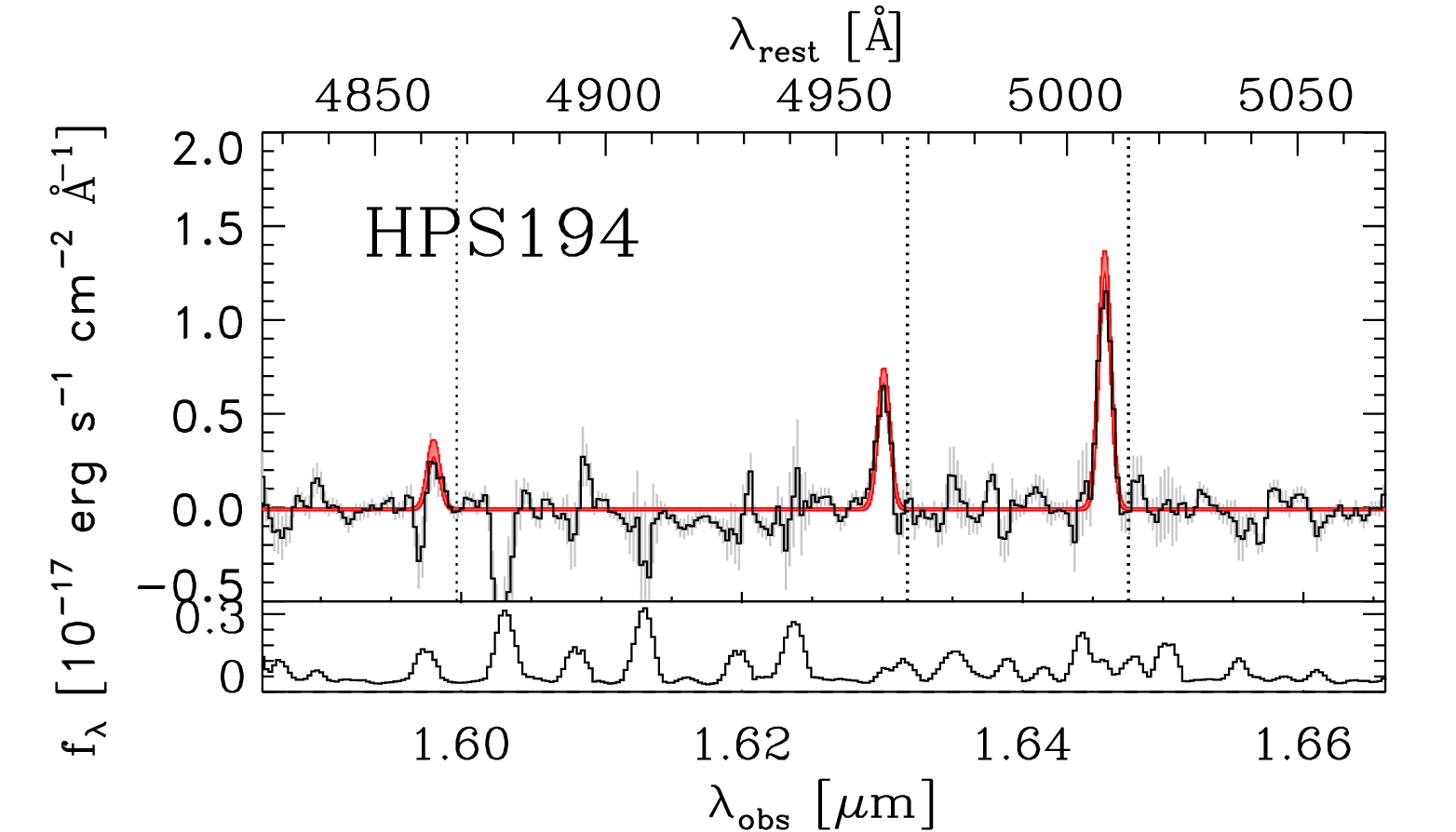}\plotone{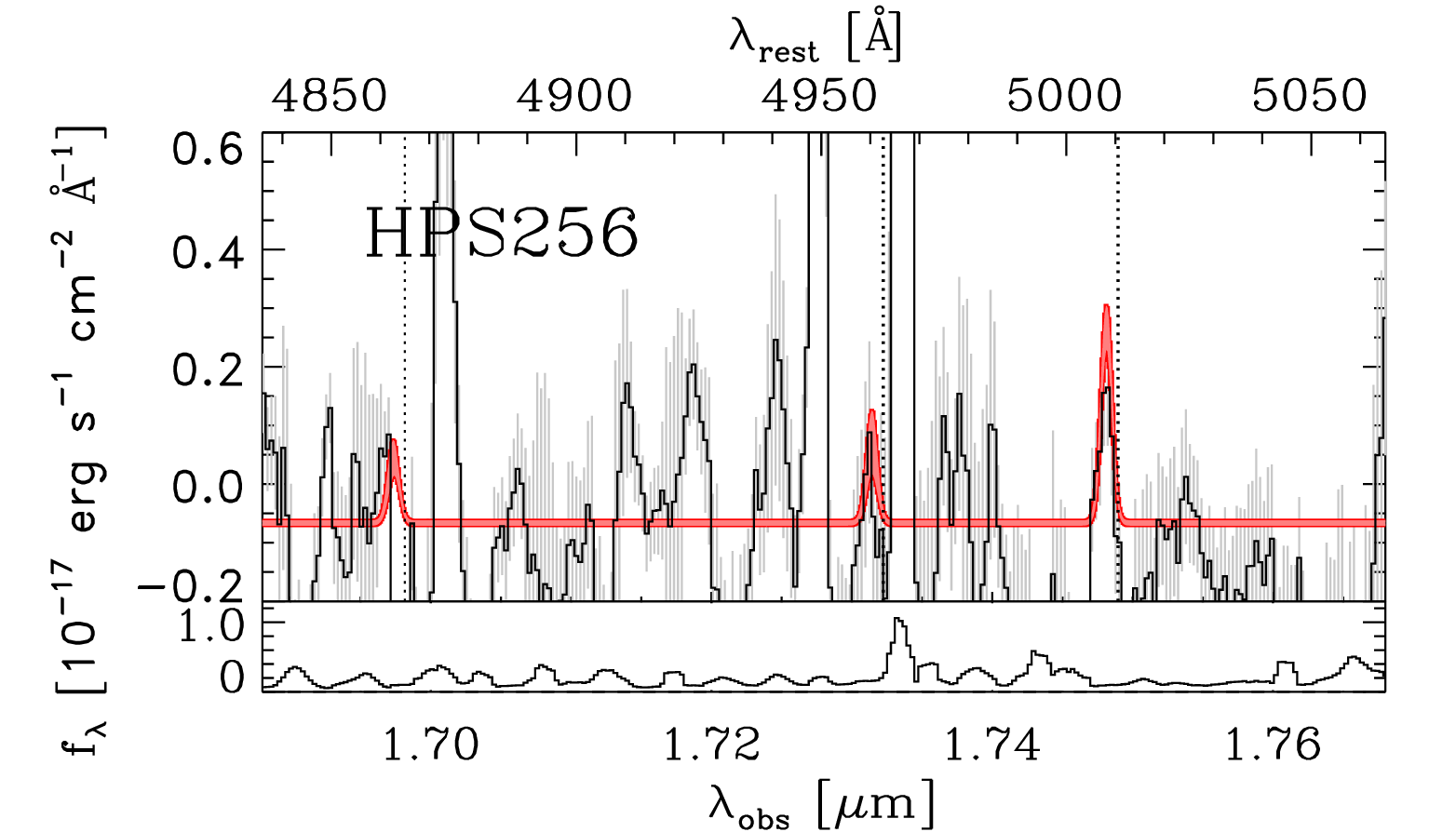}
   \caption{\label{fig:spec_H}  Similar to Figure \ref{fig:spec_K}, except for our $H$-band observations, showing the region around the H$\beta$ and \OIII$\lambda\lambda$4959,5007 lines.  The first two rows show our VLT/SINFONI observations, while the last row shows the Keck/NIRSPEC $H$-band spectra (a reprocessing of the spectra shown in \citet{fin11b}. The best-fit triple Gaussian is overplotted in red, and the expected wavelengths of H$\beta$, \OIII$\lambda\lambda$4959,5007 lines are shown as vertical dotted lines.} 
 \end{figure*}

\subsubsection{NIRSPEC}
 
The NIRSPEC data reduction proceeded in a nearly identical fashion to that described in \citet{fin11b}, thus we refer the reader there for more details.  To ensure consistency, we reprocessed the data from that paper, as we wished to include those two LAEs in our sample (HPS\,194 and HPS\,256).  Briefly, we used the NIRSPEC IDL reduction pipeline $\tt{red spec}$ to perform the wavelength calibration and rectification.  We used our own specialized routines for the remainder of the reduction, including cosmic ray rejection (using the IRAF task L.A.Cosmic; \citealt{van01}), sky subtraction, and one-dimensional spectral extraction.  During the extraction and subsequent combination of individual frames, we only included frames in the stack which increased the resultant SNR, which typically resulted in 4 frames being used.  The total exposure times in the final spectra are listed in Table \ref{tab:linedet}.  The same procedure was applied to the telluric standard star.  After the 1D extraction, the analysis was identical to that described above for our SINFONI data.  
 
 
 \subsection{Line Detection} \label{sec:line_detection}
 
We measured the emission line flux, FWHM, and redshift by fitting a double Gaussian to the  \textit{K}-band spectra (for H$\alpha$ and \NII$\lambda$6583) and a triple Gaussian to the \textit{H}-band spectra (for H$\beta$, \OIII$\lambda$4959, and \OIII$\lambda$5007) with Gaussian weighting from the error spectrum calculated in the previous section. Since the \NII\ line is weak compared to the noise level of our spectra, it is difficult to constrain its properties from these fits. Therefore, we fix its redshift and FWHM to be the same as H$\alpha$ as no other strong forbidden line is available in the \textit{K}-band spectral range, while leaving its flux as a free parameter: we iteratively fit a double Gaussian until the redshift and FWHM of two lines satisfies $\Delta z<$ 0.00001 and $\Delta$FWHM $<$ 0.01 \AA. Similarly, the redshift and FWHM of the H$\beta$ and \OIII$\lambda$4959 lines are fixed to those of \OIII$\lambda$5007.  We furthermore impose a constraint on the \OIII$\lambda$4959 flux such that the ratio of \OIII$\lambda$5007/\OIII$\lambda$4959 is equal to the theoretical value of 2.98 \citep{storey00}. All line fluxes are constrained to be positive. The uncertainties for line flux, FWHM, and redshift were quantified as the 68\% confidence interval from 10$^3$ Monte Carlo realizations of the data, where the input spectrum is given as the observed spectrum perturbed by Gaussian random noise, with the Gaussian $\sigma$ equal to the noise spectrum value at a given wavelength.

For the SINFONI data, we measured the line fluxes for each of the OB combinations discussed above in Section \ref{sec:data_reduction_sinfo}.  These line fluxes and associated uncertainties measured from individual and co-added OBs were used to select the final spectrum for each object to be used for further analysis, using the one with the highest H$\alpha$ or \OIII$\lambda$5007 signal-to-noise ratio.  In summary, imposing a 3$\sigma$ detection limit, we detect H$\alpha$ in 5 out of 10 SINFONI-observed LAEs and \OIII$\lambda$5007 in 5 out of 10, among which 4 have both H$\alpha$ and \OIII$\lambda$5007 detections. One object (HPS\,194) is detected in H$\beta$ with more than 3$\sigma$ significance, and no galaxy has significant \NII\ detection.  

We measured line fluxes in a similar way on the spectra of the Keck/NIRSPEC observed LAEs.  Of the newly observed LAEs, we find a $\geq$ 3$\sigma$ significant detection of H$\alpha$ for HPS\,251, HPS\,286 and HPS\,306, for a total sample of five LAEs (including the previously published HPS\,194 and HPS\,256) in the COSMOS field with detected H$\alpha$.  None of these five objects have significant \NII\ emission.  Of these five LAEs, only HPS\,194 and HPS\,256 were observed in the $H$-band, as discussed in \citet{fin11b}; both objects have detected \OIII$\lambda$5007 emission, and marginal H$\beta$ emission.

We emphasize that these 10 LAEs originally selected via strong Ly$\alpha$ emission have secure spectroscopic redshifts confirmed by these detections of rest-frame optical nebular emissions.

Meanwhile, the non-detection in NIR in 6 out of 16 targets may be attributed to the following:
first, they may be LAEs with high Ly$\alpha$ escape fraction, as to be 
discussed in Section \ref{sec:lya_escape}. 
Another possibility is that their Ly$\alpha$ may be false detections caused by statistical noise fluctuations. \citet{adams11} predict a 4--10\% contamination fraction due to spurious sources in the HPS LAE sample based on simulations and empirical tests they performed. While sources with high Ly$\alpha$ SNR are free from this possibility, two sources (HPS\,160 and HPS\,223) have low enough SNR that their Ly$\alpha$ detection could be spurious.
Lastly, there is a possibility of a misidentification of \OII\ line as Ly$\alpha$, and they may in fact be low redshift \OII\ emitters. If this is true, we would have detected other recombination lines through \OIII$\lambda$5007 for these sources (except HPS\,419 of which \OIII\ falls out of spectral coverage) in their HETDEX Pilot Survey data. 
 However, we do not find any hint of other line detections in any of them. 
Under the hypothesis they are \OII\ emitters, we also checked our NIR datacubes if Br$\delta$ ($\lambda_{\rm rest}$=1.9447 $\mu$m) or Br$\gamma$ ($\lambda_{\rm rest}$=2.1657 $\mu$m) line, which should be bright if they are \OII\ emitters, is detected. However, we found no indication of them, and thus we believe the chance of them being low redshift interlopers to be low. 


\subsection{Upper limit on \NII\ and H$\beta$ flux}\label{sec:upplimit}

A robust measurement of the \NII\ line fluxes is critical to constrain the metallicities of our sample.  As none of our LAEs have a detected \NII\ line, we quantified the upper limits via simulations, inserting a mock line at the \NII\ wavelength with varying flux and fixing the line FWHM to be equal to that of the H$\alpha$ line. We measured the SNR of the mock line by performing the same fitting procedure described in the previous section. We input lines at a range of fluxes, resulting in recovered SNRs of $\sim$ 3 -- 50.  The 1$\sigma$ upper limit is estimated as one fifth of the input flux which has a SNR of 5.

A possible caveat of this approach is that when there is an underlying weak line in a noisy spectrum, the resulting flux limit estimated from the simulation would be underestimated, as the underlying weak line contributes to the SNR measurements. Therefore, we performed a second simulation, inserting mock lines at multiple wavelengths around the \NII\ line (rather than directly on top).  Mock lines with varying flux were inserted at 30 different wavelengths around the expected location of the \NII\ wavelength with $\Delta\lambda_{\rm rest}$=5\AA~(excluding the H$\alpha$ and \NII\ wavelengths).   From this simulation the 1$\sigma$ flux limit was estimated as the median of the 1$\sigma$ limits measured at these 30 wavelengths.  If the observed \NII\ line fell on a sky line (where any weak \NII\ emission would be thoroughy washed out by the sky noise), we used the result from the first simulation. Otherwise, the final 1$\sigma$ limit for \NII\ flux for each object was then determined as the one from the second simulation. Excluding the former cases, the results of two simulations show a qualitative agreement (the mean difference in the resulting 1$\sigma$ limits of $\langle {\Delta} \rangle$=1.4 $\times$ 10$^{-18}$ erg s$^{-1}$ cm$^{-2}$). Upper limits of the H$\beta$ fluxes were measured in the same manner.

\begin{deluxetable*}{crccccc}
\tablewidth{0pt}
\tablecaption{\label{tab:linedet} Emission line detection}
\tablehead{
\colhead{Object} & \colhead{Line} & \colhead{$\lambda_{\rm rest}$\,\tablenotemark{a}}   & \colhead{F$_{\rm line}$\,\tablenotemark{b}} & \colhead{SNR} 
& \colhead{$z$\,\tablenotemark{c}} & \colhead{EXPTIME\,\tablenotemark{d}} \\
\colhead{$ $} & \colhead{$ $} & \colhead{(\AA)} & \colhead{(10$^{-17}$ erg s$^{-1}$ cm$^{-2}$)} & \colhead{$ $} & \colhead{$ $} & \colhead{(min)} \\
}
\startdata
\smallskip
\textbf{VLT/SINFONI} &&&&&& \\

HPS\,182 & H$\beta$ & 4861 & $<$ 2.14 & --- & &\\
& [O\,{\sc iii}] & 5007 &   8.18     $\pm$ 0.72    & 11.4 & 2.43422   $\pm$  0.00011 & 80\\
\smallskip
& & & & & ($z_{\rm sys}$ = 2.43422 $\pm$ 0.00011 $\pm$ --- ) & \\
\tableline\\

HPS\,183 
& H$\alpha$ & 6563  & 11.74 $\pm$   1.70   &   6.89 &  2.16210  $\pm$  0.00024 & 45 \\
 & [N\,{\sc ii}] & 6583 & $<$ 1.32 & ---  & &\\
\smallskip
 &&&&&  ($z_{\rm sys}$ = 2.16210 $\pm$ 0.00024 $\pm$ ---) & \\
\tableline\\

HPS\,189 & H$\beta$ & 4861 & $<$ 0.74 & ---  & &\\
& [O\,{\sc iii}] & 5007 &  10.47  $\pm$   0.95 &    11.1 &  2.45039   $\pm$ 0.00010 & 40  \\
& H$\alpha$ & 6563 &  5.47   $\pm$  1.04    &  5.27 &   2.44994  $\pm$  0.00010 & 80 \\
& [N\,{\sc ii}] & 6583 & $<$ 1.72 & --- &  & \\
\smallskip
&&&&& ($z_{\rm sys}$ = 2.45017 $\pm$ 0.00007  $\pm$ 0.00022) & \\
\tableline\\
  
HPS\,194 & H$\beta$ & 4861 & 3.49 $\pm$ 1.15  & 3.04 &  & \\
& [O\,{\sc iii}] & 5007 & 15.02   $\pm$  1.39    & 10.8 &   2.28699    $\pm$ 0.00008 & 80 \\
& H$\alpha$ & 6563 & 10.28    $\pm$ 1.16   &  8.85 &  2.28675   $\pm$  0.00011 & 80\\
& [N\,{\sc ii}] & 6583 & $<$ 0.68 & --- &  & \\
\smallskip
&&&&& ($z_{\rm sys}$ = 2.28690 $\pm$ 0.00007 $\pm$ 0.00012) & \\
\tableline\\

HPS\,313 & H$\beta$ & 4861 & $<$ 1.92 & --- &  \\
& [O\,{\sc iii}] & 5007 & 9.46    $\pm$ 1.83    &  5.16 &   2.09711   $\pm$  0.00022 & 25 \\
& H$\alpha$ & 6563 &  14.41    $\pm$ 4.43   &   3.25 &  2.09726  $\pm$   0.00061 & 40\\
& [N\,{\sc ii}] & 6583 & $<$ 1.78 & --- & &\\
\smallskip
&&&&& ($z_{\rm sys}$ = 2.09713 $\pm$ 0.00021 $\pm$ 0.00005) & \\
\tableline\\

HPS\,318 & H$\beta$ & 4861 & $<$ 0.46 & --- & & \\
& [O\,{\sc iii}] & 5007 & 12.65  $\pm$   0.90  &   14.1 &  2.45406  $\pm$  0.00007 & 80 \\
& H$\alpha$ & 6563 & 7.29    $\pm$ 1.00    &  7.26 &  2.45313  $\pm$  0.00025 & 40 \\
& [N\,{\sc ii}] &6583 & $<$ 0.94 & --- &  &\\
\smallskip
&&&&& ($z_{\rm sys}$ = 2.45399 $\pm$ 0.00007 $\pm$ 0.00025) & \\
\tableline\\
\smallskip
\textbf{Keck/NIRSPEC} &&&&&& \\

HPS\,194 & H$\beta$ & 4861 & 3.64 $\pm$ 0.42 & 8.72 &  &\\
& [O\,{\sc iii}] & 5007 & 14.56 $\pm$ 0.48       & 30.3 &  2.28632 $\pm$ 0.00003 & 90 \\
& H$\alpha$ & 6563  & 8.99 $\pm$ 0.30 & 30.4 & 2.28667 $\pm$ 0.00004 & 60 \\
& [N\,{\sc ii}] & 6583  & $<$ 0.14 & ---  &  & \\
\smallskip
&&&&& ($z_{\rm sys}$ = 2.28646 $\pm$ 0.00003 $\pm$ 0.00018) & \\
\tableline\\

HPS\,251  
& H$\alpha$ & 6563  & 3.06 $\pm$ 0.15  & 20.3 &  2.28500 $\pm$ 0.00009 & 60 \\
& [N\,{\sc ii}] & 6583  & $<$ 0.08 & ---  & & \\
\smallskip
&&&&& ($z_{\rm sys}$ = 2.28500 $\pm$ 0.00009 $\pm$  ---)  & \\
\tableline\\

HPS\,256 & H$\beta$ & 4861 &  1.30 $\pm$ 0.30 & 4.29 &  &\\
& [O\,{\sc iii}] & 5007 & 3.27 $\pm$ 0.38       & 8.58 & 2.49048 $\pm$ 0.00012 & 20 \\
& H$\alpha$ & 6563  & 8.58 $\pm$ 0.37 & 23.3 & 2.49029 $\pm$ 0.00006 & 60  \\
& [N\,{\sc ii}] & 6583  & $<$ 0.37 & ---  &  & \\
\smallskip
&&&&& ($z_{\rm sys}$ = 2.49032 $\pm$ 0.00005 $\pm$ 0.00008) & \\
\tableline\\

HPS\,286 
& H$\alpha$ & 6563  & 1.98 $\pm$ 0.12   & 16.1 &  2.22970 $\pm$ 0.00006 & 45  \\
& [N\,{\sc ii}] & 6583  & $<$ 0.11 & ---  &  & \\
\smallskip
&&&&& ($z_{\rm sys}$ = 2.22970 $\pm$ 0.00006 $\pm$ ---) & \\
\tableline\\

HPS\,306 
& H$\alpha$ & 6563  & 2.26 $\pm$ 0.13   & 17.1 & 2.43905 $\pm$ 0.00006 & 60 \\
& [N\,{\sc ii}] & 6583  & $<$ 0.08 & ---  &  & \\
&&&&&  ($z_{\rm sys}$ = 2.43905 $\pm$ 0.00006 $\pm$ ---) & \\
\enddata
\tablecomments{Dash bars mean non-detection, while blank fields indicate non-independent quantities: redshifts of H$\beta$ and \OIII $\lambda$4959 (\NII) are fixed 
to that of \OIII $\lambda$5007 (H$\alpha$). \OIII $\lambda$4959 flux is determined by flux of \OIII $\lambda$5007, 
by $f$(\OIII $\lambda$5007)/$f$(\OIII $\lambda$4959) = 2.98 \citet{storey00}.}
\tablenotetext{a}{Wavelength in air. For redshft estimation, we use vacuum wavelengths; $\lambda$(H$\beta$)= 4862.7\AA, $\lambda$([O\,{\sc iii}])=  5008.2\AA, $\lambda$(H$\alpha$)= 6564.6\AA, $\lambda$([N\,{\sc ii}])= 6585.2\AA.}
\tablenotetext{b}{For non-detection ($<$ 3$\sigma$), the 1$\sigma$ limit is listed. }
\tablenotetext{c}{Listed in parentheses are $z_{\rm sys} \pm \delta$(phot) $\pm \delta$(sys) -- i.e., systemic redshift (an inverse-variance weighted mean of $z$(H$\alpha$) and $z$([O\,{\sc iii}])), photometric error, and systematic error (see \S \ref{sec:dv}).}
\tablenotetext{d}{Total on-source integration time used for analysis selected  based on SNR of the H$\alpha$ or \OIII\ line (see Section \ref{sec:line_detection}).}
\end{deluxetable*}

Figures \ref{fig:spec_K} and \ref{fig:spec_H} show the final \textit{K}- and \textit{H}-band spectra of our samples 
with H$\alpha$ and/or \OIII\ line detections, and Table \ref{tab:linedet} summarizes the measured emission line 
wavelength, flux, 1$\sigma$ limit of \NII\ flux, redshift inferred from H$\alpha$ and/or \OIII, and the total integration time of the data used for analysis.


\section{Physical Properties}\label{sec:prop}

\subsection{Spectral Energy Distribution Fitting} \label{sec:sed}

Using broadband photometry, one can measure several physical properties of galaxies using spectral energy distribution (SED) fitting.  In this method, one compares the measured photometry to a suite of stellar population models while varying several parameters; typically stellar mass, dust content, stellar population age, stellar metallicity, and star formation history.  Depending on the rest-frame wavelengths probed, there can be several degeneracies between these parameters, thus not all can be well-constrained.  The stellar mass is typically the best-constrained parameter, since although differing values of dust or age can reproduce a given color, the possible fractional range of mass-to-light ratios is typically less (e.g., \citealt{shapley01, papovich01}).  
Additionally, when photometry is measured redward of rest-frame 4000 \AA, the dust attenuation can be reasonably well constrained.  For our analysis, we wish to measure the stellar masses of our LAEs (such that we can explore our LAEs on a stellar mass and gas-phase metallicity plane), as well as the dust extinction, to determine dust-corrected star formation rates (SFRs) as well as to explore the escape of Ly$\alpha$ photons.

We utilize archival multi-wavelength photometry from a total of 25 bands from observed optical to the mid-infrared in the COSMOS field; 12 are broad-bands from \textit{V}-band to \textit{Spitzer}/IRAC 4.5 $\mu$m, and 13 are Subaru/Suprime-Cam optical medium and narrow bands. Most of the photometric measurements were taken from the COSMOS Intermediate and Broad Band Photometry Catalog.  
We add to these recently obtained first-year UltraVISTA \textit{Y}- and \textit{Ks}-band imaging \citep{mccracken12}, as well as \textit{Hubble Space Telescope/Wide Field Camera 3} ($HST$/WFC3) F125W (\textit{J}) and F160W (\textit{H}) imaging from the Cosmic Assembly Near-infrared Deep Extragalactic Legacy Survey (CANDELS; \citealt{grogin11,koekemoer11}) and $Spitzer$/IRAC 3.6 $\mu$m and 4.5 $\mu$m imaging from the $Spitzer$ Very Deep Survey of the $HST$/CANDELS Fields (S-CANDELS; \citealt{fazio11}). We measure our own photometry from these new UltraVISTA and CANDELS data using the Source Extractor package \citep{bertin96}, using the FLUX\_AUTO measurement for the UltraVISTA data, and using the techniques from \citet{fin10} for the CANDELS data. 

For reliable photometry of MIR (rest-frame NIR) imaging data, which is crucial for the stellar mass determination but is often challenging due to severe source confusion, we utilize the Template-fitting software, TFIT \citep{laidler07}, for our own photometry on the S-CANDELS data. 
 Briefly, we performed our photometry on the first two IRAC channels (3.6 and 4.5 $\mu$m) using
 the CANDELS \textit{HST}/F160W (or COSMOS F814W for two objects -- HPS\,182 and HPS\,183 -- lying out of the F160W field) data as a high-resolution image. 
This high-resolution detection image is smoothed to construct low-resolution (MIR) models of each object, from which the best-fit fluxes are determined as the ones when best reproduce the low-resolution data.
Our photometry is confirmed to be consistent within 0.2 and 0.1 dex with the S-COSMOS IRAC Photometry Catalog \citep{sanders07} and the TFIT SEDS photometry (Nayyeri et al. 2014, in prep.), respectively, for objects not contaminated by nearby sources and of which fluxes are above the shallower depths of the S-COSMOS (23.9 AB in 3.6 $\mu$m, 5$\sigma$) or the SEDS (26.0 AB in 3.6 and 4.5 $\mu$m, 3$\sigma$; \citealt{ashby13}) catalogs.

For one object of particular interest (HPS\,194; to be discussed in Section \ref{sec:mzr} and Figure \ref{fig:mzr}), we also utilize the CANDELS COSMOS TFIT multi-wavelength catalog (Nayyeri et al. 2014, in prep.), in which photometry of all bands except \textit{HST} data is performed with TFIT. 
As a consequence, each component in \textit{HST} images that HPS\,194 consists of, but is blended together in other ground-based or \textit{Spitzer} images, could be analyzed seperately.

\begin{figure*}[t]
  \epsscale{0.38}
  \plotone{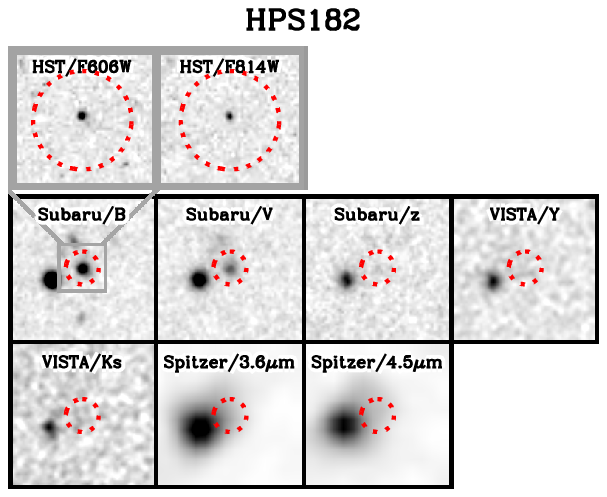}\plotone{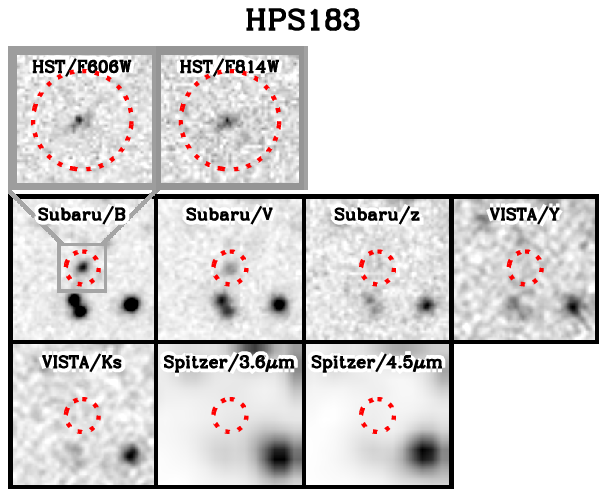}\plotone{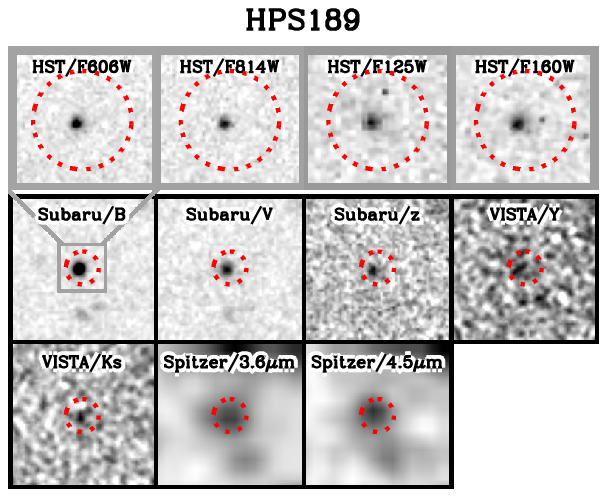}
  \vspace{-30pt}
  \plotone{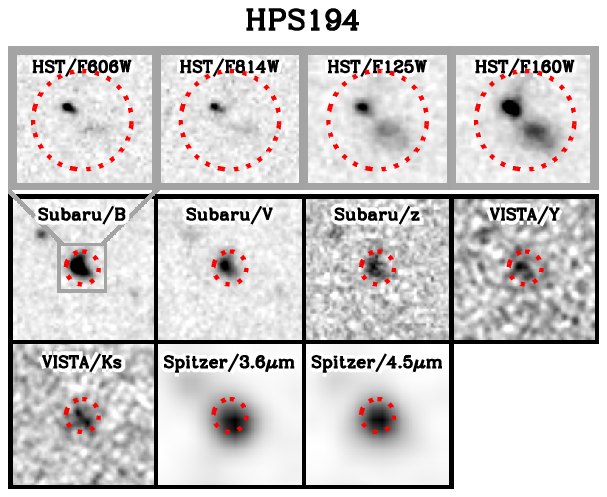}\plotone{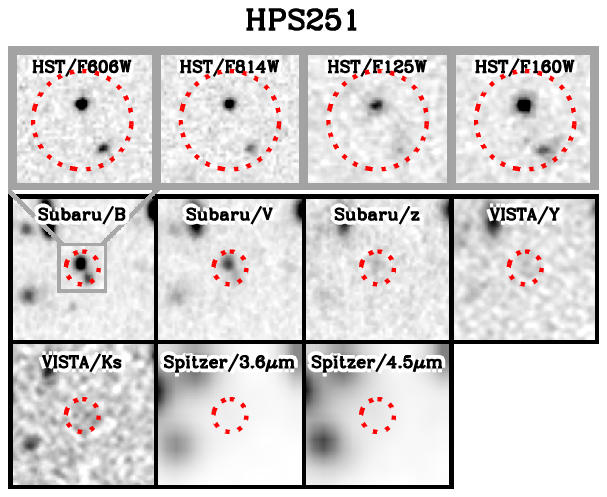}\plotone{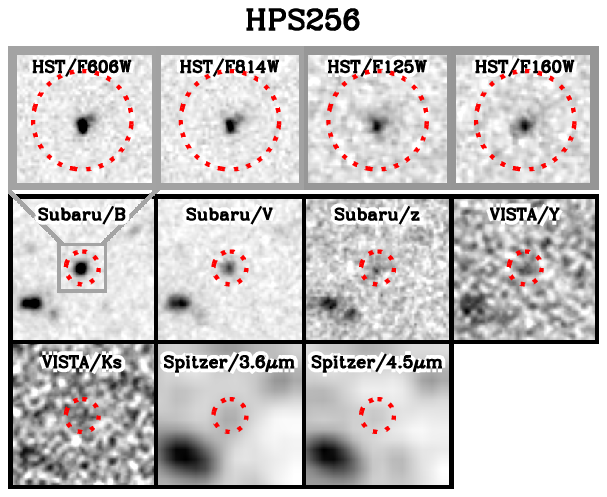}
  \vspace{-30pt}
  \plotone{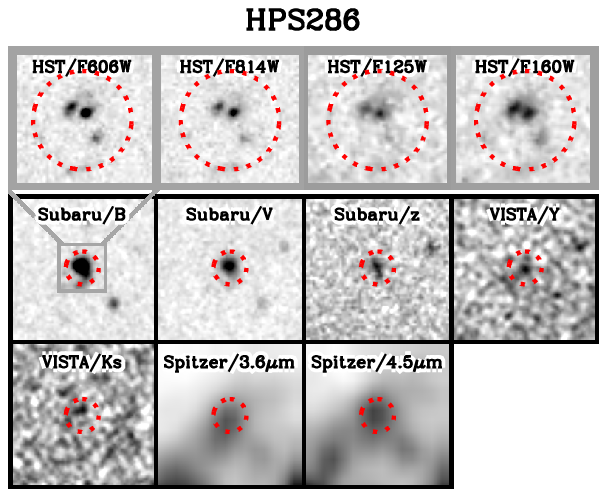}\plotone{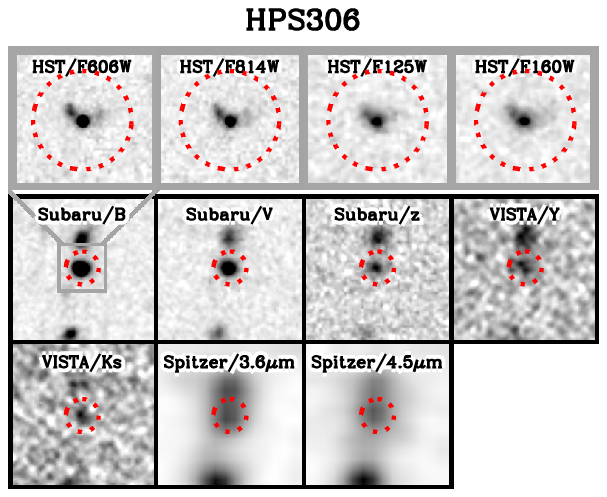}\plotone{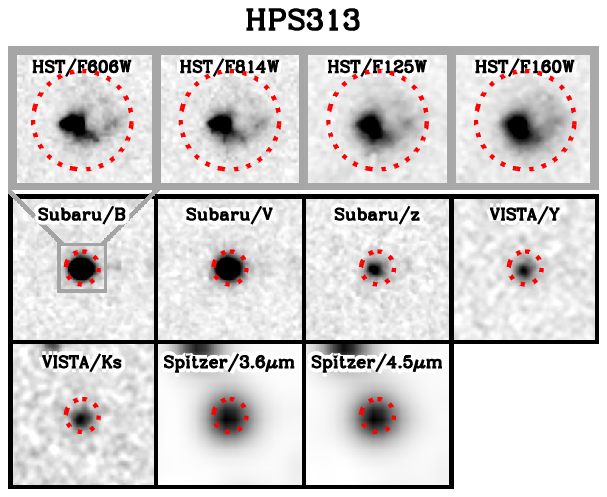}
  \vspace{-30pt}
  \plotone{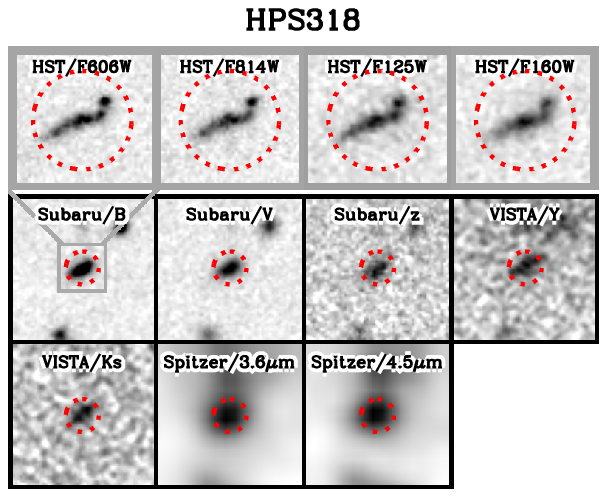}
  \vspace{-30pt}
  \caption{\label{fig:stamp} Cutout stamp images of our LAEs.  From top left to bottom right, \textit{HST}/F606W ($V$), F814W ($I$), F125W ($J$), F160W ($H$), Subaru/$B_J$, $V_J$, $z^+$, 
  VISTA/\textit{Y}, \textit{K$_s$}, \textit{Spitzer}/3.6, 4.5 $\mu$m.
  The \textit{HST} images are 3\arcsec $\times$ 3\arcsec, and others are 9\arcsec $\times$ 9\arcsec (indicated by the inset in the Subaru/$B_J$ panel). For the two components that consists of HPS\,194, we name the upper-left one as HPS\,194A and the lower-right one as HPS\,194B in our subsequent analysis.
  }
\end{figure*}

Table \ref{tab:phot} lists the filter sets and photometry used in the SED fitting, and Figure \ref{fig:stamp} shows ``postage stamp'' images from the \textit{B}-band to \textit{Spitzer}/IRAC 4.5~$\mu$m for each object.

\begin{turnpage}
\renewcommand\tabcolsep{3pt}
\begin{deluxetable*}{cccccccccccccc}
\tabletypesize{\scriptsize}
\tablecaption{\label{tab:phot} Multi-wavelength photometry of the sample}
\tablewidth{0pt}  
\tablehead{
\multicolumn{14}{c}{Broad Band}\\
\colhead{Object ID}  & \colhead{$V_J$} & \colhead{$g^+$} & \colhead{$r^+$} & \colhead{$i^+$} & \colhead{F814W} & \colhead{$z^+$} & \colhead{$Y$} & \colhead{F125W} &
 \colhead{F160W} & \colhead{$Ks$} & \colhead{3.6 $\mu$m} & \colhead{4.5 $\mu$m}\\
}
\startdata

HPS\,182 &  25.30 (0.08) &  25.56 (0.09) &  25.44 (0.09) &  25.44 (0.09) &  --- &  25.53 (0.24) &  24.78 (0.11) &  --- &  --- &  24.64 (0.24) &  24.46 (0.12) &  24.77 (0.15)  \\
HPS\,183 &  25.36 (0.10) &  25.64 (0.13) &  25.45 (0.11) &  25.53 (0.13) &  --- &  25.45 (0.28) &  26.94 (0.34) &  --- &  --- &  25.62 (0.27) &  25.35 (0.27) &  26.26 (0.51)  \\
HPS\,189 &  25.09 (0.09) &  25.28 (0.09) &  25.20 (0.09) &  25.23 (0.10) &  25.31 (0.14) &  25.43 (0.28) &  25.12 (0.11) &  25.10 (0.06) &  25.07 (0.05) &  24.63 (0.17) &  24.82 (0.15) &  24.99 (0.17)  \\
HPS\,194 &  24.07 (0.05) &  24.24 (0.06) &  24.10 (0.05) &  24.18 (0.06) &  23.82 (0.10) &  23.90 (0.10) &  23.85 (0.07) &  23.49 (0.03) &  22.84 (0.01) &  22.51 (0.05) &  22.46 (0.04) &  22.26 (0.04)  \\
HPS\,251 &  24.70 (0.07) &  24.93 (0.09) &  24.82 (0.08) &  24.83 (0.08) &  25.03 (0.11) &  24.53 (0.15) &  25.39 (0.22) &  24.94 (0.07) &  24.21 (0.03) &  23.68 (0.12) &  24.77 (0.22) &  24.76 (0.24)  \\
HPS\,256 &  25.07 (0.10) &  24.99 (0.09) &  25.18 (0.10) &  25.31 (0.13) &  25.71 (0.13) &  25.26 (0.28) &  27.21 (0.54) &  25.55 (0.08) &  25.69 (0.08) &  25.11 (0.24) &  25.64 (0.25) &  26.38 (0.48)  \\
HPS\,286 &  24.46 (0.06) &  24.48 (0.06) &  24.39 (0.06) &  24.30 (0.06) &  24.39 (0.13) &  24.44 (0.13) &  24.75 (0.10) &  24.39 (0.07) &  23.87 (0.04) &  24.44 (0.18) &  23.97 (0.27) &  23.97 (0.26)  \\
HPS\,306 &  24.08 (0.05) &  24.24 (0.05) &  24.24 (0.05) &  24.10 (0.05) &  24.15 (0.09) &  24.17 (0.10) &  24.17 (0.05) &  24.19 (0.04) &  24.17 (0.03) &  24.03 (0.11) &  24.09 (0.08) &  24.07 (0.08)  \\
HPS\,313 &  22.83 (0.03) &  23.10 (0.03) &  22.86 (0.03) &  22.78 (0.03) &  22.70 (0.03) &  22.66 (0.04) &  22.86 (0.03) &  22.21 (0.01) &  22.03 (0.01) &  21.75 (0.03) &  21.68 (0.01) &  21.62 (0.01)  \\
\smallskip
HPS\,318 &  23.84 (0.04) &  24.04 (0.05) &  23.84 (0.04) &  23.75 (0.04) &  23.72 (0.10) &  23.67 (0.08) &  23.60 (0.04) &  23.53 (0.03) &  23.40 (0.02) &  22.75 (0.05) &  22.78 (0.02) &  22.94 (0.03)  \\

\hline\\
\multicolumn{14}{c}{Medium and Narrow Band}\\
\smallskip
Object ID  & IA464 & IA484 & IA505 & IA527 & IA574 & IA624 & IA679 & IA709 & NB711 & IA738 & IA767 & NB816 & IA827\\
\hline\\

HPS\,182 & 25.37 (0.13) & 25.56 (0.13) & 25.53 (0.17) & 25.50 (0.13) & 25.43 (0.14) & 25.70 (0.18) & 25.10 (0.12) & 25.56 (0.16) & 25.40 (0.31) & 25.39 (0.18) & 25.67 (0.26) & 25.32 (0.15) & 25.49 (0.17)  \\
HPS\,183 & 25.85 (0.24) & 25.81 (0.21) & 25.90 (0.29) & 25.37 (0.15) & 25.40 (0.19) & 25.65 (0.22) & 25.80 (0.27) & 26.05 (0.30) & 25.49 (0.40) & 25.98 (0.36) & 25.82 (0.41) & 25.72 (0.26) & 26.39 (0.44)  \\
HPS\,189 & 25.72 (0.22) & 25.23 (0.14) & 25.28 (0.19) & 25.38 (0.15) & 25.62 (0.23) & 25.22 (0.16) & 25.16 (0.16) & 25.33 (0.17) & 25.34 (0.36) & 25.46 (0.23) & 25.20 (0.21) & 25.57 (0.24) & 25.70 (0.26)  \\
HPS\,194 & 24.35 (0.09) & 24.13 (0.07) & 24.25 (0.09) & 24.16 (0.07) & 24.30 (0.09) & 24.14 (0.08) & 23.99 (0.07) & 24.10 (0.08) & 24.05 (0.16) & 24.24 (0.10) & 24.23 (0.10) & 24.04 (0.09) & 24.07 (0.08)  \\
HPS\,251 & 25.14 (0.15) & 24.71 (0.10) & 24.81 (0.13) & 25.03 (0.13) & 25.33 (0.19) & 25.01 (0.15) & 25.14 (0.18) & 25.01 (0.15) & 25.14 (0.34) & 24.83 (0.16) & 24.79 (0.16) & 24.80 (0.14) & 24.87 (0.15)  \\
HPS\,256 & 25.29 (0.18) & 25.09 (0.15) & 24.87 (0.15) & 25.45 (0.19) & 25.32 (0.21) & 25.18 (0.18) & 25.39 (0.24) & 25.43 (0.23) & --- & 26.19 (0.49) & 25.22 (0.25) & 25.15 (0.19) & 25.52 (0.27)  \\
HPS\,286 & 24.35 (0.08) & 24.41 (0.08) & 24.51 (0.10) & 24.49 (0.08) & 24.71 (0.11) & 24.31 (0.08) & 24.16 (0.08) & 24.54 (0.10) & 24.86 (0.27) & 24.60 (0.12) & 24.48 (0.12) & 24.66 (0.12) & 24.67 (0.12)  \\
HPS\,306 & 24.20 (0.07) & 24.12 (0.06) & 24.13 (0.07) & 24.18 (0.06) & 24.16 (0.07) & 24.03 (0.07) & 23.84 (0.06) & 24.14 (0.07) & 24.16 (0.14) & 24.08 (0.08) & 24.30 (0.09) & 24.27 (0.08) & 24.29 (0.08)  \\
HPS\,313 & 23.12 (0.04) & 23.07 (0.03) & 22.93 (0.04) & 23.15 (0.04) & 23.02 (0.04) & 22.85 (0.03) & 22.64 (0.03) & 22.82 (0.03) & 22.78 (0.05) & 22.87 (0.04) & 22.77 (0.04) & 22.80 (0.04) & 22.73 (0.03)  \\
HPS\,318 & 24.32 (0.10) & 24.10 (0.07) & 24.15 (0.08) & 23.99 (0.06) & 23.83 (0.06) & 23.81 (0.07) & 23.66 (0.06) & 23.89 (0.07) & 23.80 (0.13) & 23.90 (0.08) & 23.76 (0.08) & 23.73 (0.06) & 23.96 (0.10)  \\

\enddata
\tablecomments{
 Magnitudes and magnitude errors for Subaru/SuPrimeCam \textit{V$_J$, g$^+$, r$^+$, i$^+$, z$^+$} and medium \& narrow bands are from the COSMOS Intermediate and Broadband Photometry Catalog. 
The rest are from our photometry on the CANDELS v1.0 data for \textit{HST}/ACS F814W (\textit{I}), \textit{HST}/WFC3 F125W (\textit{J}), F160W (\textit{H}), on the first-year UltraVISTA data \citet{mccracken12} for UVISTA/\textit{Y, Ks}, and on the S-CANDELS data for Spitzer/IRAC 3.6 and 4.5 $\mu$m. \\
}
\end{deluxetable*}
\end{turnpage}

\begin{figure*}[t]
  \epsscale{1.1}
  \plotone{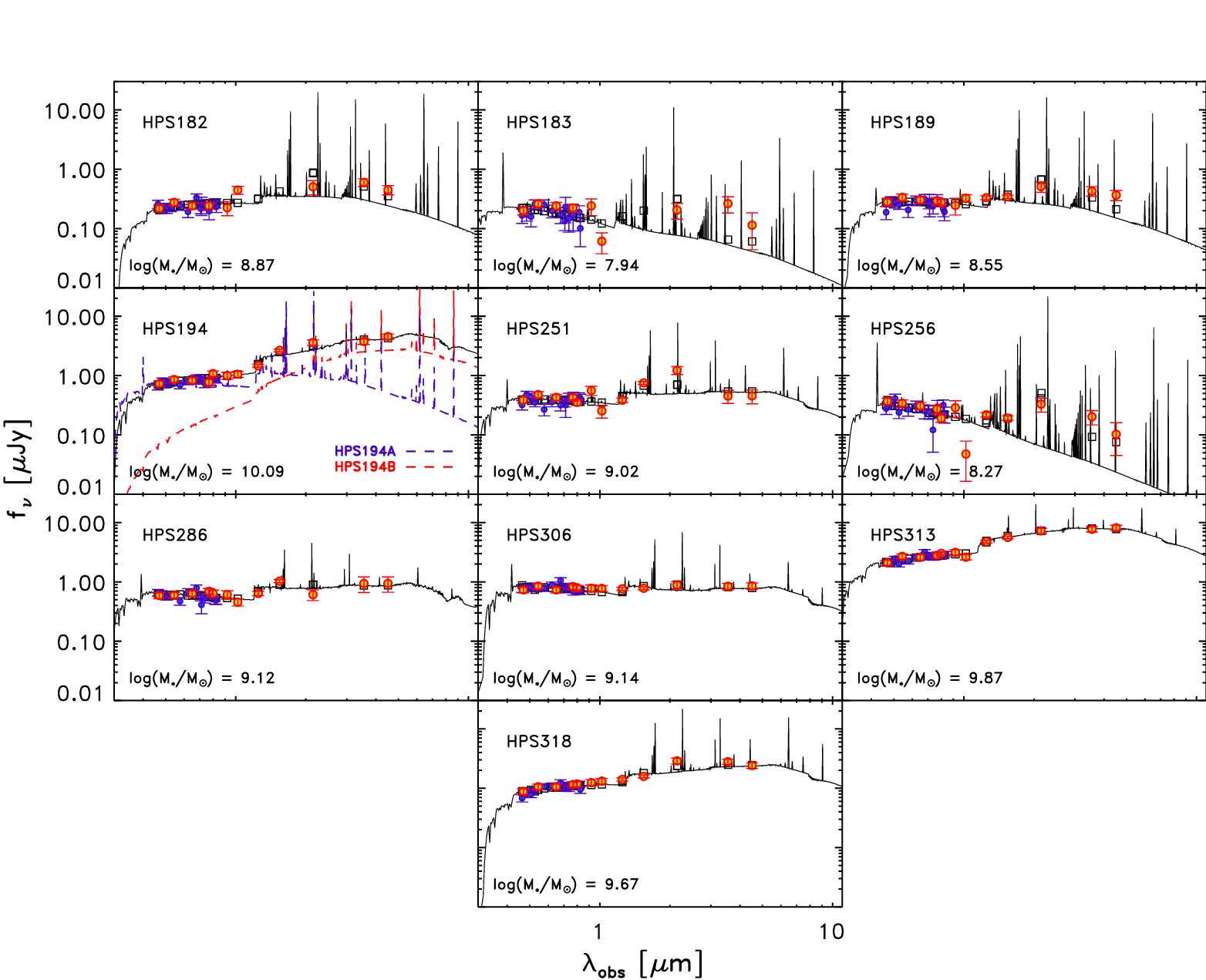}
  \caption{\label{fig:sedfit}
    The best-fit stellar population model template overlaid on the observed SEDs. The filled orange and small blue circles represent 
    observed broad-band and medium/narrow-band photometry points, respectively. 
    The solid line and black squares indicate the best-fit template model and model bandpass-averaged flux, respectively. 
    We impose constraints on a set of emission line models by fitting the observed \OIII\ and H$\alpha$ fluxes simultaneously with photometric data. The red and blue dashed lines in the HPS\,194 panel are the best-fit models for each component that consists of HPS\,194.}
\end{figure*}

The model templates were generated using the updated version of the \citet{bruzual03} stellar population synthesis models (the 2007 version; hereafter CB07).  In order to take into account the contribution of nebular emission, which has proven to be important by several recent studies on high redshift galaxies \citep{schaerer10,fin11b,labbe13,stark13}, nebular emission line spectra with extra attenuation of $E(B-V)_{\rm neb} = 2.27~ E(B-V)_{\rm stellar}$ (\citealt{calzetti00}; see Section \ref{sec:sfr} for more discussion) were added, following the prescription of Salmon et al. (2014, in prep.).  In brief, the line strengths depend on the number of ionizing photons, which is set by the stellar population age and metallicity, and the ionizing escape fraction: the H$\beta$ luminosity is given by the number of ionizing photons and ionizing escape fraction, and the strengths of other lines ($\lambda$= $\lambda_{\rm Ly\alpha}$ -- 1 $\mu$m) are determined by the H$\beta$ luminosity and metallicity using the table in \citet{inoue11} calculated with the photoionization code {\sc cloudy} 08.00 \citep{ferland98}, in addition to Paschen and Brackett series taken from \citet{osterbrock06} and \citet{storey95}. We assumed the ionizing escape fraction to be zero, since constraints from observations searching for escaping Lyman continuum photons at $z \lesssim$ 3 suggest a low ionizing escape fraction of at most a few percent (e.g., \citealt{malkan03,cowie09,fynbo09,siana10,vanzella10}).\footnote{Although at $z \sim$ 3 there are signatures that faint galaxies have a large ionizing escape fraction, a significant fraction of them might be from foreground contaminators (\citealt{vanzella12}; Siana et al. 2013, in prep.).} 

As we have already measured the H$\alpha$ and \OIII\ emission line fluxes for most of our objects, we add these line fluxes as constraints during the SED fitting, along with the 25 photometric data points. 
By treating the H$\alpha$ and \OIII\ lines as very-narrow-bands with a signal-to-noise equal to the ratio of line flux to line flux error, we can fold the observed line flux errors into the estimation of uncertainties of physical parameters.

We assume a Salpeter initial mass function (IMF; \citealt{salpeter55}) with a lower and an upper stellar mass cut-off of 0.1 and 100 M\sol\ (to convert the resultant stellar mass to one from a Chabrier IMF, one can multiply by a factor of 0.55). Star formation histories are parameterized as to be exponentially decreasing (with timescale $\tau$ of [$1 ,10 ,100, 500, 10^3, 2\times10^3, 3\times10^3, 5\times10^3, 10\times10^3$] Myr), 
effectively constant ($\tau$ = 100 Gyr), and exponentially increasing ($\tau$ = [$-300, -10^3, -10\times10^3$] Myr) as recent studies show an indication that the average star formation history above $z \sim$ 2--3 is rising with time \citep{papovich11,  finlator11, reddy12, jaacks12}.  The model ages span from 1 Myr to the age of the universe at the redshift of each object, metallicity ranges from $Z = 0.02 Z$\sol~to $Z = 2.5 Z$\sol, and internal dust extinction varies from $E(B-V) =$ 0 to $E(B-V) =$ 0.6, assuming the extinction law from \citet{calzetti00}.  Intergalactic medium (IGM) attenuation is included using the prescription from \citet{madau95}, but due to large uncertainties in modeling the Ly$\alpha$ line and IGM attenuation, we restrict the SED fitting to wavelengths longward of the Ly$\alpha$ line.

The best-fit model is determined by maximizing a log likelihood, $L \propto e^{-\chi^2/2}$, assuming data points with Gaussian errors. The redshift of model templates is fixed to the systemic redshift measured from the observed H$\alpha$ and/or \OIII\ lines.
 To account for potential zeropoint uncertainties, we add an error proportional to the flux of 5\% for \textit{HST} and 10\% for ground-based and \textit{Spitzer} bands in quadrature to the photometric flux errors in each band.  The uncertainties of the derived physical properties are obtained from 10$^3$ Monte Carlo simulations with the observed photometry perturbed within the corresponding errors.  Table \ref{tab:sedfit} lists the physical properties inferred from SED fitting, and Figure \ref{fig:sedfit} shows the best-fit model and observed SED for each object.

\begin{deluxetable}{lcccc}
\tabletypesize{\scriptsize}
\tablecaption{\label{tab:sedfit} SED Fitting Results}
\tablewidth{0pt}
\tablehead{
\colhead{Object ID} & \colhead{log Stellar Mass} & \colhead{log Age} & \colhead{E(B-V)} & \colhead{$\chi_r^{2}$}\\
\colhead{} & \colhead{(M\sol)} & \colhead{(yr)} & \colhead{} & \colhead{}\\
}
\startdata

HPS\,182 &  \phantom{1}8.87 $^{ +0.07}_{ -0.08}$ &   6.6 $^{ +0.0}_{ -0.0}$ &  0.28 $^{ +0.02}_{ -0.02}$ &    1.3 \\
HPS\,183 &  \phantom{1}7.94 $^{ +0.14}_{ -0.18}$ &   6.7 $^{ +0.1}_{ -0.2}$ &  0.08 $^{ +0.04}_{ -0.04}$ &    2.1 \\
HPS\,189 &  \phantom{1}8.55 $^{ +0.11}_{ -0.01}$ &   6.7 $^{ +0.0}_{ -0.1}$ &  0.18 $^{ +0.02}_{ -0.00}$ &    1.4 \\
HPS\,194 & 10.09 $^{ +0.11}_{ -0.02}$ &   8.3 $^{ +0.3}_{ -0.0}$ &  0.20 $^{ +0.02}_{ -0.06}$ &    1.5 \\
HPS\,251 &  \phantom{1}9.02 $^{ +0.07}_{ -0.06}$ &   8.0 $^{ +0.1}_{ -0.1}$ &  0.10 $^{ +0.02}_{ -0.02}$ &    1.5 \\
HPS\,256 &  \phantom{1}8.27 $^{ +0.07}_{ -0.06}$ &   6.5 $^{ +0.0}_{ -0.2}$ &  0.08 $^{ +0.02}_{ -0.00}$ &    1.7 \\
HPS\,286 &  \phantom{1}9.12 $^{ +0.01}_{ -0.09}$ &   9.5 $^{ +0.0}_{ -1.9}$ &  0.04 $^{ +0.04}_{ -0.02}$ &    1.2 \\
HPS\,306 &  \phantom{1}9.14 $^{ +0.01}_{ -0.58}$ &   7.5 $^{ +0.0}_{ -0.8}$ &  0.04 $^{ +0.00}_{ -0.02}$ &    0.9 \\
HPS\,313 &  \phantom{1}9.869 $^{ +0.421}_{ -0.002}$ &   6.9 $^{ +1.1}_{ -0.0}$ &  0.24 $^{ +0.00}_{ -0.08}$ &    0.6 \\
HPS\,318 &  \phantom{1}9.67 $^{ +0.01}_{ -0.02}$ &   7.3 $^{ +0.2}_{ -0.0}$ &  0.22 $^{ +0.00}_{ -0.00}$ &    1.4 \\

\enddata
\end{deluxetable}

To verify that our Monte Carlo-based parameter uncertainties are robust, we also perform a Bayesian likelihood analysis following \citet{kauffmann03}. Using flat priors in parameter grids, we compute the 4-dimensional posterior probability density function (PDF) of each parameter (dust extinction, age, SFH, and metallicity) using the $\chi^2$ array that fully samples the model parameter space.
Then, we compute 1 dimensional posterior PDFs for each parameter by marginalizing over all other parameters. 
The median value and the central 68\% confidence level (by excluding the 16\% tails at each end) for each parameters are then estimated from these marginalized PDFs.  We find that the Bayesian-derived parameter uncertainties agree well with our original values from the Monte Carlo simulations.  Throughout the paper, we quote values from our original Monte Carlo analysis.

\subsubsection{Stellar Mass}

From the best-fit model obtained as in the previous section, we calculate the stellar mass for each object as the normalization from the observed SED to the best-fit model spectrum which is normalized to a current stellar mass of 1 M\sol.  The inferred stellar masses show a wide range, 7.9 $<$ log $(M_{*}/\rm M_{\odot}) <$ 10.1. The typical uncertainty of our estimated stellar mass is 0.1 dex.

\subsubsection{Dust Extinction}

From our SED fitting, the color excess ranges from $E(B-V)=$ 0.04 to $E(B-V)=$ 0.28, which corresponds to a visual extinction range from $A_{\rm V}$ = 0.16 to $A_{\rm V}$ = 1.13 mag. This is comparable to the dust obscuration of $\langle E(B-V) \rangle=$ 0.22 ([0.00, 0.31]) found from SED fitting analysis for $\sim$ 200 $z \sim$ 2.1 LAEs from the narrowband MUSYC survey \citep{guaita11}. 

To test the validity of the color excess inferred from the SED fitting, we derive the color excess from the Balmer decrement measurements for the two objects with a $>$\,3$\sigma$ H$\beta$ detection assuming an intrinsic H$\alpha$/H$\beta$ ratio of 2.86 (Case B recombination at T=10$^4$ K and n$_{\rm e}$=10$^2$--10$^4$ cm$^{-3}$; \citealt{brocklehurst71}).  We find $E(B-V)_{\rm Balmer}=$ 0.00 $\pm$ 0.09 and 0.71 $\pm$ 0.46 for HPS\,194 and HPS\,256, respectively. Meanwhile, applying the extra attenuation factor of 2.27 toward \HII\ regions (see Section \ref{sec:sfr} for more discussion) to the color excess inferred from the SED fitting for these objects yields $E(B-V)=$ 0.45$^{+0.05}_{-0.14}$ and 0.18$^{+0.05}_{-0.00}$ for HPS\,194 and HPS\,256, respectively, implying 3.2$\sigma$ and 1.1$\sigma$ deviation. But the H$\beta$ emission for HPS\,194 is contaminated by a sky line, and the H$\beta$ emission for HPS\,256 is detected at only 4$\sigma$ significance. 
Since the H$\beta$ line is only detected for two LAEs, and neither with high significance ($\langle$SNR$_{\rm H\beta}\rangle$ $\sim$ 4), we use the dust reddening ($E(B-V)$) of the best-fit model derived from the SED fitting analysis throughout our study, rather than the observed Balmer decrement.


\subsection{Line Diagnostics}

\subsubsection{Gas-phase Metallicity}\label{sec:metallicity}

Although we do not detect the \NII\ line for any of our LAEs, we can use the upper limits on the \NII\ fluxes estimated in Section \ref{sec:upplimit} to place constraints on the gas-phase metallicities of our LAEs, using the N2 index of \citep{pettini04}. The metallicity (oxygen abundance) is given by
\begin{equation}
12 + \log (O/H) = 8.90 + 0.57 \times N2
\end{equation}
where N2 $\equiv$ log(\NII$\lambda$6583/H$\alpha$). Using the measured H$\alpha$ flux and the 1$\sigma$ upper limit of \NII\ flux for each object, we estimate the 1$\sigma$ upper limit of the metallicity for individual LAEs. The estimated 1$\sigma$ upper limit of metallicity for our LAEs ranges from 12 + log(O/H) = 7.87 to 8.61, with a median upper limit of 8.23.  As we will discuss later, these upper limits are set by the quality of the spectra; thus the higher limits are likely not indicative of higher metallicities. Rather, the objects are fainter, thus their H$\alpha$ lines are less well-detected (and the upper limit on the N2 index is higher), and therefore the resultant metallicity limit is higher.  Better limits will require deeper spectroscopy, available now with the new generation of multi-object NIR spectrographs such as KMOS \citep{sharples12} and MOSFIRE \citep{mclean12}.  

 \begin{figure}[t]
  \epsscale{1.2}
 \plotone{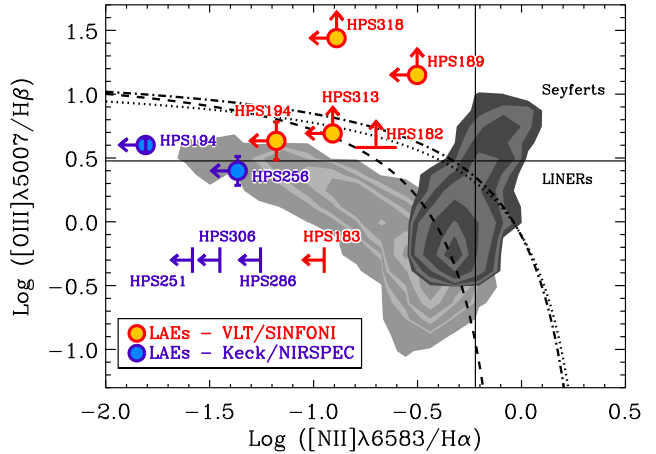}
  \epsscale{1.0}
  \caption{\label{fig:bpt}
 The BPT emission line diagnostic diagram.  Local SDSS star-forming and AGN host galaxies are plotted together as light and dark contours, respectively.  Orange and blue filled circles and lines are our $z \sim$ 2.3 LAEs, 
 where arrows denote 1$\sigma$ limits for objects for which either the H$\alpha$ or \OIII\ line is unavailable (either not observed or undetected).  The black dashed curve represents the boundary between pure star-forming galaxies and star-forming/AGN composites from \citet{kauffmann03}, and the black dotted and dash-dotted curves are the maximum starburst from \citet{kewley01} and its updated version at $z \sim$ 2.5 \citep{kewley13}, respectively. Although many of our objects lie above the SDSS star-forming sequence, similar to other populations of high-redshift galaxies, none are constrained to lie on the AGN sequence, 
 thus we conclude that the chance of AGN contamination in our sample is not high.}
 \end{figure}

\subsubsection{AGN Contamination}\label{sec:agn}

Gas-phase metallicities measured from emission lines are unreliable when there is significant contribution to the emission line flux from active galactic nucleus (AGN) activity, as the AGN ionizing spectra are quite different from those in star-forming regions.  To identify possible AGN contamination, we first search for X-ray counterparts for our objects; we find no associated X-ray detection (down to a flux limit of 0.73 $\times$ 10$^{-15}$ erg s$^{-1}$ cm$^{-2}$ in the 2--10 keV band; $L_{\rm 2-10\,keV} > 3 \times 10^{43}$ erg s$^{-1}$ at $z$ = 2.3; \citealt{adams11}).
For the subset of our LAEs which has detections in all 4 IRAC channels (5/10 LAEs) in the SEDS TFIT catalog, we use the MIR AGN diagnostic proposed by \citet{donley12} and confirm that none of our LAEs falls in the region of color space expected for AGN. Finally, we search for the presence of AGN via an optical emission line diagnostic diagram (BPT diagram; \citealt{baldwin81}), as shown in Figure \ref{fig:bpt}.  Our samples have elevated \OIII/H$\beta$ ratios compared to low redshift ($z \sim$ 0.1) star-forming galaxies from the SDSS. This elevated \OIII/H$\beta$ ratio has been reported by several studies for some LBGs at high redshift and also for local starbursts with no indication of AGN (e.g., \citealt{erb06b, brinchmann08, liu08}). It is often claimed that their higher SFRs compared to their stellar masses and the associated high ionization parameter is responsible for this shift in the BPT diagram (e.g., \citealt{brinchmann08,liu08,hainline09}). 
We conclude that while we cannot exclude the presence of low-luminosity AGNs which are obscured or undetected, there are no confirmed AGNs in our sample. 
As excluding the two LAEs with large \OIII/H$\beta$ ratios (HPS\,189 and HPS\,318) from our analysis is confirmed not to influence our results qualitatively, we include all the sample in our subsequent study.


\subsection{Star Formation Rate}\label{sec:sfr}

The relative extinction suffered by the stellar continuum and nebular emission is not a settled issue: some studies of local star-forming galaxies and starbursts and high redshift star-forming galaxies have found evidence of additional attenuation toward \HII\ regions (e.g., \citealt{calzetti00, forster09,wuyts11}), while others favor the same amount of dust extinction for nebular emission as for the stellar continuum (e.g., \citealt{erb06c,reddy10}).  We test these two scenarios, the first one of $E(B-V)_{\rm stellar} = E(B-V)_{\rm neb}$ and the other of $E(B-V)_{\rm stellar} = 0.44 ~E(B-V)_{\rm neb}$, by comparing two SFR indicators, based on the UV continuum and H$\alpha$ emission strength \citet{kennicutt98b}, and correcting both for our measured dust extinction.  We find that assuming a greater extinction toward the \HII\ regions produces more consistent results ($\langle SFR_{\rm H\alpha} \rangle/\langle SFR_{\rm UV} \rangle=$ 0.77 vs. $\langle SFR_{\rm H\alpha} \rangle/\langle SFR_{\rm UV} \rangle=$ 0.39)  for our samples, 
and thus we correct the observed H$\alpha$ fluxes for internal dust extinctions assuming the ionized gas suffers a greater extinction as suggested by \citet{calzetti00}.

Then H$\alpha$ fluxes are converted to H$\alpha$ luminosities as
\begin{equation}
L(H\alpha)_{\rm corr} = f(H\alpha)_{\rm obs} \times 10^{0.4 E(B-V)_{neb} k(\lambda_{\rm H\alpha})} \times 4\pi D_{\rm L}^2
\end{equation} 
where $k(\lambda)$ is the Calzetti extinction curve, and $D_{\rm L}$ is the luminosity distance for  the systemic redshift inferred from the H$\alpha$ and/or \OIII\ line.
We derive SFRs using the \citet{kennicutt98b} prescription, SFR(H$\alpha$) (M$_{\odot} \rm ~yr^{-1}$)= 7.9 $\times 10^{-42}~ L(\rm H\alpha)_{\rm corr}$ ($\rm erg~s^{-1}$), assuming a Salpeter IMF and solar metallicity.

\begin{figure}[t]
  \epsscale{1.2}
 \plotone{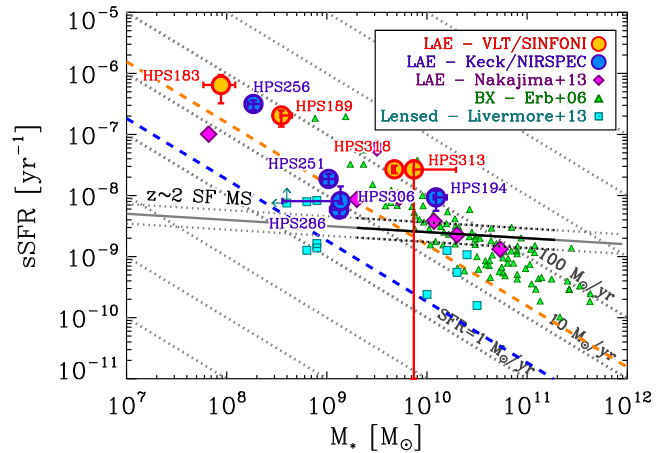}
  \epsscale{1.0}
  \caption{\label{fig:ssfr_vs_M}
  The specific star formation rate, sSFR $\equiv$ SFR(H$\alpha$)/$M_*$, versus the stellar mass for our LAE sample (orange and blue filled circles).  
  The 3$\sigma$ flux limits of our VLT/SINFONI and Keck/NIRSPEC data and the median color excess are converted to 
  the lower limit of SFRs at the median redshift of our samples (15.5 M\sol yr$^{-1}$ for VLT/SINFONI, 1.8 M\sol yr$^{-1}$ for Keck/NIRSPEC) and are shown as orange and blue dashed diagonal lines, highlighting that the lower bound of the observed trend is likely due to selection effects.  LAEs from \citet{nakajima13}, continuum-selected star-forming galaxies (BX galaxies) from \citet{erb06c}, and lensed galaxies from \citet{livermore13} 
  at similar redshifts are plotted as magenta diamonds, green triangles, and cyan squares, respectively.  
  Black lines indicate the $z \sim$ 2 star-forming ``main-sequence'' (black solid line) defined from BzK star-forming galaxies (sBzKs) by \citet{daddi07}, its extrapolation to higher/lower masses than those probed (grey solid line), and the interquartile range of 0.32 dex in sSFR (black dotted lines). All points and lines plotted are converted to a Salpeter IMF. Also shown are grey dotted diagonal lines of constant SFRs.  The massive LAEs appear consistent with the ``main sequence'', as well as with continuum-selected star-forming galaxies at the same redshift.  Lower-mass LAEs ($M_* \lesssim 10^{9}$ M\sol) have elevated sSFRs, indicating that low-mass LAEs may be star-bursting sources, although our selection renders us unable to see ``main-sequence'' galaxies at these masses.
}
 \end{figure}

The sample is characterized by a mean (median) SFR value of 74 (58) M\sol~yr$^{-1}$, ranging between 8 and 197 M\sol~yr$^{-1}$.\footnote{When a mean or median value is quoted, we use a simple arithmatic mean and median, which means that the PDF for each object is assumed to be symmetric about its mean and have a similar width.}
 This is comparable to the average SFR of $\langle SFR_{H\alpha} \rangle=$ 78 M\sol~yr$^{-1}$ for four LAEs at similar redshift with H$\alpha$ detection in \citet{hashimoto13},  but larger than that of $\langle SFR_{\rm SED} \rangle=$ 35 M\sol~yr$^{-1}$ inferred from SED fitting analysis for narrowband selected LAEs in \citet{guaita11}. 
Using the UV SFR indicator and the \citet{kennicutt98b} conversion, SFR(UV) (M$_{\odot} \rm ~yr^{-1}$) = 1.4 $\times 10^{-28}~L_{\rm \nu, UV}$ ($\rm erg~s^{-1} Hz^{-1}$), we find a mean (median) value of 92 (37) M\sol~yr$^{-1}$.  
However, these indicators probe different regimes; the H$\alpha$ SFR is sensitive to the instantaneous SFR, while the UV indicator probes the average SFR over the past 100 Myr.  
For galaxies younger than 100 Myr, SFR(UV) derived from the Kennicutt conversion which is based on an assumption of constant star formation history over the past 100 Myr underestimates the ``true'' (time-averaged) SFR. Although our attenuation test showed that H$\alpha$ and UV-based SFR values are consistent, the result from our SED fitting analysis implies that many galaxies in our sample are young ($t <$ 100 Myr). Combined with the young age for LAEs in the literature (e.g., \citealt{gawiser06, fin09a}), the H$\alpha$ SFR is likely more indicative of the true SFR, and it has a lower dust correction, thus we use that estimate in our subsequent analysis.

With our derived SFRs, we investigate the relation between the specific star formation rate, sSFR $\equiv$ SFR/$M_*$, versus stellar mass for our sample in Figure \ref{fig:ssfr_vs_M}. Massive LAEs have sSFRs similar to those of continuum-selected galaxies at the same redshift from \citet[][a few $\times$ 10$^{-9}$ yr$^{-1}$]{erb06c}, following the $z \sim$ 2 star-forming ``main-sequence'' \citep{daddi07}, while low-mass LAEs appear to be undergoing a star-bursting phase with a stellar mass-doubling timescale of as short as a few million years.  The lower bound of the diagonal distribution in sSFR and $M_*$ is likely a combined effect of the LAE selection and detection limit of H$\alpha$, i.e., the high sSFR for low-mass LAEs is attributed to the LAE selection method, which requires a bright Ly$\alpha$ emission which (roughly) correlates with SFR, as well as the H$\alpha$ detection limit of our data.


\subsection{Size}\label{sec:size}

As the H$\alpha$ emission is not spatially resolved in our NIR seeing-limited data, we utilize \textit{HST} rest-frame UV imaging to measure the half-light radius and surface areas associated with star formation activity in our sample of LAEs, since, as mentioned above, the rest-frame UV also probes recent star formation (albeit on a longer timescale).
The half-light radius of each object is measured in the \textit{HST}/F814W image from the ACS parallels to the CANDELS survey (v1.0). 
Most of our sample is spatially-resolved at the $\sim$ 0\farcs06 ($\sim$ 0.5 kpc at $z$=2.3) pixel scale of the CANDELS data, including a marginally-resolved HPS\,182 ($r$ = 0.6 kpc). 
Using the redshift of each object and the pixel scale of the image, we then convert the measured size to the physical size and surface area, $SA=\pi r^2$.  
When two or more clumps or galaxies appear blended in the ground-based imaging data that we utilized in our SED fitting, we first calculate the total surface area, $SA=\Sigma (\pi r_i^2)$, and obtain the equivalent half-light radius as $r= (SA/\pi)^{1/2}$. The half-light radius ranges from 0.6 kpc to 2.9 kpc, with a mean (median) of 1.5 (1.5) kpc (Table \ref{tab:prop}). This is comparable (or slightly larger) than the mean half-light radius of 1.3 $\pm$ 0.2 kpc  \citep{malhotra12} or the median of 1.4 kpc \citep{bond12} for narrowband selected LAEs at $z \sim$ 2, but our sample displays a wider distribution in the size, as our sample includes more massive (and larger) LAEs than those from narrowband surveys.

The assumption we are making here, that the H$\alpha$-emitting star-forming regions are identical to UV-emitting star-forming regions, breaks when there exist short-term ($<$ 100 Myr) spatial fluctuations in the recent star formation history, since the star formation timescales traced by UV and H$\alpha$ are $\sim$ 100 Myr and $\sim$ 10 Myr, respectively.  However, given the typically young nature of LAEs (see Section \ref{sec:sfr}), the H$\alpha$ flux is likely a better SFR estimator, but the UV morphology should also well-represent the size of the star-forming regions.

 \begin{figure}[t]
  \epsscale{1.2}
 \plotone{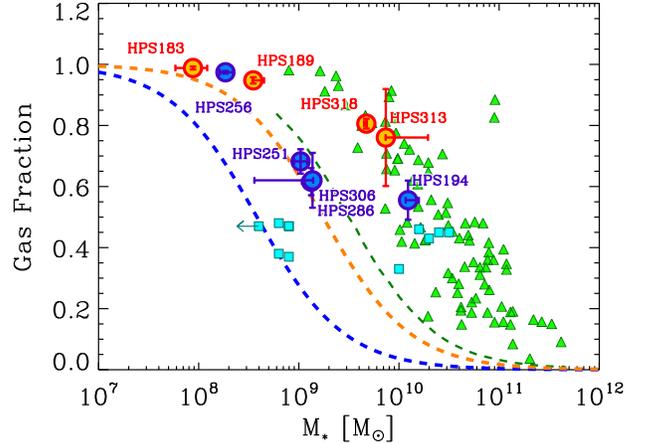}
  \epsscale{1.0}
  \caption{\label{fig:gasfrac_vs_M}
  The gas-mass fraction versus stellar mass, where the gas-mass fraction is estimated from the inversion of Kennicutt-Schmidt law using the SFR(H$\alpha$) and size measured from \textit{HST} rest-frame UV imaging. Vertical error bars include
  the uncertainties in the observed H$\alpha$ flux and the color excess described in Sections \ref{sec:line_detection} and \ref{sec:sed}, respectively, 
  but not the systematic uncertainties associated with, e.g., the $L(\rm H\alpha)$--SFR conversion and the relation between SFR and gas density.
  Orange and blue dashed lines represent our observational limit of the gas fraction for the VLT/SINFONI and Keck/NIRSPEC data, respectively (see text). 
  Green triangles and cyan squares denote $z \sim$ 2.2 BX galaxies from \citet{erb06b} and lensed galaxies at $1.5<z<3.0$ from \citet{livermore13}, respectively, converted into a Salpeter IMF. Gas fractions from these studies are inferred using the same methodology as ours. 
 }
 \end{figure}

\subsection{SFR Surface Density and Gas Fraction}\label{sec:gasfrac}

Optical/NIR imaging of galaxies does not reveal the whole picture, as high-redshift galaxies have substantial gas reservoirs fueling their active star formation.
Direct gas measurements are challenging for high-redshift galaxies and are biased toward 
luminous and massive galaxies (\citealt{tacconi10, geach11}; but see \citealt{tacconi13}) except for a few lensed galaxies (e.g., \citealt{livermore13}). 
Consequently, the gas fraction for high-redshift galaxies is often inferred assuming the Kennicutt-Schmidt Law \citep{kennicutt98a} 
which relates the gas surface density to the SFR surface density (e.g., \citealt{erb06b,fin09b, weinzirl11}).
Although we have no direct measurements for the gas content of our sample, 
we can obtain a rough estimate of the gas properties of our sample using the same methodology. We caution however that the results obtained via this method can be uncertain by a factor of 2--3, as the relation between gas surface density and SFR surface density for various galaxy types and redshifts show systematic deviation from the original relation (e.g., \citealt{daddi10,genzel10,kennicutt12,tacconi13}). Other sources of error includes the assumption that the spatial extent of star formation is related to that of the gas.

We first combine the dust-corrected SFR inferred from H$\alpha$ emission in Section \ref{sec:sfr} and surface area in Section \ref{sec:size} to estimate the SFR surface density $\Sigma_{\rm SFR} = {\rm SFR}/SA$.  The measured SFR surface density of our sample ranges from 0.7 to 32.8 M\sol\ $\rm yr^{-1}~ kpc^{-2}$ and is characterized by a mean value of
 $\langle \Sigma_{\rm SFR} \rangle = 10.3 $ M\sol\ $\rm yr^{-1} ~kpc^{-2}$. This is comparable to the typical SFR surface density seen in local starbursts (excluding the high end tail) and Giant Molecular Clouds in the Milky Way, but much higher than that seen in local blue compact dwarfs (see Figure 9 in \citealt{kennicutt12}). Compared to galaxies at similar redshifts, the range overlaps that for BX galaxies from \citeauthor{erb06b} ([0.5, 10.7] M\sol\ $\rm yr^{-1} ~kpc^{-2}$), but the mean SFR surface density of our LAEs is higher ($\langle \Sigma_{\rm SFR} \rangle  _{\rm BX}= 2.8 \pm 2.4 $ M\sol\ $\rm yr^{-1} ~kpc^{-2}$).

The (total \HI +H$_2$) gas surface density is measured by applying the inversion of the Kennicutt-Schmidt law \citep{kennicutt98a}:
\begin{equation}
\Sigma_{\rm gas} = 2.4 \times 10^{2}  \left(\frac{\Sigma_{\rm SFR}}{\rm M_{\odot}~yr^{-1} ~kpc^{-2}}\right)^{0.71}~ {\rm M_{\odot} ~pc^{-2}}
\end{equation}
We do not apply a factor 1.36 in mass to account for helium in order to remain consistent with other works to which we compare our results.  We can now obtain the gas mass 
$M_{\rm gas} =  \Sigma_{\rm gas} \times SA$ and gas fraction $\mu = M_{\rm gas}/(M_{*}+M_{\rm gas})$.

\begin{figure}
  \epsscale{1.2}
 \plotone{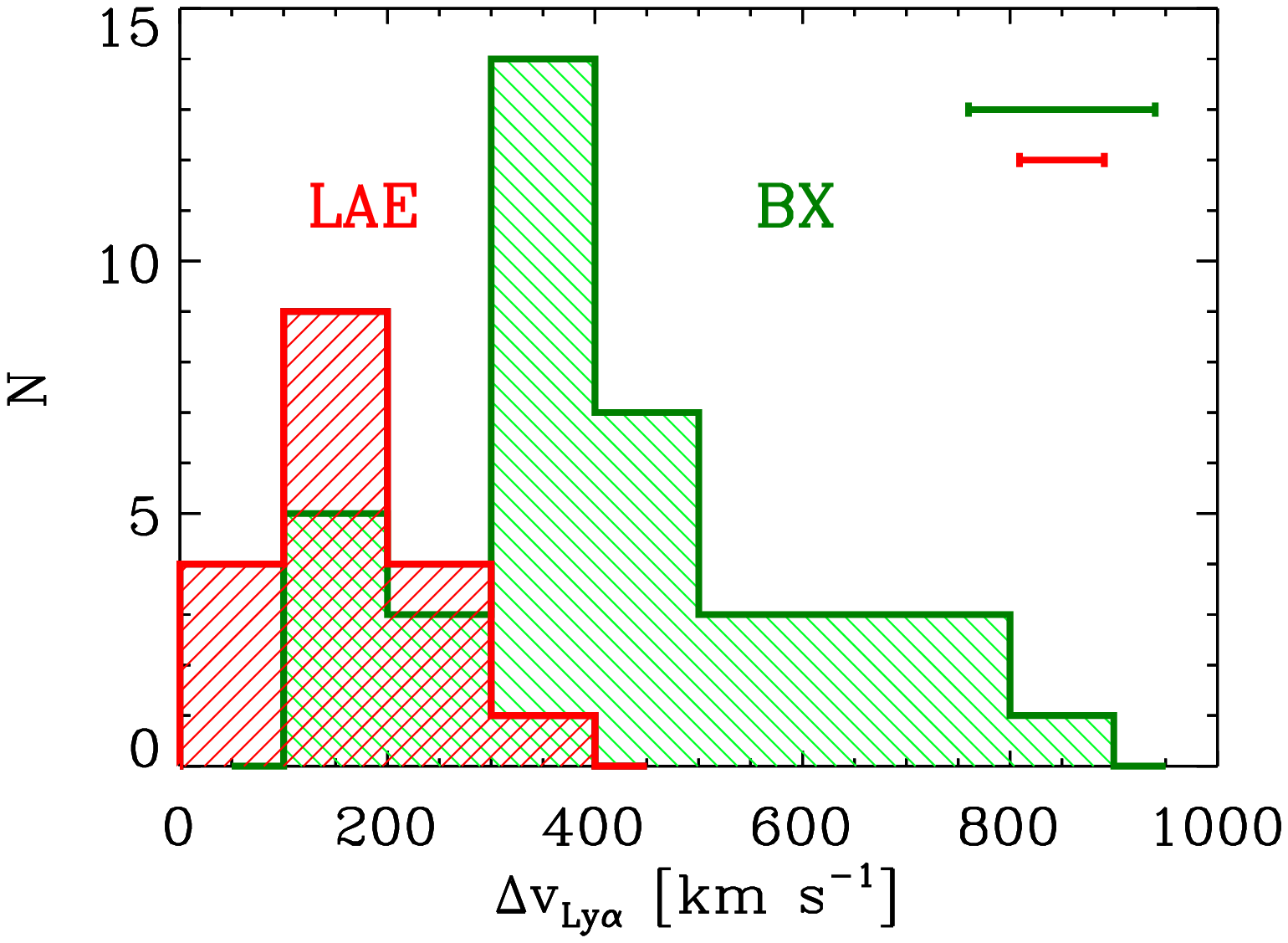}
  \epsscale{1.0}
  \caption{\label{fig:veloffset}
  Compilation of Ly$\alpha$ velocity offsets for 18 LAEs at $z$=2--3 (10 from this study, 2 from \citealt{mclinden11}, 4 from \citealt{hashimoto13}, 2 from \citealt{guaita13}) 
  of which Ly$\alpha$ velocity offsets are measured from the centroids of Ly$\alpha$ lines and 
  nebular (H$\alpha$ and/or \OIII) lines, where the latter represent the systemic redshifts of these galaxies. Ly$\alpha$ offsets of $z \sim$ 2.2 BX galaxies from \citet{steidel10} are shown together for comparison. On the upper right corner are shown the typical uncertainties in velocity offsets for LAEs and BX galaxies.
  We find that all our LAEs show redshifted Ly$\alpha$ emission compared to their systemic redshifts, but 
  have systematically smaller Ly$\alpha$ velocity offsets than BX galaxies. 
 }
 \end{figure}

In Figure \ref{fig:gasfrac_vs_M}, we plot the derived gas-mass fraction as a function of the stellar mass. 
We find in general $M_{\rm gas} > M_*$ for our LAEs.  
The inferred gas-mass fraction reaches near unity for low-mass LAEs and decreases overall with stellar mass, 
following similar trends found in other studies on more massive galaxies at similar redshifts. 
This observed trend, however, is likely dominated by the selection bias. Using the median redshift and color excess of our samples and adopting the minimum radius of 0.6 kpc (comparable to the minimum detection area adopted in our size measurement), the median 3$\sigma$ flux limits of our VLT/SINFONI ($\sim$ 2.8 $\times$ 10$^{-17}$ erg s$^{-1}$ cm$^{-2}$) and Keck/NIRSPEC ($\sim$ 3.3 $\times$ 10$^{-18}$ erg s$^{-1}$ cm$^{-2}$) data estimated in Section \ref{sec:upplimit} are translated into the lower limits in gas-mass fraction, and are shown as orange and blue dashed lines, respectively. 
The lines of our lower limits in gas fraction are located well below the data points by $\gtrsim$ 0.3 in gas fraction except the low mass end ($M_* \lesssim 10^{8.5}$ M\sol), but following the observed trend of our data points and \citet{erb06b}'s.  This result suggests that the trend seen in gas fraction versus stellar mass could be due in part to the observational limits, especially at the low-mass end, since we may be probing only the upper envelope of the distribution between the gas fraction and stellar mass as suggested by the existence of lensed galaxies with lower gas fraction than our limits in Figure \ref{fig:gasfrac_vs_M}. Although our sample size is small, however, the lack of massive objects with high gas fraction in the LAE population as well as in BX galaxies is suggestive.

The inferred gas-mass fractions for massive LAEs are 
comparable to the average gas mass fraction of $\sim$ 50\% for $z \sim$ 2.3 BX galaxies found in \citet{erb06b}, and much higher than the average gas mass fraction of local star-forming galaxies of $\sim$ 5\% \citep{saintonge11}. Although the uncertainty in gas-mass fraction 
estimated from the indirect method is large, Erb et al. used the same methodology and 
\citet{tacconi10} reported a similar results: an average molecular gas fraction of 44\% from CO observations for a subset of the \citeauthor{erb06b} sample.

\begin{figure}[t]
  \epsscale{1.2}
 \plotone{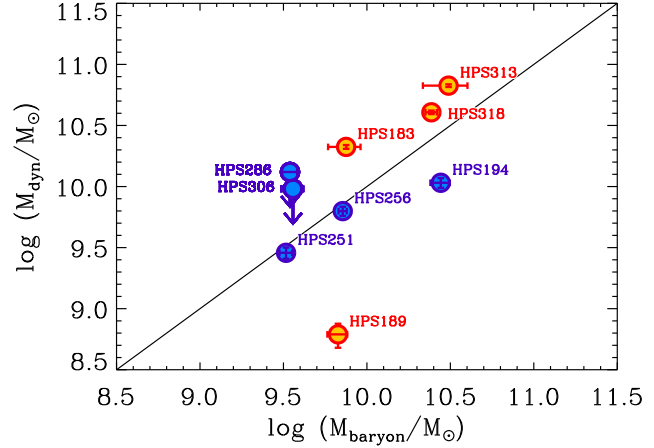}
  \epsscale{1.0}
  \caption{\label{fig:mdyn_vs_mbar}
  The dynamical mass versus baryonic (stellar + gas) mass. The dynamical mass is estimated from the H$\alpha$ line widths and half-light radius, as $M_{\rm dyn} \sim 5 \sigma^2 r/G$.
A 1:1 line is shown for reference.
 }
 \end{figure}

 \begin{deluxetable*}{ccccccccc}
\tabletypesize{\scriptsize}
\tablecaption{\label{tab:prop} Physical Properties of LAEs}
\tablewidth{0pt}
\tablehead{
\colhead{Object ID} 
& \colhead{12+log(O/H)$_{\rm N2}$\,\tablenotemark{a}} & %
\colhead{$\Delta v_{\rm Ly\alpha}$\,\tablenotemark{b}} 
& \colhead{SFR\,\tablenotemark{c}} & 
\colhead{size\,\tablenotemark{d}} & \colhead{$\mu$\,\tablenotemark{e}} & \colhead{EW(H$\alpha)_{\rm rest}$\,\tablenotemark{f}} 
& \colhead{$\sigma$\,\tablenotemark{g}}
& \colhead{log $M_{\rm dyn}$\,\tablenotemark{h}}  \\
\colhead{} & \colhead{} & \colhead{(km s$^{-1}$)} & \colhead{(M\sol yr$^{-1}$)} & \colhead{(kpc)} & \colhead{} & \colhead{(\AA)} & \colhead{(km s$^{-1}$)} & \colhead{(M\sol)}\\
}
\startdata

\smallskip
\textbf{VLT/SINFONI} &&&&&& \\

HPS\,182  & --- & \phantom{1}85 $\pm$   36 $\pm$   --- &    ---  &    0.6 &   ---  &     ---  & \phantom{1}67 $\pm$ 3 &   \phantom{1}9.47 $^{ +0.04}_{ -0.04}$   \\
HPS\,183  & $<$ 8.36 & 161 $\pm$  44 $\pm$  --- &   \phantom{1}56.7 $\pm$ 17.8 &    1.5 &   0.988 $\pm$   0.005 & 6849 $^{ +3489}_{ -3168}$ & 112 $\pm$ 2 &  10.32 $^{ +0.02}_{ -0.02}$   \\
HPS\,189 &  $<$ 8.61 & 254 $\pm$   35 $\pm$   28 &   \phantom{1}71.8 $\pm$  14.5 &    0.8 &   0.95 $\pm$   0.01 &  1067 $^{ +215\phantom{1}}_{ -234\phantom{1}}$ & \phantom{1}25 $\pm$ 3 &  \phantom{1}8.79 $^{ +0.09}_{ -0.11}$   \\
HPS\,194  & $<$ 8.23 & 255 $\pm$  37 $\pm$  16 &   131.2 $\pm$ 56.8 &    2.2 &   0.58 $\pm$  0.07 &  \phantom{1}180 $^{ +20\phantom{1}\phantom{1}}_{ -22\phantom{1}\phantom{1}}$ & \phantom{1}74 $\pm$ 2 & 10.14 $^{ +0.02}_{ -0.02}$   \\
HPS\,313 & $<$ 8.38 & 171 $\pm$  44 $\pm$  10 &  197.0 $\pm$ 81.7 &    2.3 &   0.76 $\pm$ 0.16 &   \phantom{1}\phantom{1}97 $^{ +32\phantom{1}\phantom{1}}_{ -30\phantom{1}\phantom{1}}$ & 158 $\pm$ 2 &  10.83  $^{ +0.01}_{ -0.01}$   \\
\smallskip
HPS\,318  & $<$ 8.39 & 296 $\pm$  35 $\pm$  57 &  127.0 $\pm$ 17.5 &    2.9 &   0.81 $\pm$ 0.02 &    \phantom{1}189 $^{ +28\phantom{1}\phantom{1}}_{ -26\phantom{1}\phantom{1}}$ & 110 $\pm$ 2 &  10.61 $^{ +0.01}_{ -0.01}$   \\

\tableline \\
\smallskip
\textbf{Keck/NIRSPEC} &&&&&& \\

HPS\,194 & $<$ 7.87 & 296 $\pm$   37 $\pm$  23 &   114.8 $\pm$ 48.1 &    2.2 &   0.56 $\pm$  0.06 &   \phantom{1}157 $^{ +5\phantom{1}\phantom{1}\phantom{1}}_{ -8\phantom{1}\phantom{1}\phantom{1}}$ & \phantom{1}65 $\pm$ 3 & 10.03 $^{ +0.04}_{ -0.04}$   \\
HPS\,251 & $<$ 8.00 &  146 $\pm$   37 $\pm$   --- &   \phantom{1}19.5 $\pm$  \phantom{1}2.9 &    0.7 &   0.68 $\pm$  0.04 &   \phantom{1}294 $^{ +25\phantom{1}\phantom{1}}_{ -23\phantom{1}\phantom{1}}$ & \phantom{1}60 $\pm$ 3 &  \phantom{1}9.46 $^{ +0.04}_{ -0.05}$   \\
HPS\,256 & $<$ 8.12 & 161 $\pm$  35 $\pm$ 12 &   \phantom{1}58.5 $\pm$ \phantom{1}4.8 &    1.3 &   0.974 $\pm$ 0.004 &  7327 $^{ +1109}_{ -1820}$ & \phantom{1}66 $\pm$ 3 &  \phantom{1}9.80 $^{ +0.04}_{ -0.04}$   \\
HPS\,286 & $<$ 8.18 &  \phantom{1}93 $\pm$   38 $\pm$   --- &    \phantom{1}\phantom{1}7.8 $\pm$ \phantom{1}2.2 &    1.9 &   0.62 $\pm$  0.04 &  \phantom{1}119 $^{ +9\phantom{1}\phantom{1}}_{ -12\phantom{1}\phantom{1}}$ & $<$ 77 &  $<$ 10.12   \\
HPS\,306 & $<$ 8.07 &  126 $\pm$   35 $\pm$  --- &    \phantom{1}11.1 $\pm$ \phantom{1}1.0 &    1.4 &   0.62 $\pm$  0.09 &  \phantom{1}155 $^{ +32\phantom{1}\phantom{1}}_{ -10\phantom{1}\phantom{1}}$ & $<$ 78 &   $<$ 9.98 \\

\enddata
\tablenotetext{a}{1$\sigma$ upper limit of oxygen abundance from the N2 index of \citet{pettini04}.}
\tablenotetext{b}{$\Delta v_{\rm Ly\alpha} \pm \delta(\rm phot) \pm \delta(\rm sys)$, where the systematic error is only available when we have 2 measurements of $z_{\rm sys}$.}
\tablenotetext{c}{Dust-corrected SFR derived by applying the Kennicutt conversion \citep{kennicutt98b} to the H$\alpha$ luminosity. 
 The observed H$\alpha$ emission is corrected for dust extinction using $E(B-V)$ of the best-fit model from the SED modeling, 
assuming $A_{V,\rm stellar} = 0.44~ A_{V, \rm nebular}$ and the Calzetti extinction law \citep{calzetti00}.}
\tablenotetext{d}{Half-light radius measured from rest-frame UV imaging.}
\tablenotetext{e}{Gas-mass fraction $\mu= M_{\rm gas}/(M_{*}+M_{\rm gas})$, estimated from the inversion of Kennicutt-Schmidt Law, 
using the measured SFR (column 4), size (column 5), and stellar mass (column 2 in Table \ref{tab:sedfit}).}
\tablenotetext{f}{Rest-frame H$\alpha$ equivalent widths estimated from the observed H$\alpha$ flux and the continuum of the best-fit stellar synthesis model.}
\tablenotetext{g}{Line-of-sight velocity dispersion derived from H$\alpha$ line widths.}
\tablenotetext{h}{Dynamical mass derived from the line-of-sight velocity dispersion (column 8) and inferred size (column 5).}
\end{deluxetable*}


\subsection{Kinematics: Ly$\alpha$ Velocity Offsets}\label{sec:dv}

In this section, we measure the difference between the redshift of Ly$\alpha$ and the systemic redshift (called the Ly$\alpha$ velocity offset), using the systemic redshift as measured from H$\alpha$ and/or \OIII.  We will investigate in Section \ref{sec:dv_vs_prop} correlations between the Ly$\alpha$ velocity offsets and other physical properties.

To calculate Ly$\alpha$ velocity offsets properly, we first correct our observed data for Earth's motion during our observations.
We utilize an IDL translation (written by D. Nidever) of the IRAF task {\tt rvcorrect} to calculate the radial heliocentric velocity of the observer with respect to the heliocentric frame, 
$v_{\rm helio}$, for each object. Using the median time of the observation, we find $v_{\rm helio}$ to range between 
[$-$18.6, $+$23.9] km s$^{-1}$. 
Wavelengths of \OIII\ and H$\alpha$ (in vacuum) are adjusted to the heliocentric frame of reference using the estimated values.
The systemic redshift for each object is calculated as the weighted mean of the redshifts measured from H$\alpha$ and \OIII, or either of the two when only one is detected (Table \ref{tab:linedet}).

We find for each LAE in our sample that the Ly$\alpha$ line is observed at a slightly higher redshift than the systemic redshift.  Figure \ref{fig:veloffset} shows the histogram of Ly$\alpha$ velocity offsets, compiled from LAEs at $z$ = 2--3, which is the only epoch where this quantity has been measured for a sizeable number of LAEs (including the 10 LAEs from this study, as well as two from \citealt{mclinden11}, four from \citealt{hashimoto13}, and two from \citealt{guaita13}).  For comparison, we also show the distribution of Ly$\alpha$ velocity offsets for continuum-selected star-forming galaxies at $z$ = 2--3 from \citet{steidel10}.  Interestingly, we find the mean Ly$\alpha$ velocity offsets of LAEs to be +180 km s$^{-1}$, 
a factor of $\sim$2--3 smaller than that of continuum-selected star-forming galaxies at similar redshifts.  This trend, that LAEs show systematically smaller Ly$\alpha$ velocity offsets than those of continuum-selected star-forming galaxies, will be discussed further in Section \ref{sec:dv_vs_prop}.

\subsection{Equivalent Widths: Ly$\alpha$ and H$\alpha$}\label{sec:ew}

For objects such as LAEs with faint continuum levels, it is difficult to measure the equivalent width (EW) of emission lines from spectroscopy alone.  Often, the broad-band flux is used to determine the continuum flux near the line, but the associated uncertainty is large.  Thus, we utilize the best-fit stellar population models to estimate the EWs of the Ly$\alpha$ and H$\alpha$ lines of our LAEs.
We estimate EWs of the Ly$\alpha$ line using the observed Ly$\alpha$ flux and the mean continuum flux density of the best-fit model in a $\Delta \lambda_{\rm rest}$= 100 \AA\ region at wavelengths redward of the Ly$\alpha$ line (as the region blueward is affected by IGM absorption).
EWs of the H$\alpha$ line are obtained in a similar way, with the observed H$\alpha$ flux and 
the mean best-fit model flux density in two $\Delta \lambda_{\rm rest}$= 100 \AA\ bands, one at wavelengths shortward and one longward of the H$\alpha$ line. 
Uncertainties in the EWs are estimated from the error of the observed line flux (Section \ref{sec:line_detection}) 
and the 68\% range of the model continnum flux density from the Monte Carlo realizations described in Section \ref{sec:sed}.
The derived Ly$\alpha$ and H$\alpha$ rest-frame EWs are tabulated in Table \ref{tab:target} and \ref{tab:prop}, respectively. 
All of our LAEs (with NIR spectroscopic detections) have estimated rest-frame Ly$\alpha$ EWs of EW$_{\rm Ly\alpha}$ $\lesssim$ 240 \AA, which can be explained with normal stellar populations \citep{charlot93}.
Compared to the values in \citet{adams11} and \citet{blanc11} where Ly$\alpha$ EWs are measured for the HETDEX Pilot Survey sample utilizing \textit{R}-band flux or power-law extrapolation of broad-band photometry for determining the continuum flux near Ly$\alpha$, we find their measurements yield $\sim$ 30\% lower values than ours 
($\langle$ EW$_{\rm Ly\alpha} \rangle$ (Adams+11, Blanc+11)/$\langle$ EW$_{\rm Ly\alpha} \rangle$ (this study) = 0.73, 0.69, respectively).

\subsection{Dynamical Masses}\label{sec:mdyn}

The typical FWHM of the H$\alpha$ line for objects observed with VLT/SINFONI is $\sim$ 17 \AA, significantly larger than the $K$-band instrumental resolution of $\sim$ 5 \AA\ at 2.2 $\mu$m ($R \sim$ 4,000). For objects observed with Keck/NIRSPEC, with a $K$-band resolution of $\sim$ 14 \AA\ at 2.2 $\mu$m ($R \sim$ 1,500), we find H$\alpha$ lines for 3 out of 5 objects are resolved.  We utilize these resolved H$\alpha$ (or \OIII\ for one case, HPS\,182) lines to constrain the dynamical mass of our sample.  For objects with unresolved lines, we calculate conservative upper limits on the dynamical masses by assigning the observed line width assuming zero width for the instrumental profile.


\begin{figure*}[t]
  \epsscale{1.2}
 \plotone{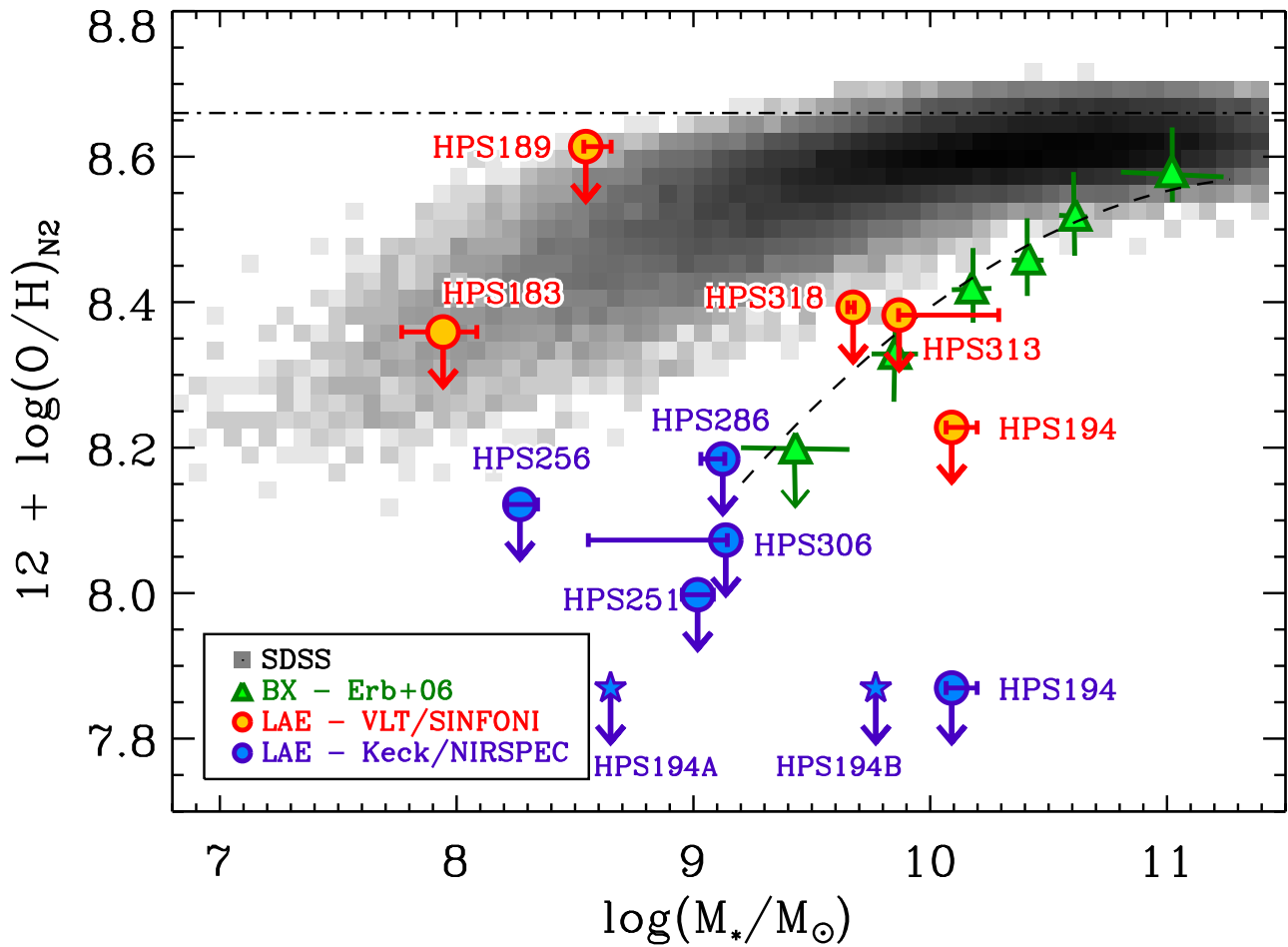}
  \epsscale{1.0}
  \caption{\label{fig:mzr}
  A plot of galaxy stellar mass versus gas-phase metallicity.  Orange (VLT/SINFONI) and blue (Keck/NIRSPEC) filled circles represent 1$\sigma$ upper limits on the metallicities for the LAEs at $z$= 2.1--2.5 from this study.  Each LAE has an 84\% likelihood of lying below these points.  Blue filled stars represent each component of HPS\,194, assuming all the H$\alpha$ emission originates from one or the other. Grey 2-D histogram displys the density of $z \sim$ 0.1 star-forming galaxies (\citealt{tremonti04}; darker regions represent higher density), and green triangles are $z \sim$ 2.3 continuum-selected star-forming galaxies (BX galaxies; \citealt{erb06a}). The dashed line represents the $z \sim$ 0.1 MZR \citep{tremonti04} shifted downward by 0.56 dex, to match the observations by Erb et al.  All points on this figure have their metallicities derived via the N2 index \citep{pettini04}.   
 For reference, we denote the solar metallicity by a dash-dotted line. Our NIR spectra are not deep enough to probe the low-mass end ($\lesssim 10^9$ M\sol) of the MZR of LAEs, but from higher-mass LAEs ($\gtrsim 10^9$ M\sol), where 1$\sigma$ upper limits for nearly all of them lie either on or below the MZR for continuum-selected star forming galaxies, we find hints that galaxies selected on the basis of strong Ly$\alpha$ emission may lie below the MZR for continuum-selected star forming galaxies at the same epoch.
}
 \end{figure*}

 \begin{figure}[t]
  \epsscale{1.2}
 \plotone{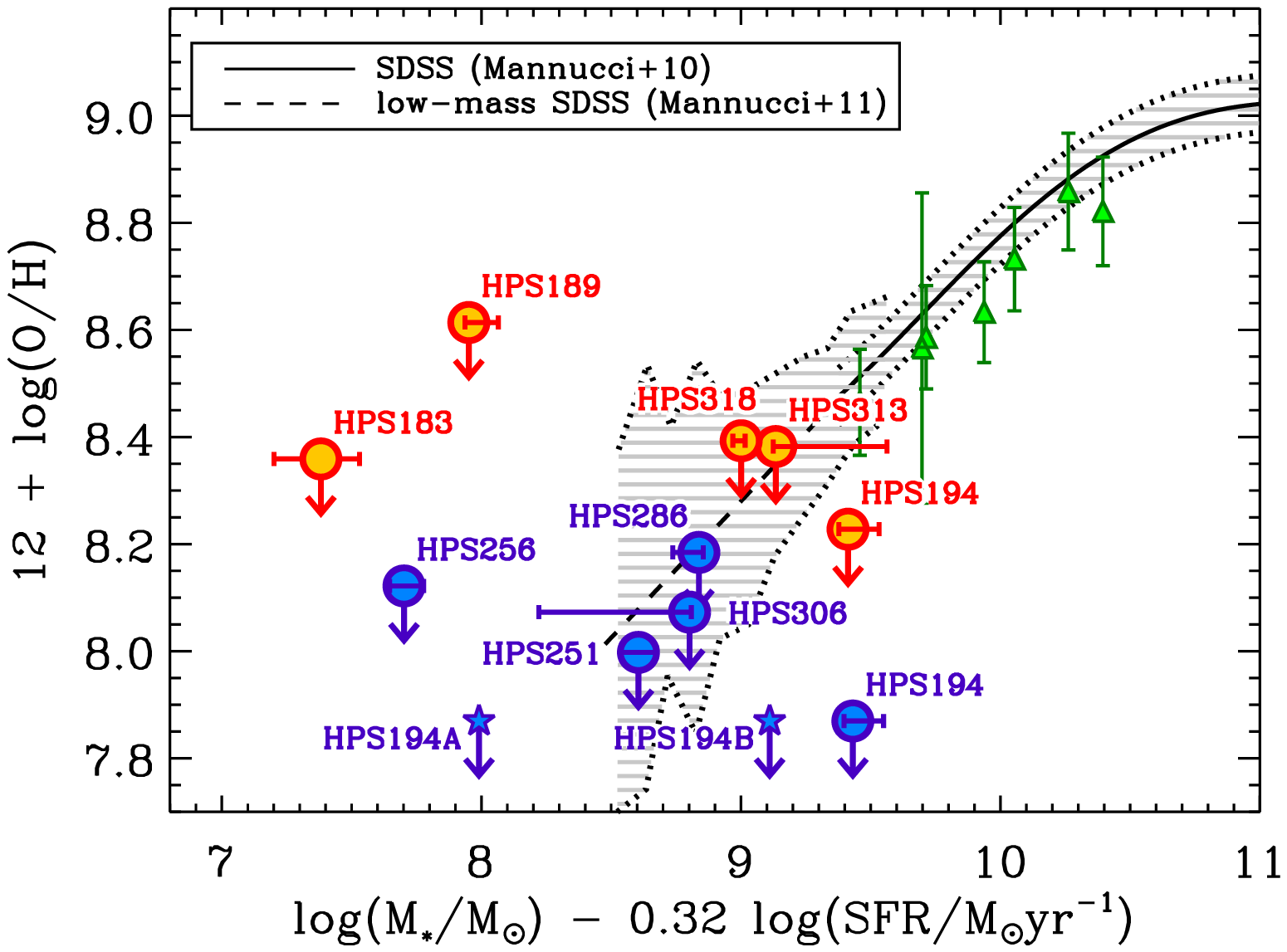}
  \epsscale{1.0}
  \caption{\label{fig:fmr}
  The fundamental metallicity relation (FMR) proposed by \citet{mannucci10,mannucci11}. 
  Solid and dashed black lines are the original FMR and its extension toward lower mass, respectively, and shaded regions indicate the 1$\sigma$ dispersion. For reference, $z \sim$ 2.2 star-forming galaxies from \citet{mannucci10} are shown together in green triangles. 
 }
 \end{figure}

Gaussian FWHMs of the H$\alpha$ line (or \OIII\ when H$\alpha$ was not available) are corrected for the instrumental resolution and then are converted to line-of-sight velocity dispersions.  Then we estimate dynamical masses as $M_{\rm dyn} \sim 5 \sigma^2 r/G$, under the assumption of the virial theorem and a uniform sphere, which has been frequently used in the previous estimates of dynamical masses of high-redshift galaxies (e.g., \citealt{erb03,shapley04}). Here, $\sigma$ is the line-of-sight velocity dispersion, $r$ is the half-light (effective) radius measured in Section \ref{sec:size}, and $G$ is the gravitational constant.  

The measured dynamical masses (thus within the central $r$ kpc traced by the H$\alpha$ emission, and a lower limit on the {\it true} dynamical mass) range from log($M_{\rm dyn}$/M\sol)= 8.8 to 10.8, with a mean of log($M_{\rm dyn}$/M\sol)= 9.9. The derived dynamical mass and the total baryonic (stellar and gas) mass are in overall agreement (Figure \ref{fig:mdyn_vs_mbar}), with the Spearman's rank correlation coefficient \citep{spearman04} of $r_s=$ 0.86 (2.1$\sigma$ deviation from the null hypothesis).

Our analysis above assumes that the nebular line broadening is mainly due to the gravitational potential, and the contribution of inflow and/or outflow is negligible. Although this might not be true, we do not find strong evidence against it: we do not find any correlation between the velocity dispersion and the SFR or Ly$\alpha$ velocity offset which might be expected if the contribution of outflowing material on nebular line broadening is significant \citep{erb03, green10}.

Table \ref{tab:prop} lists the physical properties obtained in this section, i.e., gas-phase metallicity, Ly$\alpha$ velocity offset, SFR, half-light radius, gas fraction, H$\alpha$ equivalent width,  line-of-sight velocity dispersion, and dynamical mass.

\section{Discussion}\label{sec:discussion}

\subsection{Mass--Metallicity Relation \& Fundamental Metallicity Relation}\label{sec:mzr}

There are suggestions that our LAEs may lie below the MZR for continuum-selected star-forming galaxies at a given redshift. This result is shown in Figure \ref{fig:mzr}, where we place our LAE sample on the stellar mass -- gas-phase metallicity plane. Vertical arrows represent the inferred 1$\sigma$ upper limit of the metallicity from the N2 index calibrated by \citet{pettini04}, and the horizontal error bars illustrate the 68\% confidence interval in stellar mass. For reference, we also plot continuum-selected (BX) galaxies at $z \sim$ 2.3 from the stacking analysis by \citet[green triangles]{erb06a} and local star-forming galaxies from the SDSS \citep[grey 2-D histogram]{tremonti04}.  For consistency with our work, the stellar masses of SDSS galaxies for which a Kroupa IMF \citep{kroupa01} is assumed are converted to those with a Salpeter IMF (multiplied by 1.6). We leave the points from \citeauthor{erb06a} uncorrected, 
since they used the integral of the SFR over the lifetime of the galaxy (thus the total stellar mass ever formed which is not equal to stellar mass due to gas recycling) as stellar mass, 
and the correction to the current stellar mass ($\sim$ 10\% -- 40\%) depends on the star-formation history of each galaxy. Correction to the current stellar mass and the conversion from a Chabrier IMF \citep{chabrier03} that they assumed to a Salpeter IMF will yield a combined systematic offset of 0.05 -- 0.15 dex to the right. 
  All points on this figure have their metallicities derived via the N2 index.  

As can be seen in Figure \ref{fig:mzr}, our observations are not deep enough to probe the low-mass end ($M_{*} \lesssim$ 10$^9$ M\sol) of the mass-metallicity relation since the true metallicities of these low-mass LAEs can be any value below these (relatively high) upper limits.  However, although our results only provide upper limits, 
for galaxies with $M_{*} \gtrsim$ 10$^9$ M\sol, nearly all of them lie either on or below the MZR for  continuum-selected star-forming galaxies. As these are 1$\sigma$ upper limits, each galaxy has a 84\% chance of lying below the currently drawn data point.
Thus, the trend observed by deep Keck/NIRSPEC data for LAEs with $M_* \gtrsim$ 10$^9$ M\sol\ implies that our sample of LAEs may have a systematically lower metallicity than continuum-selected star-forming galaxies at a common stellar mass.

Among our sample, HPS\,194 is of particular interest since its location in Figure \ref{fig:mzr} implies that it is less chemically-enriched by at least a factor of 4 than the typical continuum-selected SFGs with the same stellar mass and redshift. As noted earlier, however, the interpretation is complicated since this object consists of two components. While the spectroscopic redshift for each component is unknown, deep CANDELS \textit{HST}/WFC3 imaging (Figure \ref{fig:stamp}) showing a tidal bridge connecting them, together with SED fitting analysis for each component (described below and in Section \ref{sec:sed}), suggest that this system is likely a merger. If the observed optical nebular lines  (which determines the metallicity) of HPS\,194 originated from one or the other, the stellar mass for HPS\,194 is likely to be overesimated, leading to a misplacement of HPS\,194 in Figure \ref{fig:mzr}. Therefore, we estimated stellar mass for each component (see Section \ref{sec:sed}) and placed them as blue filled stars in Figure \ref{fig:mzr}. 
Our analysis shows both remain lying below the MZR, and the trend seen above persists.

This result is not terribly surprising, as LAEs were originally thought to be young and metal-poor systems \citep{partridge67}.  A number of studies of LAEs support this; while some appear moderately dusty (e.g., \citealt{fin09a,pentericci09}), the majority are relatively blue, and thus likely have minimal dust attenuation, and by extension, lower metallicities. Thus, the majority of studied LAEs appeared less evolved than continuum-selected galaxies at the same redshifts. However, typical narrowband-selected LAE studies probe lower masses; $M_* \lesssim$ 10$^9$ M\sol, thus these previous comparisons were comparing two galaxy samples selected in different ways, in different mass regimes. Due to the large volume probed with the HETDEX Pilot Survey, we have been able to compile a sample of LAEs with comparable masses to continuum-selected star-forming galaxies, and while their masses are the same, our results imply that their metallicities may be systematically lower.  This could imply that LAEs reside on their own MZR, shifted downward in metallicity.  More likely, however, is that there is significant scatter in the high-redshift MZR, and that galaxies on the lower-metallicity end of that scatter have less dust, and thus are more likely to exhibit Ly$\alpha$ in emission.

In any case, evidence for different galaxy populations occupying different locations in the stellar mass -- metallicity plane is now emerging from other studies: locally, \citet{pilyugin13} recently reported that irregular SDSS galaxies characterized by their high sSFR form a different MZR in that they are more metal-poor than normal spirals for a given mass. 
\citet{ly14} also found suggestion that galaxies at 0.1 $< z <$ 0.9 selected by their strong emission lines populate the lower-side of the MZR, with a direct measurement of metallicity from the \OIII$\lambda$4363 auroral line. 
  A similar result was found by \citet{xia12} for emission-line selected galaxies at 0.6 $< z <$ 2.4. These galaxies share some characteristics in common with our low-mass LAEs (e.g., mass, SFR, age).  Finally, \citet{fin11a} studied the MZR of Ly$\alpha$ emitting galaxies at $z \sim$ 0.3, selected from {\it GALEX} spectroscopy, and found that they too resided below the MZR for SDSS galaxies at similar redshifts (see also \citealt{cowie10}).

This observed trend of LAEs being relatively more metal-poor than continuum-selected star-forming galaxies (Figure \ref{fig:mzr}) can, however, be affected by a number of systematic uncertainties.  First, it is known that the absolute metallicity derived from different calibrations can differ up to 0.7 dex \citep{kewley08}.  As we mentioned above, all points in Figure \ref{fig:mzr} are derived using the same metallicity calibration, the N2 index.  Therefore, there exist no systematics from using different calibrators. 

Second, as noted in Section \ref{sec:agn}, some studies indicate that high-redshift galaxies have a higher ionization parameter compared to the local ones and suggest caution when using the locally-calibrated metallicity indicators such as the N2 index used in this study (e.g., \citealt{kewley02}). If LAEs and continuum-selected star-forming galaxies have different physical conditions in their star-forming regions, this issue could vertically move points differently for two populations: 
since the N2 index increases with metallicity but decreases with ionization parameter, 
the inferred metallicity for an object with higher ionization parameter would be underestimated if a constant ionization parameter is assumed in calibration. This effect of high ionization parameters will be discussed further in Section \ref{sec:lae}.

Third, we focus on the fact that HPS\,194 was observed both with Keck/NIRSPEC and VLT/SINFONI, and the 1$\sigma$ upper limit for metallicity of this object from the Keck/NIRSPEC data is much lower, as shown in Figure \ref{fig:mzr}.  This is because the Keck/NIRPSEC spectrum of HPS\,194 is much deeper than the VLT/SINFONI spectrum, thus the superior SNR leads to a much more stringent limit on the \NII\ flux, and in turn on the metallicity.  Therefore, if the VLT/SINFONI spectra were deeper, we would expect the points to move lower (or \NII\ to be detected), resulting in a greater difference in metallicity between the two populations.  Clearly a more uniform, deeply observed sample of LAEs is required to make progress on this issue.

Since the establishment of the MZR by \citet{tremonti04}, observational efforts on investigating its dependence on a second parameter have found that for a given mass, a galaxy with a higher SFR has a lower metallicity (\citealt{mannucci10, lara10, yates12}; but see \citealt{sanchez13}). 
This result is in qualitative agreement with the expectation from theoretical models where galaxies are in an equilibrium state, with their metallicity set by gas inflows, star formation, and outflows \citep{dave11}.  Based on these data, \citet{mannucci10} proposed that the MZR is a 2D manifestation of the thin plane that galaxies form in stellar mass -- gas-phase metallicity -- SFR space.  This ``fundamental metallicity relation'' (FMR) is reported to be valid and show no evolution at least up to $z \sim$ 2.5 using continuum-selected star-forming galaxies.
Figure \ref{fig:fmr} shows our LAEs plotted with the originally proposed FMR \citep{mannucci10} and its extension toward lower mass \citep{mannucci11} converted into the common IMF of Salpeter and metallicity indicator of the N2 index.   
Unfortunately, our data are not deep enough to probe if the FMR is applicable to our population of LAEs, suggesting the need for deeper spectroscopy in the future.

\begin{figure*}[t]
\centering
\epsscale{1.1}
\plotone{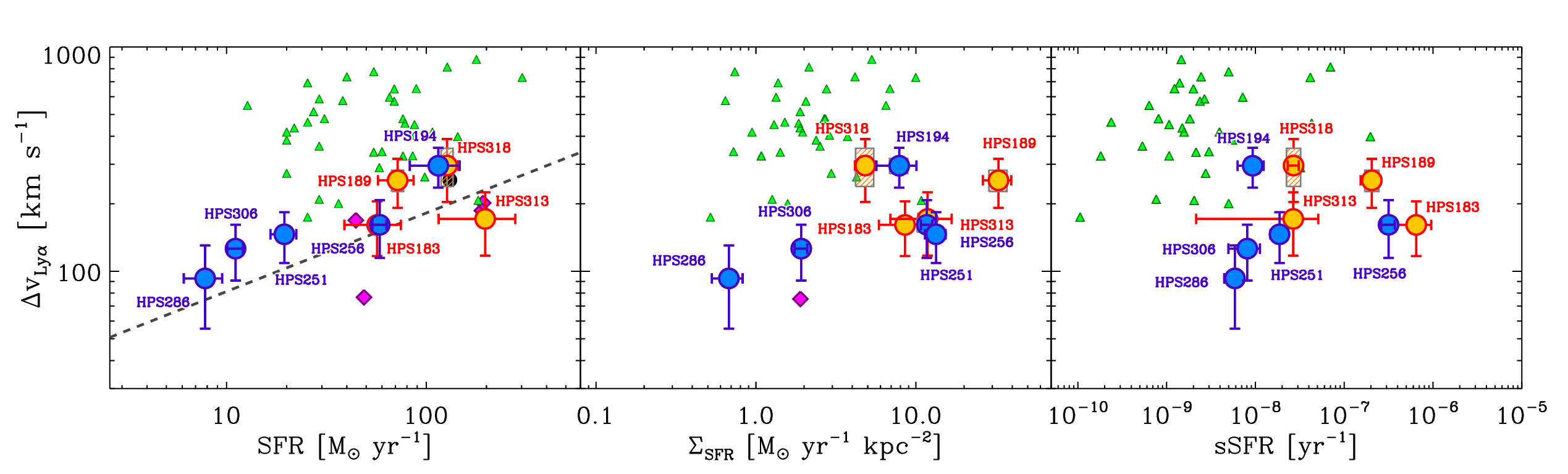}
\plotone{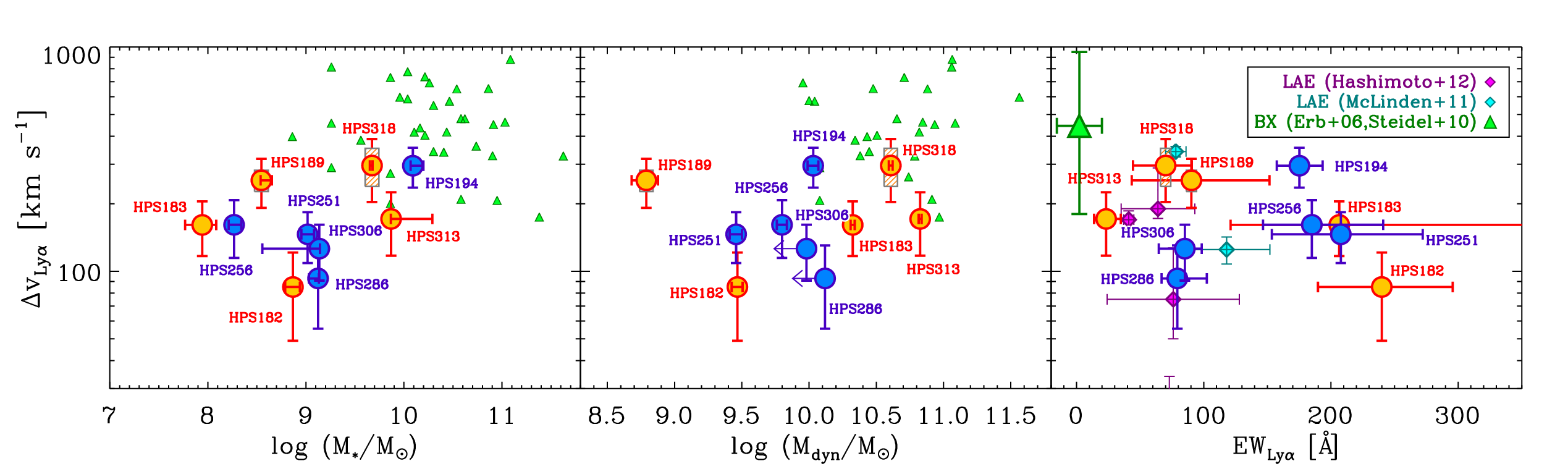}
\epsscale{1.0}
  \caption{\label{fig:dv_vs_prop}
 Ly$\alpha$ velocity offsets versus various physical properties. From top-left to bottom-right, $\Delta v_{\rm Ly\alpha}$ versus SFR, SFR surface density, specific SFR, stellar mass, dynamical mass, and rest-frame Ly$\alpha$ EW.
 The vertical boxes on orange/blue filled circles represent the systematic errors of Ly$\alpha$ velocity offsets for the 4 objects detected in H$\alpha$ and \OIII, whereas the vertical bars illustrate 
 the 1$\sigma$ photometric errors (see Section \ref{sec:dv} for details). For comparison, $z \sim$ 2.2 BX galaxies from \citet{erb06b, erb06c} and \citet{steidel10} are plotted as green triangles, corrected to a \citet{salpeter55} IMF. The error bar of the green triangle in the lower-right panel (Ly$\alpha$ velocity offsets vs. EW$_{\rm Ly\alpha}$) represents the typical EW$_{\rm Ly\alpha} (< 20$ \AA) range and the range of Ly$\alpha$ velocity offsets in BX sample from \citet{steidel10}. The grey dashed line in the upper left panel is the local trend in outflow velocity ($\Delta v_{\rm Na{\sc I}}$) vs. SFR from \citet{martin05}.
 In general, there exist moderate correlations of Ly$\alpha$ velocity offsets with SFR ($r_s=$ 0.90), but no correlation with $\Sigma_{\rm SFR}$, sSFR, $M_*$, $M_{\rm dyn}$, or EW$_{\rm Ly\alpha}$ ($r_s=$ 0.30, 0.42, 0.48, 0.38, -0.43). 
}
\end{figure*}

\subsection{Ly$\alpha$ Velocity Offset vs. Physical Properties}\label{sec:dv_vs_prop}

Outflows are believed to be ubiquitous both in local and high-redshift galaxies (e.g., \citealt{pettini01, shapley03,martin05}), as directly probed by neutral gas in outflows (e.g., \citealt{martin05}) or indirectly implied by studies on metal-line absorption systems along the line-of-sight of high-redshift quasars or galaxies showing metal-enriched intergalactic/intercluster medium, indicating metals must be transported in large scale by galactic winds (e.g., \citealt{ellison00,adelberger03,simcoe06}).
Although both observations and theories indicate that outflows are comprised of multiple velocity components (e.g., \citealt{heckman90,pettini02}), in observational studies, especially of high-redshift galaxies, the outflow is often assumed to have a single velocity component, for example an expanding shell model \citep{verhamme06,schaerer11}, for simplicity and due to the limited spectral resolution and SNR achievable (but see \citealt{steidel10, barnes10} for different outflow models with velocity gradients).
This outflow velocity can be traced by UV interstellar (IS) absorption lines, which are observed to be blueshifted by a few hundred km s$^{-1}$ relative to the systemic velocity \citep{shapley03}.  Blueshifted absorption lines are an unambiguous signature of outflows, probing the material in front of a continuum source moving towards us.

Arguably, stronger outflows are predicted in galaxies with more intense star formation \citep{leitherer95, veilleux05, murray11}, as it is assumed that the primary energy source behind the outflows is supernovae-driven winds.  Indeed, some observational studies found a positive correlation between the outflow velocity traced by interstellar absorption lines and the SFR (e.g., \citealt{martin05}), as well as with other physical parameters, such as dynamical mass from local starbursts \citep{martin05,rupke05}.  This correlation is also observed at higher redshift, in continuum-selected star-forming galaxies at intermediate ($z \sim$ 1) redshift \citep{weiner09,bradshaw13}, although \citet{kornei12} found no correlation with SFR but with SFR surface density in star-forming galaxies at similar redshifts.  At $z \sim$ 2, \citet{law12} reported a correlation between outflow velocity and SFR surface density, while \citet{steidel10} found that for continuum-selected star-forming galaxies there exists no significant correlation between outflow velocity and SFR or SFR surface density, but a negative correlation is seen between outflow velocity and dynamical mass.

For faint objects such as high-redshift LAEs, the interstellar absorption lines are extremely difficult to measure since high a SNR continuum is often impractical to obtain, so the Ly$\alpha$ line is often the only tracer for ISM kinematics.  
In the simple scenario of an expanding shell, Ly$\alpha$ should also probe the outflow velocity, as Ly$\alpha$ photons will preferentially escape towards the observer after gaining additional redshift backscattering off of the receding edge of the shell.  
However, the interpretation is not straightforward: Ly$\alpha$ photons are resonantly scattered in spatial and frequency space by neutral hydrogen and thus suffer from selective absorption by dust. Consequently, it is not clear if the Ly$\alpha$ velocity offset (discussed in Section \ref{sec:lya_escape}) 
correlates with outflow velocity, since the radiative transfer and emergent spectrum of Ly$\alpha$ involves many parameters, such as \HI\ column density, Doppler parameter, dust, ISM kinematics, and geometry.  Moreoever, the multiple peaks of the Ly$\alpha$ emission expected from an expanding shell model \citep{verhamme06} are often blended under low instrumental resolution \citep{chonis13}.

Nevertheless, studies on continuum-selected star-forming galaxies at high redshift have often found both blueshifted interstellar absorption lines and redshifted Ly$\alpha$ emission compared to the systemic redshift (e.g., \citealt{steidel10}), suggesting there may be a connection between the  Ly$\alpha$ velocity offset and the outflow velocity. 
The outflow shell model also predicts that for a neutral column density of N(\HI) $\gtrsim$ 10$^{20}$ cm$^{-2}$, the primary peak of the Ly$\alpha$ emission will be redshifted by twice the outflow velocity: this shift is produced by photons backscattering from the receding side of the shell \citep{verhamme06, schaerer11}. For low column density, this is no longer true due to blending with redward emerging photons that are scattered through the front of the shell, which become more dominant than the backscattered photons \citep{verhamme06,chonis13}.

If the Ly$\alpha$ velocity offsets of LAEs are indeed related to outflow velocities, we would expect them to correlate with other physical properties. In Figure \ref{fig:dv_vs_prop}, from top-left to bottom-right, we compare the Ly$\alpha$ velocity offset ($\Delta v_{\rm Ly\alpha}$) with a number of quantities defined in the previous section: the star formation rate (SFR), star formation rate surface density ($\Sigma_{\rm SFR}$), specific star formation rate (sSFR), stellar mass ($M_*$), dynamical mass ($M_{\rm dyn}$), and rest-frame Ly$\alpha$ EW (EW$_{\rm Ly\alpha}$).  
Overall, there appears to be a correlation between $\Delta v_{\rm Ly\alpha}$ and SFR, but no clear correlation is seen in $\Delta v_{\rm Ly\alpha}$ versus $\Sigma_{\rm SFR}$, sSFR, $M_*$, $M_{\rm dyn}$, or EW$_{\rm Ly\alpha}$.  Spearman's rank correlation coefficients, a statistical measure of the strength of a monotonic correlation beween two variables, for these physical properties and Ly$\alpha$ offsets are 
$r_s=$ 0.90, 0.30, 0.42, 0.48, 0.38, -0.43 (a significance of 2.5, 0.8, 1.2, 1.5, 1.0, and 1.3-$\sigma$ 
from the null hypothesis of no correlation), respectively, suggesting tenuous ($\sim$2.5$\sigma$) monotonic correlations between $\Delta v_{\rm Ly\alpha}$ and SFR, but no significant correlations between $\Delta v_{\rm Ly\alpha}$ and $\Sigma_{\rm SFR}$, sSFR, $M_*$, $M_{\rm dyn}$, or EW$_{\rm Ly\alpha}$.

Noticeable differences between our LAEs and continuum-selected star-forming galaxies are seen in Figure \ref{fig:dv_vs_prop}.  Compared to our sample, these continuum-selected star-forming galaxies have a similar SFR, but lower SFR surface density and sSFR and higher stellar mass and dynamical mass. As discussed in Section \ref{sec:dv}, continuum-selected star-forming galaxies have higher Ly$\alpha$ velocity offsets than our LAEs for a given physical parameter.

In Figure \ref{fig:dv_vs_prop}, the bottom-right panel shows Ly$\alpha$ equivalent widths versus Ly$\alpha$ velocity offsets, a compilation of our data with other studies at $z$=2--3 (4 LAEs from \citealt{hashimoto13}; 2 LAEs from \citealt{mclinden11}; BX galaxies from \citealt{steidel10}).  Qualitatively, the trend between Ly$\alpha$ velocity offsets and physical properties is consistent with the picture that Ly$\alpha$ velocity offsets are related with outflow velocities.  We find, however, no positive correlation (but a marginal anti-correlation) between Ly$\alpha$ equivalent widths and Ly$\alpha$ velocity offsets, contrary to what is expected in the expanding shell model in which outflow velocity aids the escape of Ly$\alpha$ photons (e.g., \citealt{verhamme06, verhamme08, schaerer11}).

What are these Ly$\alpha$ velocity offsets tracing, and why do LAEs display smaller Ly$\alpha$ velocity offsets than those of continuum-selected star-forming galaxies at similar redshifts?  Recently \citet{hashimoto13} reported that the outflow velocity traced by blueshifted interstellar absorption lines ($\Delta v_{\rm IS ~(LAE)} = -180$ km s$^{-1}$) in the composite FUV spectrum of their sample of four LAEs at $z \sim$ 2.2 is comparable to that of continuum-selected star-forming galaxies ($\Delta v_{\rm IS ~(BX)}\sim -150$ km s$^{-1}$) as well as the mean value of redshifted Ly$\alpha$ velocity offsets ($\Delta v_{\rm Ly\alpha ~(LAE, BX)} = +180$ km s$^{-1}$), although they do not find any significant correlations between physical parameters and Ly$\alpha$ velocity offsets, possibly due to their small sample size of 4 objects.\footnotemark \footnotetext{\citet{berry12} report a $\sim$ 600 km s$^{-1}$ velocity offset between Ly$\alpha$ and the low-ionization state absorption lines for UV-bright LAE composite at $z \sim$ 2, which is larger than that reported in \citeauthor{hashimoto13} The reason for this difference is uncertain, although EW(Ly$\alpha$) might be a factor as the LAE sample of \citeauthor{berry12} has a low EW(Ly$\alpha$) (30 $\pm$ 6 \AA), while \citeauthor{hashimoto13}'s sample has a larger EW(Ly$\alpha$) of 75 $\pm$ 9 \AA.}
The latter is in contrast to what is observed in continuum-selected star-forming galaxies which display $| \Delta v_{\rm Ly\alpha}| \sim 2-3 \times |\Delta v_{\rm IS}|$.  They also report that Ly$\alpha$ EWs decrease with increasing Ly$\alpha$ velocity offsets.  Combining these results with numerical modeling of Ly$\alpha$ radiative transfer of \citet{verhamme06}, they claim that LAEs have low H\,{\sc i} column densities.  Intriguingly, an independent study by \citet{chonis13} on three LAEs among our sample modeling high resolution Ly$\alpha$ line profiles with the expanding shell model finds that all of the best-fit expanding shell models have low column densities (N(\HI) $\gtrsim$ 10$^{18}$ cm$^{-2}$), although the failure at reproducing the shape of the observed Ly$\alpha$ profile suggests the limitation of the simplified expanding shell model.  Taken at face value, Ly$\alpha$ velocity offsets in LAEs trace the outflow velocity within a factor of two, at least for the sample of \citet{hashimoto13}. If we assume Ly$\alpha$ velocity offsets are approximate to the outflow velocity, $\Delta v_{\rm Ly\alpha}$ versus SFR in our sample is consistent with local trend seen in \citet{martin05}, which is shown as a grey dashed line in the upper left panel of Figure \ref{fig:dv_vs_prop}.

 \begin{figure}[t]
  \epsscale{1.2}
 \plotone{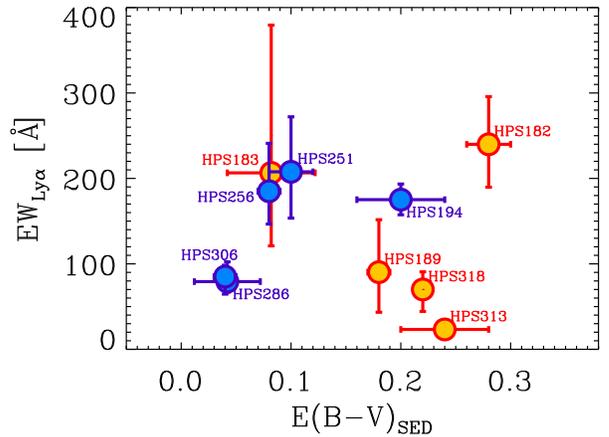}
  \caption{\label{fig:ew_vs_ebv}
  Rest-frame Ly$\alpha$ equivalent widths vs. $E(B-V)$.
 }
 \end{figure}

\subsection{The Role of Dust, ISM Geometry and Kinematics on Ly$\alpha$ Escape}\label{sec:lya_escape}

Dust likely plays an important role in regulating the escape of Ly$\alpha$ photons. In a homogeneous static medium where dust and gas are well mixed, the long path lengths of Ly$\alpha$ photons due to their resonant nature allow dust to effectively quench Ly$\alpha$ emission.  Therefore, the Ly$\alpha$ flux should drop significantly with increasing dust content \citep{charlot93}.  
However, in Figure \ref{fig:ew_vs_ebv} our measured Ly$\alpha$ rest-frame EWs show no clear correlation with $E(B-V)$, in agreement with the results of \citet{blanc11} who found no correlation using the whole LAE sample from the HETDEX Pilot Survey. 
The absence of an anti-correlation suggests the existence of other factors governing Ly$\alpha$ escape. 

In addition to kinematics, the geometry of the ISM likely also influences the escape of Ly$\alpha$.  If the ISMs in these galaxies were clumpy, with the dust confined in the high density regions (i.e., \HI\ clouds), Ly$\alpha$ would freely travel the optically-thin inter-clump medium and resonantly scatter off of the clump surfaces, while non-resonant photons (e.g., UV continuum, H$\alpha$ photons) must penetrate through the clouds \citep{neufeld91}. In this scenario, only the continuum (and non-resonant line) photons suffer dust attenuation, while Ly$\alpha$ photons 
bounce through the inter-clump medium and eventually escape the galaxy seeing no (or little) dust.  This scenario would result in a Ly$\alpha$ EW which is enhanced over the value intrinsic to the star-forming regions.  In this idealized case, the Ly$\alpha$ EW should positively correlate with dust content.  The realistic case would probably be something in between these two cases in which the inter-clump medium is not entirely optically thin for Ly$\alpha$ photons.  In this case, the clumpy ISM would reduce the effect of dust on the decrease of the Ly$\alpha$ EW (with the reduction depending on the clumpiness).  We investigate this effect by parameterizing the clumpiness of the ISM as $q \equiv \tau_{\rm Ly\alpha}/\tau_{c}$, following the definition of \citet{fin08}. Here $\tau_{\rm Ly\alpha}$ and $\tau_{c}$ are the optical depth due to dust seen by Ly$\alpha$ and continuum photons, respectively. The case $q=0$ is the idealized clumpy ISM, while $q=\infty$ represents a homogeneous (dusty) medium.  This model was first observationally studied by \citet{fin08, fin09a}, who, through including Ly$\alpha$ emission in their SED fitting process, found that $q \approx$ 1 was consistent with their observations.

\begin{figure}[t]
  \epsscale{1.2}
 \plotone{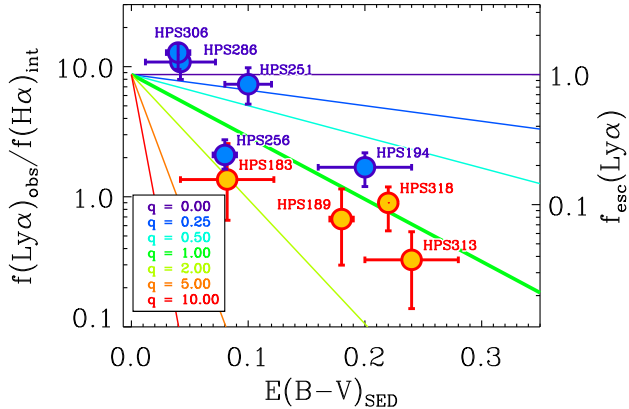}
  \caption{\label{fig:fesc_vs_ebv}
 Observed Ly$\alpha$ flux-to-intrinsic H$\alpha$ flux ratio (or Ly$\alpha$ escape fraction) vs. inferred dust reddening. Solid lines represent the ISM clumpiness parameter, $q \equiv \tau_{\rm Ly\alpha}/\tau_{c}$, where $q=0$ indicates the idealized clumpy medium while $q=\infty$ imply a homogeneous medium. A majority of our LAEs locate near $q \sim$ 1 line, suggesting no selective attenuation or enhancement of Ly$\alpha$ photons compared to non-resonant (e.g., continuum or H$\alpha$) photons.
 }
 \end{figure}

Observationally, we can test this scenario by examining how the ratio of Ly$\alpha$ to a non-resonant line, such as H$\alpha$, varies with dust attenuation.  Figure \ref{fig:fesc_vs_ebv} shows the Ly$\alpha$/H$\alpha$ ratio versus the SED-fitting-derived $E(B-V)$ with lines of constant $q$ values overplotted.  The $y$-axis on the right side indicates the corresponding escape fraction of Ly$\alpha$, $f_{\rm esc}$(Ly$\alpha$), which scales with the observed Ly$\alpha$ and H$\alpha$ flux ratio with the assumption of Case B recombination, as \begin{eqnarray}
  f_{\rm esc}(\rm Ly\alpha) &=& \frac{L_{\rm obs}(\rm Ly\alpha)}{L_{\rm int}(\rm Ly\alpha)} = \frac{L_{\rm obs}(\rm Ly\alpha)}{8.7 L_{\rm int}(\rm H\alpha)} \nonumber\\
  &=& \frac{L_{\rm obs}(\rm Ly\alpha)}{8.7 L_{\rm obs}({\rm H\alpha}) 10^{0.4 E(B-V)_{\rm neb}k(\lambda_{\rm H\alpha})}} \end{eqnarray}
The median Ly$\alpha$ escape fraction is $\sim$ 19\%, comparable with the median value of 29\% found by \citet{blanc11}.

Figure \ref{fig:fesc_vs_ebv} has two interesting features: first, we see an anti-correlation between the escape fraction of Ly$\alpha$ and $E(B-V)$, although the escape fraction of Ly$\alpha$ for dusty LAEs is non-zero and still significant. This suggests that the emergent Ly$\alpha$ emission from our sample cannot be explained via a homogeneous ISM.  Second, Figure \ref{fig:fesc_vs_ebv} implies a $q$ value of around unity for the majority of our LAEs.  Considering that $q$ traces the effective dust opacity seen by Ly$\alpha$ photons compared to non-resonant photons, this implies that while we cannot break the degeneracy between the role of ISM geometry and kinematics on the escape of Ly$\alpha$ photons, either of the two, or both, effectively reduce the attenuation Ly$\alpha$ would otherwise experience by dust, as the Ly$\alpha$ and non-resonant photons appear to suffer a similar extinction.  Our results are in general agreement with those of \citet{blanc11}, who conducted a similar analysis using the UV continuum to infer SFR on a larger sample of LAEs from the HETDEX Pilot Survey \citep{adams11}. 

Figure \ref{fig:fesc_vs_ebv} also shows that two LAEs, HPS\,286 and HPS\,306, have the Ly$\alpha$/H$\alpha$ ratio above the theorical value for Case B recombination, yet the discrepancy is not significant: HPS\,286 is consistent with $f_{\rm esc}$=1 within the error bar, and HPS\,306 has deviation from $f_{\rm esc}$=1 at 1.3$\sigma$. Although explanations for these objects are not trivial, one possibility is that the underlying stellar Ly$\alpha$ and H$\alpha$ absorption is significant and our absorption-uncorrected Ly$\alpha$/H$\alpha$ ratio might be overestimated. Indeed, the inferred ages of these 2 LAEs from our SED fitting analysis are $>$ 10$^7$ Myr, at which both the Ly$\alpha$ and Balmer absorption become  non-negligible (see Figure 3 in \citealt{schaerer08, gonzalez99}). 
However, since our LAEs have large EWs of Ly$\alpha$ ($>$ 70 \AA) and H$\alpha$ ($>$ 100 \AA), the effect of stellar absorption on Ly$\alpha$/H$\alpha$ ratio may not be sufficient to explain the observed deviation.

Another possibility is that the origin of the Ly$\alpha$ emission for these objects is not associated with star formation but (or additionally) with other mechanisms, such as collisional excitation by shocks from supernova-driven winds (e.g., \citealt{taniguchi00}) or cooling radiation by gravitational heating (e.g., \citealt{haiman00, dijkstra06, dijkstra09,faucher10}), which have been suggested to explain the origin of Lyman-alpha blobs. As the morphologies of these objects suggest, interaction/merger induced shocks is also a possible origin.

Another explanation could be that the geometry and velocity field of the ISM is more complicated than the assumed model in which isotropic Ly$\alpha$ emission is expected, in such a way that the H$\alpha$ emission is preferentially attenuated along the line-of-sight (e.g., by dusty clumps in between the source of H$\alpha$ emission and the observer) but the non-isotropic Ly$\alpha$ emission is preferentially scattered into the line-of-sight, resulting in enhanced Ly$\alpha$ flux. Among our original targets, those with no H$\alpha$ detection are candicates of a high Ly$\alpha$ escape fraction, and even with $f_{\rm esc} >$ 1, if not all of them are dusty. Their observed Ly$\alpha$ fluxes and the typical H$\alpha$ detection limit of our observations suggest a lower limit of Ly$\alpha$ escape fraction to be $\langle f_{\rm esc} \rangle$= 0.72 $\pm$ 0.48 at $E(B-V)$=0.15 (the mean value for the H$\alpha$-detected sample), with two (HPS\,145 and HPS\,419) with $f_{\rm esc}$ above unity.

\subsection{Nature of LAEs}\label{sec:lae}

In the previous section we showed that the effect of kinematics and ISM geometry on Ly$\alpha$ escape in LAEs may not be dramatically different from that in continuum-selected star-forming galaxies, as the low column density might explain the systemically small Ly$\alpha$ velocity offsets observed in LAEs.  

In Section \ref{sec:dv_vs_prop}, we compared a wide variety of physical properties between our observed LAEs, and continuum-selected star-forming galaxies at the same redshift.  Restricting our comparison to galaxies at log~($M_*$/M\sol) $>$ 9, where our NIR spectra have sufficient depth, we find that these two populations are remarkably similar in a number of physical properties, including stellar mass (set by our definition of this comparison), but also gas-mass fraction, dynamical mass, and SFR.  However, there are also several key differences.  First, due to their selection, LAEs have larger Ly$\alpha$ EWs.  We also find that LAEs have lower Ly$\alpha$ velocity offsets, and, perhaps most importantly, LAEs are likely to have lower metallicities.

One suggestion, recently proposed by \citet{hashimoto13} is that the galaxies we select as LAEs have lower H\,{\sc i} column densities than their continuum-selected brethren.  This suggestion is consistent with an earlier study of \citet{schaerer08}, who argue, based on Ly$\alpha$ transfer modeling on LBGs, that LBGs are not a different population of LAEs but are intrinsically LAEs, and the column density and dust content (which is proportional to column density) is mainly responsible for the observed Ly$\alpha$ diversity.  The lower column density is consistent with observations in three ways.  First, it will result in less spectral diffusion by Ly$\alpha$ photons before escape, resulting in lower Ly$\alpha$ velocity offsets.  Second, Ly$\alpha$ photons have less gas to traverse before escape, reducing the chance of absorption by any dust that is present.  Finally, UV photons will likewise undergo less dust attenuation, resulting in a lower measured dust reddening for these galaxies. However, column density cannot be the only difference, as our results suggest that metallicity might play an important role in the escape of Ly$\alpha$ photons.

 \begin{figure}
  \epsscale{1.1}
 \plotone{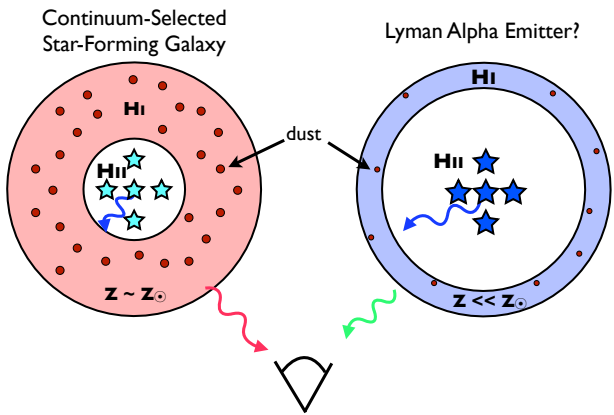}
  \caption{\label{fig:laefig}
 A schematic of our toy model for the differences between LAEs and continuum-selected galaxies.  
 Each galaxy has a similar amount of star formation occurring, but the LAE began with lower metallicity gas, resulting in hotter stars, 
 and thus a larger ionized region.  The neutral hydrogen column density between the observer and the star-forming region is thus less for the LAE, 
 such that Ly$\alpha$ photons traversing through the H\,{\sc i} suffer less resonant scattering (and have a lower probability of dust absorption), 
 and thus escape closer to line center.  Additionally, the lower metallicity galaxy will create less dust, with the higher gas-to-dust ratio further reducing the attenuation experienced by Ly$\alpha$ photons.
 }
 \end{figure}

A simple toy model may be able to explain these similarities and differences between Ly$\alpha$ and continuum-selected galaxies, with metallicity as the key parameter.  If one takes two collapsing objects, each in a similar mass dark-matter halo and with a similar amount of gas, but one has $\sim$ solar metallicity, and one has $\sim$1--10\% $Z$\sol, we may recreate the observed population (Figure \ref{fig:laefig}).  Star formation will proceed in both objects, but in the object with lower metallicity, the ensuing massive stars will have hotter stellar photospheres than their solar-metallicity counterparts \citep{tumlinson00}, and thus will produce more ionizing photons per unit SFR \citep{sternberg03, leitherer10}.  
The enhanced rate of ionizing photon production will in turn create larger \HII\ regions, reducing the neutral gas column density. Low metallicity environments also imply less dust, as well as more effective dust destruction by sublimation or evaporation. 
As a result, Ly$\alpha$ photons suffer less scattering and chance of absorption by dust in the lower-metallicity object, and the probability of Ly$\alpha$ escape can be enhanced.

However, we caution that this is simply a toy model which hides a number of uncertainties.  First, our metallicities, while lying below those of continuum-selected star-forming galaxies at the same mass, are still only 1$\sigma$ upper limits.  Clearly deeper spectroscopy is needed to measure the true metallicities for these systems, or, at the least, to push their metallicity limits much lower.  
Assuming the mean H$\alpha$ flux for our sample and an undetected \NII\ line, 
an emission line flux limit of $2 \times 10^{-18}$ erg s$^{-1}$ cm$^{-2}$ will place LAEs with $M_* >10^8$ M\sol\ below the MZR for continuum-selected galaxies (at 5$\sigma$), which is achievable with $\sim$ 6 hours of integration with MOSFIRE.

Additionally, the metallicities of continuum-selected galaxies that we compare against are based on stacks of multiple objects, thus there is significant uncertainty in the scatter of the relation.  Second, if LAEs do in fact have much lower metallicities, the hotter stars mentioned above will likely result in a higher ionization parameter.  As we discussed earlier, the metallicity calibrations can be affected by varying ioniziation parameters, necessitating more detailed photoionization modeling to measure an accurate metallicity for a given system.  \citet{nakajima13} report an elevated ionization parameter for LAEs.  As shown in their Figure 7, one of their two measured LAEs does appear to have an ionization parameter higher than  continuum-selected galaxies at the same redshift.  However, the second LAE has a similar ionization parameter to the comparison sample.  Thus, observationally, it is difficult to make definite conclusions on the ionization parameter for LAEs.

Finally, the toy model does not address the role of interactions and/or mergers in the escape of Ly$\alpha$ photons.  Obviously, interactions/mergers or the effect of viewing angle cannot explain the origin of the entire population of LAEs, as suggested by clustering analysis on LAEs and continuum-selected galaxies: these studies show that they have different clustering properties \citep{gawiser07}, implying the two populations have different relations with the underlying dark matter halos, although comparing clustering strengths on sample of same mass range in a future study would be desirable for a clearer understanding of the relation between LAEs and other galaxy populations. 
Nonetheless, we note that especially for high mass galaxies in which column density is expected to be high, interactions/mergers can further increase the chance of the escape by opening a low opacity channel for Ly$\alpha$ \citep{cooke10,chonis13}. The high fraction of LAEs with disturbed morphology or nearby sources in \textit{HST} imaging in our high-mass sample ($M_* > 10^9$ M\sol)  might be suggestive of this, and the complex gas velocity fields and geometries resulting from interactions could be in part why \citet{chonis13} see discrepancies 
 between the simple expanding shell model and the spectrally resolved Ly$\alpha$ emission of their LAE sample with nearby continuum source.
 However, one of these systems (HPS\,194) has our lowest metallicity upper limit, thus interactions cannot be the only course of Ly$\alpha$ escape.

We conclude that the differences in metallicity and Ly$\alpha$ velocity offset between LAEs and continuum-selected star-forming galaxies at the same redshift and stellar mass might be 
key factors discerning  these two populations.
However, future deep near-infrared spectroscopy, allowing true metallicity measurements (preferably with multiple indicators) as well as probing the ionization parameter, are needed to further explore this scenario.


\section{Summary and Conclusions}\label{sec:conclusion}

We have presented results from Keck/NIRSPEC and VLT/SINFONI NIR spectroscopy for 16 LAEs at $z$ = 2.1--2.5 discovered from the HETDEX Pilot Survey \citep{adams11}.  Among these 16 LAEs selected by their bright Ly$\alpha$ emission ($f_{\rm Ly\alpha} >$ 10$^{-16}$ erg cm$^{-2}$ s$^{-1}$), we detect rest-frame optical nebular lines (H$\alpha$ and/or \OIII) for 10 LAEs, tripling the number of LAEs at $z$ = 2.1--2.5 known to date of which rest-frame optical nebular lines have been investigated. 

The main results from our analysis combining our NIR spectroscopic data with Ly$\alpha$ and ancillary imaging data 
 can be summarized as follows:

\begin{itemize}

\item 
The inferred stellar masses of HPS LAEs show a wide range from log($M_*$/M\sol) = 7.9 to log($M_*$/M\sol) = 10.1.  Dust reddening $E(B-V)$ ranges from $E(B-V) =$ 0.04 to $E(B-V) =$ 0.28, with a median of $E(B-V) =$ 0.18.

\item We find an extra attenuation of a factor of 2 towards the \HII\ region yields greater consistency between SFR(UV) and SFR(H$\alpha$).  The extinction-corrected SFR ranges between SFR = 8 M\sol yr$^{-1}$ and SFR = 197 M\sol yr$^{-1}$, with a median of SFR = 58 M\sol yr$^{-1}$.  We find specific star formation rates of massive LAEs ($M_* \sim$ 10$^{10}$ M\sol) are similar to those of continuum-selected star-forming galaxies, and are consistent with the $z \sim$ 2 star-forming ``main-sequence'' \citep{daddi07}, while low-mass LAEs have sSFR as high as $6 \times 10^{-7}$ yr$^{-1}$.

\item 
The gas-mass fractions inferred from the inversion of the Kennicutt-Schmidt Law \citep{kennicutt98a} show a trend of decreasing gas-mass fraction with increasing stellar mass.  For a given mass and redshift, LAEs have comparable gas mass fraction to that of continuum-selected star-forming galaxies in \citet{erb06b}. However, this should be taken with caution as the uncertainty associated with indirect gas mass measurements is large, and the observed trend might be only the upper envelope of the entire distribution driven by the observational limits.

\item Dynamical masses inferred from nebular lines (H$\alpha$ or \OIII) range from log($M_{\rm dyn}$/M\sol)= 8.8 to 10.8, and are in overall agreement with our inferred total baryonic mass.

\item Combining the 1$\sigma$ upper limit on the \NII\ flux with the observed H$\alpha$ flux, we provide constraints on the gas-phase metallicities of our sample.  The location of our sample in the stellar mass -- gas-phase metallicity plane suggests that LAEs may lie below the mass-metallicity relation for continuum-selected star-forming galaxies at a given redshift. 

\item The Ly$\alpha$ line is observed to be redshifted with respect to the systemic redshift (as measured by the rest-frame optical nebular lines) for all LAEs in our sample, by $\langle v_{\rm Ly\alpha} - v_{\rm sys} \rangle \sim$ +180 km s$^{-1}$ (ranging between [+85, +296] km s$^{-1}$). However, this velocity offset is systematically lower than seen in continuum-selected star-forming galaxies at similar redshifts.  We find a moderate correlation of Ly$\alpha$ velocity offset with SFR but no clear correlation with SFR surface density, specific SFR, stellar mass, dynamical mass, or rest-frame Ly$\alpha$ EWs.  

\item We explore the contribution of ISM kinematics and geometry on the escape of Ly$\alpha$ photons.  Although there is no signature of selective attenuation of the continuum due to Ly$\alpha$-screening by a clumpy ISM, the effective dust extinction seen by Ly$\alpha$ photons is similar to that of continuum photons, implying that the ISM geometry is inhomogeneous.  This result implies that dust plays an important role. 
We do not find a correlation between Ly$\alpha$ velocity offsets or $E(B-V)$ and rest-frame Ly$\alpha$ EWs, as would be expected if outflows were the dominant factor regulating Ly$\alpha$ escape under the assumption that Ly$\alpha$ velocity offsets probe the outflow velocities.

\end{itemize}

Although outflows are surely present in our LAEs (given their ubiquity in star-forming galaxies at high redshift), the lack of correlation between Ly$\alpha$ velocity offsets and Ly$\alpha$ EW leads us to propose an alternative explanation for the difference between LAEs and continuum-selected star-forming galaxies.  The primary difference between LAEs and continuum-selected galaxies may be the metallicity -- lower metallicity gas will make hotter stellar photospheres, which will in turn ionize more of the gas in a given galaxy.  Thus, of two equally-sized galaxies, the one with an intrinsically lower metallicity will have a lower neutral hydrogen column density, enabling easier Ly$\alpha$ escape, as well as requiring less resonant scattering prior to escape, reducing the velocity offset of Ly$\alpha$. 

Of course, this work is limited by our small sample, and future studies on larger numbers of LAEs will be valuable. Starting next year, the full HETDEX survey will ultimately discover approximately one million LAEs at 1.9 $< z <$ 3.5 over 450 deg$^2$, providing a suitable sample for additional observations.  NIR observations utilizing the improved sensitivity and multiplexing of the new generation of near-infrared multi-object spectrographs such as MOSFIRE and KMOS will enable us to constrain the metallicity of LAEs down to the low-mass end of the MZR.  Meanwhile, high resolution optical spectroscopy can provide further insight into the origin of LAEs through the modeling of Ly$\alpha$ line profile and the analysis of interstellar absorption lines.\\

\noindent
\acknowledgements

We thank the anonymous referee for valuable comments. We would also like to thank J. Adams and E. Robinson for helpful discussions and comments that improved this paper. 
M.S. and S.L.F. acknowledge support from the University of Texas at Austin, the McDonald Observatory and NASA through a NASA Keck PI Data Award, administered  by the NASA Exoplanet Science Institute. 
This research is based on observations made with the European Southern Observatory Very Large Telescope, under program ID 086.A-0424(A) and 088.A-0154(A), and with the Keck Telescope 
from telescope time allocated to NASA through the agency's scientific partnership with the California Institute of Technology and the University of California. The Observatory was made possible by the financial support of the W. M. Keck Foundation. We recognize and acknowledge the cultural role and reverence that the summit of Mauna Kea has within the indigenous Hawaiian community. This work is also based in part on observations made with the NASA/ESA Hubble Space Telescope, obtained at the Space Telescope Science Institute, which is operated by the Association of Universities for Research in Astronomy, Inc., under NASA contract NAS 5-26555, as well as the Spitzer Space Telescope, which is operated by the Jet Propulsion Laboratory, California Institute of Technology under a contract with NASA. This research is supported by the National Science Foundation under grant AST-0926815.

~

{\it Facility:} \facility{HST (ACS, WFC3)}, \facility{VLT:Yepun (SINFONI)}, \facility{Keck:II (NIRSPEC)}, \facility{Subaru (Suprime-Cam)}, \facility{Spitzer (IRAC)}, \facility{ESO: VISTA}


\end{document}